\documentclass[twocolumn]{aastex631}

\usepackage{color}
\usepackage{amsmath}
\usepackage{float}

\begin{document}

\title{Properties of the Line-of-Sight Velocity Field in the Hot and X-ray Emitting Circumgalactic Medium of Nearby Simulated Disk Galaxies}

\correspondingauthor{John A. ZuHone}
\email{john.zuhone@cfa.harvard.edu}

\author[0000-0003-3175-2347]{John A. ZuHone}
\affiliation{
    Center for Astrophysics $|$ Harvard~\&~Smithsonian,
    60 Garden St. Cambridge, MA 02138, USA
}

\author[0000-0002-4962-0740]{Gerrit Schellenberger}
\affiliation{
    Center for Astrophysics $|$ Harvard~\&~Smithsonian,
    60 Garden St. Cambridge, MA 02138, USA
}

\author[0000-0003-4504-2557]{Anna Ogorza{\l}ek}
\affiliation{
    NASA Goddard Space Flight Center, Code 662, Greenbelt, MD 20771, USA
}
\affiliation{
    Department of Astronomy, University of Maryland, College Park, MD 20742, USA
}

\author[0000-0002-3391-2116]{Benjamin D. Oppenheimer}
\affiliation{
    University of Colorado, Boulder, 2000 Colorado Ave, Boulder, CO 80305, Boulder, CO 80305, USA
}

\author[0000-0002-7541-9565]{Jonathan Stern}
\affiliation{
    School of Physics \& Astronomy, Tel Aviv University, Tel Aviv 69978, Israel
}

\author[0000-0003-0573-7733]{\'Akos Bogd\'an}
\affiliation{
    Center for Astrophysics $|$ Harvard~\&~Smithsonian,
    60 Garden St. Cambridge, MA 02138, USA
}

\author[0000-0003-4983-0462]{Nhut Truong}
\affiliation{
    NASA Goddard Space Flight Center, Code 662, Greenbelt, MD 20771, USA
}
\affiliation{
    Center for Space Sciences and Technology, University of Maryland, Baltimore County, MD 21250, USA
}
\affiliation{
    Max-Planck-Institut f\"ur Astronomie, K\"onigstuhl 17, D-69117 Heidelberg, Germany
}

\author[0000-0003-0144-4052]{Maxim Markevitch}
\affiliation{
    NASA Goddard Space Flight Center, Code 662, Greenbelt, MD 20771, USA
}

\author[0000-0003-1065-9274]{Annalisa Pillepich}
\affiliation{
    Max-Planck-Institut f\"ur Astronomie, K\"onigstuhl 17, D-69117 Heidelberg, Germany
}

\author[0000-0001-8421-5890]{Dylan Nelson}
\affiliation{
    Universit\"at Heidelberg, Zentrum f\"ur Astronomie, Institut f\"ur Theoretische Astrophysik, 69120 Heidelberg, Germany
}

\author[0000-0002-1979-2197]{Joseph N. Burchett}
\affiliation{
    New Mexico State University, Department of Astronomy, Las Cruces, NM 88001, USA
}

\author[0000-0003-3701-5882]{Ildar Khabibullin}
\affiliation{
    Universit\"ats-Sternwarte, Fakult\"at f\"ur Physik, Ludwig-Maximilians-Universit\"at M\"unchen, Scheinerstr.1, 81679 München, Germany
}
\affiliation{
    Space Research Institute (IKI), Profsoyuznaya 84/32, Moscow 117997, Russia
}
\affiliation{
    Max Planck Institute for Astrophysics, Karl-Schwarzschild-Str. 1, D-85741 Garching, Germany
}

\author[0000-0001-9464-4103]{Caroline A. Kilbourne}
\affiliation{
    NASA Goddard Space Flight Center, Code 662, Greenbelt, MD 20771, USA
}

\author[0000-0002-0765-0511]{Ralph P. Kraft}
\affiliation{
    Center for Astrophysics $|$ Harvard~\&~Smithsonian,
    60 Garden St. Cambridge, MA 02138, USA
}

\author[0000-0003-0297-4493]{Paul E. J. Nulsen}
\affiliation{
    Center for Astrophysics $|$ Harvard~\&~Smithsonian,
    60 Garden St. Cambridge, MA 02138, USA
}
\affiliation{
    ICRAR, University of Western Australia, 35 Stirling Hwy, Crawley, WA 6009, Australia
}

\author[0000-0002-3158-6820]{Sylvain Veilleux}
\affiliation{
    Department of Astronomy, University of Maryland, College Park, MD 20742, USA
}

\author[0000-0001-8593-7692]{Mark Vogelsberger}
\affiliation{
    Department of Physics, Massachusetts Institute of Technology, Cambridge, MA 02139, USA
}

\author[0000-0002-9279-4041]{Q. Daniel Wang}
\affiliation{
    University of Massachusetts Amherst, Amherst, MA 01003, USA
}

\author[0000-0001-7630-8085]{Irina Zhuravleva}
\affiliation{
    Department of Astronomy \& Astrophysics
    University of Chicago, Chicago, IL 60637, USA
}

\begin{abstract}
The hot, X-ray-emitting phase of the circumgalactic medium of massive galaxies is believed to be the reservoir of baryons from which gas flows onto the central galaxy and into which feedback from AGN and stars inject mass, momentum, energy, and metals. These effects shape the velocity fields of the hot gas, which can be observed via the Doppler shifting and broadening of emission lines by X-ray IFUs. In this work, we analyze the gas kinematics of the hot circumgalactic medium of Milky Way-mass disk galaxies from the TNG50 simulation with synthetic observations to determine how future instruments can probe this velocity structure. We find that the hot phase is often characterized by outflows from the disk driven by feedback processes, radial inflows near the galactic plane, and rotation, though in some systems the velocity field is more disorganized and turbulent. With a spectral resolution of $\sim$1~eV, fast and hot outflows ($\sim$200-500~km~s$^{-1}$) can be measured, depending on the orientation of the galaxy on the sky. The rotation velocity of the hot phase ($\sim$100-200~km~s$^{-1}$) can be measured using line shifts in edge-on galaxies, and is slower than that of colder gas phases but similar to stellar rotation velocities. By contrast, the slow inflows ($\sim$50-100~km~s$^{-1}$) are difficult to measure in projection with these other components, but may be detected in multi-component spectral fits. We find that the velocity measured is sensitive to which emission lines are used. Measuring these flows will constrain theories of how the gas in these galaxies evolves.
\end{abstract}

\section{Introduction}\label{sec:intro}

The circumgalactic medium (CGM) is the gas within the dark matter (DM) halos of galaxies outside of the central galaxy and extending out to the virial radius of the halo. It is believed to be the reservoir of gas which has been populated by inflows from the intergalactic medium (IGM) which has condensed into galactic halos \citep{Tumlinson2017}. As star formation, evolution, and death enrich and transform the ISM gas, feedback from supernovae and active galactic nuclei (AGN) inject mass, momentum, energy, and metals back into the CGM \citep{Rupke2019,Burchett2021}, which alters its thermodynamic, kinematic, and chemical properties. Ultimately, these processes regulate the growth and quenching of galaxies, making the CGM one of the primary drivers of galaxy evolution as a whole. 

The CGM is multi-phase, and the cool ($T \lesssim 10^5$~K) and warm ($10^5$~K $\lesssim T < 10^6$~K) phases of the CGM in $z < 1$ galaxies can be probed via emission and absorption lines of hydrogen and metals in the UV \citep{Bertone2013,Bordoloi2011,Bordoloi2014,Burchett:2016aa,Burchett2019,Churchill2013,Johnson2015,Nielsen2013,Tumlinson2011,Tumlinson2013,Werk:2013qy,Werk2014,Werk2016}. Observations of the CGM in the UV absorption lines of background quasar spectra have been performed by Hubble's Cosmic Origins Spectrograph (COS) \citep[e.g.][]{Tumlinson2013,Stocke2013,Johnson2015}, and have shown that most of the baryons associated with galaxies are likely in the CGM \citep{Stocke2013,Werk2014} and that most of the metals released by stars are in the CGM as well \citep{Peeples2014,Prochaska2017}. 

The cool-warm CGM has been extensively studied in cosmological hydrodynamic simulations. Most often O~VI, which traces the warm CGM and the diffuse cool CGM, is studied in large volume simulations, including IllustrisTNG \citep{Nelson2018b}, EAGLE \citep{Wijers2020}, and SIMBA \citep{Appleby2021}. Lower ionization metal species (e.g. C~III, C~IV, Mg~II, Si~III) trace denser structures, which are usually poorly resolved in these large volume simulations, therefore higher resolution simulations are often used. \citet{Oppenheimer2018b} explored higher resolution EAGLE zoom-in simulations to study low ions. \citet{Li2021} studied multiple ions in FIRE zoom-in simulations. When compared to observations, these works typically find agreement with some ions but not others, as did \citet{Appleby2021} for SIMBA. Regarding the TNG50 simulation, which has a resolution that is often in excess of zoom-in simulations, \citet{Nelson2021} studied the MgII in emission, and more recently, \citet{Weng2024} explored the physical and environmental origins of simulated HI absorbers.

The hot phase ($T \gtrsim 10^6$~K) of the CGM, which is expected to be dominant in galaxies with halos more massive than $\sim 10^{12}$~M$_\odot$, can be probed via X-ray observations. In emission, the brightest X-ray signatures of the CGM are to be found in the lines of metal ions such as O~VII, O~VIII, Fe~XVII, and Ne~IX, all of which have rest-frame energies in the 0.5-1.0~keV ($\sim$12-25~\AA) band. For galaxies that are nearby and thus the easiest to detect and study, their redshifted emission lines are in the same band as the Milky Way's (MW) own CGM, which shines brightly in the same emission lines \citep{McCammon2002}. The CGM must also be distinguished from other sources of X-ray emission in galaxies, such as the hot ISM, AGN, and X-ray binaries. Some detections in emission of individual galaxies have been made \citep{Anderson2011,Humphrey2011,Bogdan2013,Bogdan2017,Das2019a,Das2020,Li2017}, and stacking analyses of galaxies from surveys can reveal the general properties of the X-ray emitting CGM \citep{Anderson2013,Anderson2015,Li2018,Chadayammuri2022,Comparat2022,Zhang2024a,Zhang2024b}.

The main obstacle to more detailed studies of the hot CGM in X-ray emission is a lack of spectral resolution. The CCD imaging arrays aboard previous and current X-ray telescopes, including \textit{Chandra}, \textit{XMM-Newton}, \textit{Suzaku}, and \textit{eROSITA}, have spectral
resolutions of $\sim$100~eV ($R \sim 10$ at 1~keV), which is far too coarse to resolve individual emission lines. For these instruments, the lines from the CGM not only blend with each other, but they blend into and are overwhelmed by the lines in the MW foreground emission. The diffraction gratings on \textit{Chandra} and \textit{XMM-Newton} have the requisite spectral resolution, but for extended sources such as the CGM, the dispersed spectrum on the CCDs is convolved with the spatial distribution of the emission from the source, smearing out spectral features. Gratings observations also lack the effective area required to detect the faint emission from the CGM.

In order to map the CGM at the required spectral resolutions, we require an integral field unit
(IFU) instrument in the X-ray band, which is a capability that can be provided by a
microcalorimeter. Microcalorimeters detect X-ray photons and measure their energies by sensing the
heat generated when they are absorbed and thermalized. To achieve the energy resolutions of
$\sim$1-5~eV required for line emission studies in X-rays, microcalorimeters must be kept at
extremely low temperatures. The capability of microcalorimeters to resolve detailed thermodynamic,
kinematic, and chemical properties of hot space plasmas was demonstrated most recently by the
observations of the Perseus cluster of galaxies taken by the Soft X-ray Spectrometer (SXS)
microcalorimeter on board the \textit{Hitomi} spacecraft
\citep{Hitomi2016,Hitomi2017,Hitomi2018,Hitomi2018b,Hitomi2018c}, and is currently being demonstated by \textit{Hitomi}'s successor \textit{XRISM} \citep{xrism2020}.

Other planned and proposed microcalorimeter instruments include the X-ray Integral Field Unit
(X-IFU) on \textit{Athena} \citep{xifu2016,xifu2018}, the Lynx X-ray Microcalorimeter (LXM) on
\textit{Lynx} \citep{Bandler2019}, the \textit{Hot Universe Baryon Surveyor (HUBS)} \citep{Cui2020}, the and
\textit{Line Emission Mapper (LEM)} \citep{LEMWhitePaper} probe concept. Of these, \textit{HUBS} and \textit{LEM}
have the large field of view necessary to study the CGM to large radius (widths of 1~degree and
0.5~degree, respectively), so that the hot gas of the entire galaxy could be imaged in a single
pointing in the case of nearby systems. The proposed angular resolution of $\sim$10 arcseconds for
\textit{LEM} is necessary to resolve the emission from bright background AGN point sources which can
contaminate the CGM signal.

The same high spectral resolution of X-ray IFUs that will enable more detailed study of the hot CGM's thermodynamic and chemical properties using emission lines will also make it possible to measure the velocity of the hot gas via the Doppler shifting and broadening of these same lines. Determining the kinematic properties of the hot CGM is essential to understanding its physics. Measuring its bulk (or mean) velocities via line shifts can detect outflows from AGN and supernovae feedback, as well as rotation and inflows. These motions can be compared to those in other gas phases as well as the stellar component. Line broadening measurements can probe gas turbulence, but may also reveal a complex of bulk flows at different velocities projected along a common sight line \citep[see][for an analysis of this phenomenon at the galaxy cluster scale]{ZuHone2016}. To make predictions for future observations of the velocity field of the hot CGM, we can use hydrodynamical simulations of the CGM in galaxies formed in a cosmological context that also have models for feedback from AGN and stars.

We have almost no observational constraints on the velocity of the hot CGM. Tangential motion of the hot phase of the MW CGM has been observed by \citet{HodgesKluck2016} using O~VII absorption line measurements against bright background AGNs with \textit{XMM-Newton}, and it was found to be comparable to the rotation velocity of the stellar disk. The turbulent velocity dispersion of diffuse hot plasma in our Galaxy has been measured to be 79$_{-17}^{+53}$ km~s$^{-1}$ (90\% confidence interval) along the line of sight to the bright X-ray binary LMC X-3 (at 50~kpc), using \textit{Chandra}/HRC-LETG observations of the absorption lines produced by highly ionized oxygen \citep{Wang2005}. Simulations of the hot CGM predict the presence of a complex combination of motions: rotation/tangential motions, turbulence, outflows, and inflows. \citet{Oppenheimer2018} and \citet{Huscher2021} have shown that the hot CGM of low-$z$ galaxies is primarily supported against gravity by tangential motions in the inner regions (within $\sim$50~kpc), while in high-$z$ galaxies the hot gas is primarily outflowing. These simulations also found that these tangential motions are primarily in the form of coherent rotation for the cold gas, while for the hot gas there is a combination of coherent rotation and uncorrelated motions. \cite{DeFelippis2020} showed that the inner hot CGM of disk galaxies from the TNG100 simulation has rotation similar to the stars and the ISM for galaxies with high angular momentum. \citet{Hafen2022} found that the inner hot CGM in disk galaxies from the FIRE simulation is largely supported by thermal pressure with a slow inflow component. The rotation velocity increases inward, reaching values comparable to stars at the disk edge, as in the idealized hot inflow solution described in \cite{Stern2020,Stern2023}. Given the lack of observational constraints, the results of simulations in this area depend strongly on the specific implementations of the underlying astrophysical processes, especially feedback. This motivates the need for future X-ray observations to confront the simulations. 

Some investigations of this type have already been carried out or are in progress. \citet{Nelson2023} investigated the possibility that resonant scattering of O~VIIr emission line photons could boost the CGM signal from this line to be significantly brighter than the intrinsic emission alone, using galaxies from the the Illustris TNG50-1 (hereafter TNG50) simulations. \citet{Bogdan2023} showed using mock \textit{LEM} observations of galaxies from the Magneticum simulations that O~VII and O~VIII absorption lines can be detected at very large radius. Comparisons between galaxies from IllustrisTNG, SIMBA, and EAGLE demonstrate that emission lines from oxygen and iron in the X-ray band can be used to distinguish between different models of AGN feedback and determine the role of feedback from supernovae and black holes in regulating star formation \citep{Truong2023}. Finally, \citet{Schellenberger2024} studied the CGM from galaxies in the IllustrisTNG, SIMBA, and EAGLE simulations (using mock X-ray observations similar to those employed in this work), demonstrating that the CGM can be traced out to large radii using spectroscopically resolved emission lines, and maps of temperature, velocity, and abundance ratios can be produced. 

Milky Way and M31-like disk galaxies in the TNG50 simulation have complex CGM structure \citep{Ramesh2023,Ramesh2023b}, in part due to fast outflows driven by feedback processes \citep{Nelson2019} that produce bubble-like features in X-ray morphology similar to the eROSITA bubbles in the MW \citep{Pillepich2019}. These are associated with velocities directed away from the disk in the several hundreds to thousands of km~s$^{-1}$. Consequently, \citet[][using the TNG50 simulation]{Truong2021} and \citet[][using the EAGLE simulation]{Nica2022} showed that strong outflows in the CGM of disk galaxies produce anisotropic signatures in the X-ray in terms of surface brightness (SB), temperature, and metallicity \citep[as well as magnetic field strength;][]{Ramesh2023c}. Anisotropies are also expected due to rotational support in the hot gas, as shown by the idealized hot rotating CGM models of \citet{Sormani2018} and \citet{Stern2023}.

In this work, we analyze the thermodynamic and kinematic properties of the hot CGM plasma from disk galaxies in the TNG50 simulation. These galaxies are part of the sample chosen by \citet{Pillepich2021} for possessing large bubbles driven by feedback processes, and our small subsample spans the mass range of this sample and focuses on galaxies with fast outflows. We first focus on projected quantities which would be observable in X-rays, such as surface brightness, temperature, line-of-sight velocity, and line-of-sight velocity dispersion. We also examine the general properties of the velocity field of the hot gas in comparison to the warm and cold phases and to the stellar disk. These will show what features to expect from high spectral resolution X-ray observations of the CGM and the physical processes that produce them. We then follow up with synthetic microcalorimeter observations of the CGM, to determine to what extent these properties would be discernible by an instrument with capabilities such as \textit{LEM}. 

This paper is organized as follows. In Section \ref{sec:methods} we describe briefly the properties of the TNG50 simulation and the galaxies that were selected for study, as well as the procedure for determining the X-ray emission from the CGM of these galaxies and simulating observations. In Section \ref{sec:results} we present the properties of the X-ray emission from the CGM and the results of the synthetic observation study. In Section \ref{sec:summary} we discuss the results and present our conclusions. As in the TNG50 simulation, throughout this work we assume a flat $\Lambda$CDM cosmology with $h$ = 0.6774, $\Omega_m$ = 0.3089, and $\Omega_\Lambda$ = 0.6911, consistent with the Planck 2015 \citep[][]{Planck2016} results. Unless otherwise noted, all error bars in this work refer to 1-$\sigma$ uncertainties.

\section{Methods} \label{sec:methods}

\subsection{Simulation: TNG50}\label{sec:tng50}

\begin{table*}[!ht]
\begin{center}
\caption{Properties of TNG50 Galaxies\label{tab:tab_gals}}
\begin{tabular}{cccccccccc}
\hline
Galaxy \# & subhalo ID & $r_{200c}$ & $r_{500c}$ & $M_{200c}$ & $M_{500c}$ & $M_*{}$ & $M_{\rm{gas,hot}}$ & $M_{\rm{gas,warm/cold}}$ & SFR \\
 &  & (kpc) & (kpc) & ($10^{12} M_\odot$) & ($10^{12} M_\odot$) & ($10^{10} M_\odot$) & ($10^{10} M_\odot$) & ($10^{10} M_\odot$) & (M$_\odot$ yr$^{-1}$) \\
\hline
1 & 372754 & 350 & 237 & 4.58 & 3.56 & 17.68 & 9.33 & 1.93 & 1.67 \\
2 & 414917 & 297 & 202 & 2.80 &	2.19 & 14.97 & 7.19 & 1.89 & 0.43 \\
3 & 502995 & 220 & 151 & 1.13 &	0.92 & 7.89 & 2.12 & 6.34 & 7.75 \\
4 & 535410 & 216 & 149 & 1.07 & 0.89 & 6.32 & 2.73 & 2.55 & 1.29 \\
5 & 571454 & 195 & 135 & 0.79 & 0.65 & 4.34 & 0.23 & 0.33 & 0.01 \\
6 & 572328 & 190 & 134 & 0.73 & 0.63 & 4.27 & 1.02 & 1.03 & 0.0 \\
\hline
\end{tabular}
\end{center}
Columns are as follows: (1) Galaxy number; (2) {\it TNG50} subhalo ID; (3-4) Radii corresponding to enclosed average densities of 200 and 500 times the critical density; (5-6) Masses corresponding to enclosed average densities of 200 and 500 times the critical density; (7) Stellar mass within a radius of 0.2$r_{200c}$; (8) Gas mass in the hot ($T \geq 5 \times 10^5$~K) phase within $r_{200c}$; (9) Gas mass in the warm and cold ($T < 5 \times 10^5$~K) phases within $r_{200c}$; (10) Star formation rate of the galaxy within a radius of 0.2$r_{200c}$.
\end{table*}

The galaxies we examine in this work were selected from the TNG50 simulation \citep{Nelson2019,Pillepich2019}, a
magnetohydrodynamics (MHD) cosmological simulation in a periodic cube $\sim$50 comoving Mpc on a side, one of the three original IllustrisTNG boxes \citep{Springel2018,Pillepich2018b,Nelson2018,Marinacci2018,Naiman2018}.\footnote{The
IllustrisTNG simulations, including TNG50, are publicly available at \url{https://www.tng-project.org/data}
\citep{Nelson2019b}.} The simulation size and mass resolution is ideal for studies of galaxy formation and evolution
\citep[for a review of cosmological simulations see][]{Vogelsberger2020}. The
simulations are performed with the \texttt{AREPO} code \citep{Springel2010}, which combines a TreePM gravity solver with
a quasi-Lagrangian, Voronoi moving-mesh based method for fluid dynamics. The TNG model includes magnetic fields \citep{Pakmor2013}, gas cooling and heating, star formation and evolution, metal
enrichment, feedback from supernovae, and for the creation, growth, and feedback of supermassive black holes (SMBHs)
\citep{Weinberger2017,Pillepich2018}.

Much of this work focuses on the effects of AGN feedback on our galaxies, so we describe the AGN feedback model
of the TNG simulations in more detail \citep[see][]{Weinberger2017}. In brief, the feedback model employs two main
modes: at high accretion rates relative to the Eddington limit, thermal energy is continuously deposited into the
surrounding gas. At low accretion rates, kinetic energy is periodically injected into the surrounding gas, where the
direction of each injection event is random and, in the time average, isotropic.

As shown in previous works, this kinetic feedback mode is capable of driving fast, multi-phase outflows
\citep{Nelson2019,Truong2021,Ramesh2023}. These outflows preferentially propagate perpendicular to the disk, expelling
mass and metals to the outer regions of the CGM, and heating the gas. They can also inflate large overpressurized
bubbles and shell features \citep{Pillepich2021}. Given that the energy injection is isotropic, the tendency of outflows
to propagate along the minor axis directions of disk galaxies is a ``path of least resistance'' effect, where motions
imparted in the disk direction encounter more confining pressure from the ISM, resulting in the outflows being directed
away from the disk.

The gas mass resolution of TNG50 is $m_{\rm gas} = 8.5 \times 10^4$~M$_\odot$. As a result, the galaxies of our
sample thus contain $\sim$10$^5$-10$^6$ gas cells. The typical cell size for the hot, X-ray-emitting gas (see Section
\ref{sec:xrays} for the definition) is of $\sim$2-4~kpc within a radius of 50~kpc from the center of the galaxy, and
6-7~kpc within a radius of 100~kpc \citep[see also][]{Nelson2020}.
As we will show, this is sufficient to resolve rotation, inflows, and feedback-driven outflows in the CGM on
the scales relevant for future X-ray microcalorimeter instruments for the galaxies considered here. Gas cells
use an adaptive gravitational softening of 2.5~$r_{\rm cell}$, where the effective cell radius $r_{\rm cell}$ is derived
from the total volume of the Voronoi cell approximating it as a sphere \citep{Pillepich2018}.

The original sample from TNG50 from which our galaxies originate was presented in \citet{Pillepich2021,Pillepich2023}
and was selected to represent MW/M31-type galaxies: having a stellar mass of $M_*(< 30~{\rm kpc}) = 10^{10.5}-10^{11.2}
M_\odot$, having a disk-like stellar morphology, having no other massive galaxy with $M_* > 10^{10.5} M_\odot$ within
500~kpc, and that the mass of their host halo is limited to $M_{200c} < 10^{13} M_\odot$ (to avoid galaxies sitting in
massive groups or clusters). All galaxies were selected from the $z = 0$ snapshot. These simulated MW/M31-like
galaxies have been demonstrated to have properties that are compatible with available observational inferences
pertaining to the MW, M31, and analog systems, including the satellite galaxy abundance \citep[][especially
Figure 6]{Engler2021}, the star formation rate vs. stellar mass relation \citep[][Figure 4]{Pillepich2021}, disk
structural properties, magnetic field strengths, and stellar kinematics \citep[][Figures 16 and 18]{Pillepich2023}. A
sizeable fraction of them exhibit a stellar bar \citep{Pillepich2023} and stellar disk flaring \citep{Sotillo2023} and,
in relation to the halo gas, their CGM can be filled with hundreds of reasonably-resolved cool clouds similar to
observed high-velocity clouds \citep{Ramesh2023b}.

From this sample of MW/M31-like systems, we have selected 6 galaxies, keeping the sample size small to be able
to focus on individual systems. Their properties are listed in Table \ref{tab:tab_gals}. These 6 galaxies are
chosen to span the mass range of the overall MW/M31 sample, and were also determined to have fast radial ouflows as
defined by \citet[][see in particular their Section 4.3 and Figure 8]{Pillepich2021}, with speeds ranging from
$\sim$400-2000~km~s$^{-1}$. All of the galaxies in our sample have a central AGN that has been active in the kinetic
mode described above since at least $z \sim 1$, and in some cases even earlier (see Appendix \ref{sec:AppendixC}). All
of these galaxies are also above the stellar mass threshold of $M_* \gtrsim 3 \times 10^{10} M_\odot$, where at late
times a stable virial shock is expected to form which heats inflowing gas to the virial temperature of the halo
\citep[][]{Birnboim2003,Keres2005,Dekel2006}.

For all of the subsequent analysis, the coordinates and velocities of the particles and cells from each galaxy have
undergone a coordinate transformation such that the origin for each is the potential minimum of the galaxy, and the
$z$-axis of the new Cartesian coordinate system points in the direction of the normalized spin axis of the galaxy's
disk. This axis is determined by computing the total angular momentum vector of the star particles within a sphere of
radius 15~kpc centered on the galaxy's potential minimum. The direction of the $x$-axis is chosen to be perpendicular to
the $z$-axis, but otherwise arbitrarily, and the direction of the $y$-axis is then determined to give the axes a
right-handed orientation. The rest frame of the particles and cells is determined by computing the mass-weighted mean
velocity of the star particles from the same spherical region and subtracting this velocity from the velocities of all
of the particles and cells (the results are not particularly sensitive to the choice of radius for the spherical
region). We have verified that our disks are correctly aligned by also applying the alignment procedure
described in Section 2 of \citet{Garcia-Conde2022} and verifying that the results are consistent.

For the purposes of this paper, we make a distinction throughout between gas cells which we designate as ``hot'' with $T
\geq 5 \times 10^5$~K, and those we designate as ``warm/cold'' with $T < 5 \times 10^5$~K. The boundary value of $T = 5
\times 10^5$~K corresponds to $kT \sim 0.04$~keV, which in terms of photon energy is a rough boundary between the
extreme UV and X-ray bands and is also close to the typical lower energy range of X-ray detectors. Since our main focus
is the hot X-ray-emitting gas in this work, we do not distinguish further between warm and cold phases. 

\subsection{X-ray Emission and Mock Observations}\label{sec:xrays}

For modeling the X-ray emission from the CGM, we assume collisional ionization equilibrium (CIE) and use the
Astrophysical Plasma Emission Code \citep[APEC][version 3.0.9]{Smith2001}. This approximation is valid for the
temperatures and densities that we are examining in this work, since we are focusing on measuring the velocities within
the inner $\sim$200~kpc of the halo where we expect the effects of hot outflows and rotation to be most significant. We
verify that the only regions for which the assumption of CIE can make a significant difference to the emitted SB are at
radii larger than those of interest to us in this work. The elemental abundance ratio table assumed for the emission
model is from \citet{angr89}. 

The \texttt{pyXSIM} code \citep{pyxsim2016} is also used to produce synthetic X-ray observations from the
CGM of the galaxies. The galaxies are placed at a redshift of $z = 0.01$, at which the radius of $r_{500c}$ for our
galaxies fits roughly within a half-degree wide field of view on the sky (at an angular diameter distance of
$\sim$44~Mpc), and at which the emission lines from the source we are interested in detecting are sufficiently
redshifted away from the bright MW foreground lines. 

Not all gas cells in the simulated galaxies will emit X-rays. Thus, we first identify a subset of gas cells
expected to have the conditions for X-ray emission, in order to make our calculations more efficient. First, we apply a
temperature floor of $T > 3 \times 10^5$~K ($kT \gtrsim 26$~eV), below which the gas is not expected to emit
significantly in X-rays. In cosmological simulations such as TNG, hot gas radiatively cools to colder and denser phases
which can form stars. The TNG simulations follow the method of \citet{Springel2003}, which models the star-forming gas
as a multiphase interstellar medium (ISM) composed of hot and cold gas following an effective equation of state. This
ISM phase is excluded from our analysis by excluding star-forming gas and applying a density threshold of $\rho < 5
\times 10^{-25}$~g~cm$^{-3}$, which is slightly higher than the star formation density threshold in the TNG simulations.
In future work, it would be instructive to examine the X-ray emission from the ISM phase as well by modeling X-ray
emission from its hot component, but for this work we neglect it to focus on the CGM.

Within these cuts, a small fraction of the cooling gas forms dense clumps of a few cells that are still hot
enough ($3 \times 10^5$~K~$\lesssim T \lesssim 10^6$~K) to emit in X-rays. These cells appear in dense regions of the
galaxies and are not expected to be present in the CGM, as they are bright enough to be observed by existing instruments
but are not seen. They also typically have extreme values of cooling time and/or thermal pressure. Such gas cells and/or
particles have been identified in previous works in different simulations using similar star formation and cooling
prescriptions \citep[e.g.][]{Rasia2012,Rasia2014,Zhuravleva2013}, and arise from gas cooling in the absence of efficient
feedback and/or inefficient mixing between these clumps and high-entropy gas. These cells comprise $\sim$1\% of the mass
of the X-ray-emitting gas and $\lesssim$~0.1\% of the volume, but appear in projections as cold and bright point sources
in the inner CGM, close to the stellar disk. Since these clumps are not expected to be present in the CGM, we also
exclude them from the analysis (as was done in the previous works cited above), mainly to improve visualization. We have
experimented with different cuts on cooling time and pressure to exclude these cells, and found that our results are not
sensitive to the exact values chosen, given the small fraction of the mass and volume that they comprise, and that they
are not present in the CGM further away from the disk ($r \gtrsim 30$~kpc) where we focus our analysis.

Having defined the subset of gas cells that we will consider for our mock X-ray observations, we use the APEC
emission model described above to generate a cosmologically redshifted spectrum for these cells. Inputs to these
spectra include the electron and proton number densities, temperatures, and metallicities \citep[assuming relative
abundances from][]{angr89} of the cells. We then use each spectrum to generate an initial sample of photons with
specified values of exposure time $t_{\rm exp}$ = 1~Ms and telescope collecting area $A$ = 0.5~m$^2$. The value for $A$
is energy-independent and is only used to ensure that the initial sample of photons that will be drawn from in the
instrument simulation step is large.

For each mock observation, the positions of this photon sample are projected onto the sky plane along a chosen sight
line, and the energies of the photons are Doppler-shifted using the component of the gas cell velocity along the sight
line. We use the T\"ubingen/Boulder neutral absorption model \texttt{TBabs} \citep{Wilms2000} with a hydrogen column
density of $N_H = 1.8 \times 10^{20}$~cm$^{-2}$ \citep[matching the value for the MW foreground from][see
below]{McCammon2002} to remove a random subset of photons that will be absorbed by neutral gas in the MW. This creates a
large initial random sample of photons for each galaxy and each sight line that will be later used by the instrument
simulator to draw subsamples of photons to create ``observed'' X-ray events. 

This step is carried out by the \texttt{SOXS} code\footnote{\url{https://hea-www.cfa.harvard.edu/soxs}}
\citep{soxs2023}, which takes the set of photons produced by \texttt{pyXSIM} and passes them through an instrument model
to produce observed X-ray ``events.'' This includes convolving with the energy-dependent effective area (auxiliary
response file or ``ARF'') of the combined telescope and instrument, convolving with the response matrix (``RMF'') that
converts photon energies into spectral channels, and scattering the photon positions by the telescope PSF. For this
work, we use an instrument model with capabilities similar to the \textit{LEM} probe concept, with a field of view of 32
arcminutes, 0.9~eV spectral resolution, and an effective area of $\sim$0.2~m$^2$ in the 0.5-2.0~keV ($\sim$6-25~\AA)
band \citep{LEMWhitePaper}.
    
\begin{figure*}[!hp]
\centering
\includegraphics[width=0.92\textwidth]{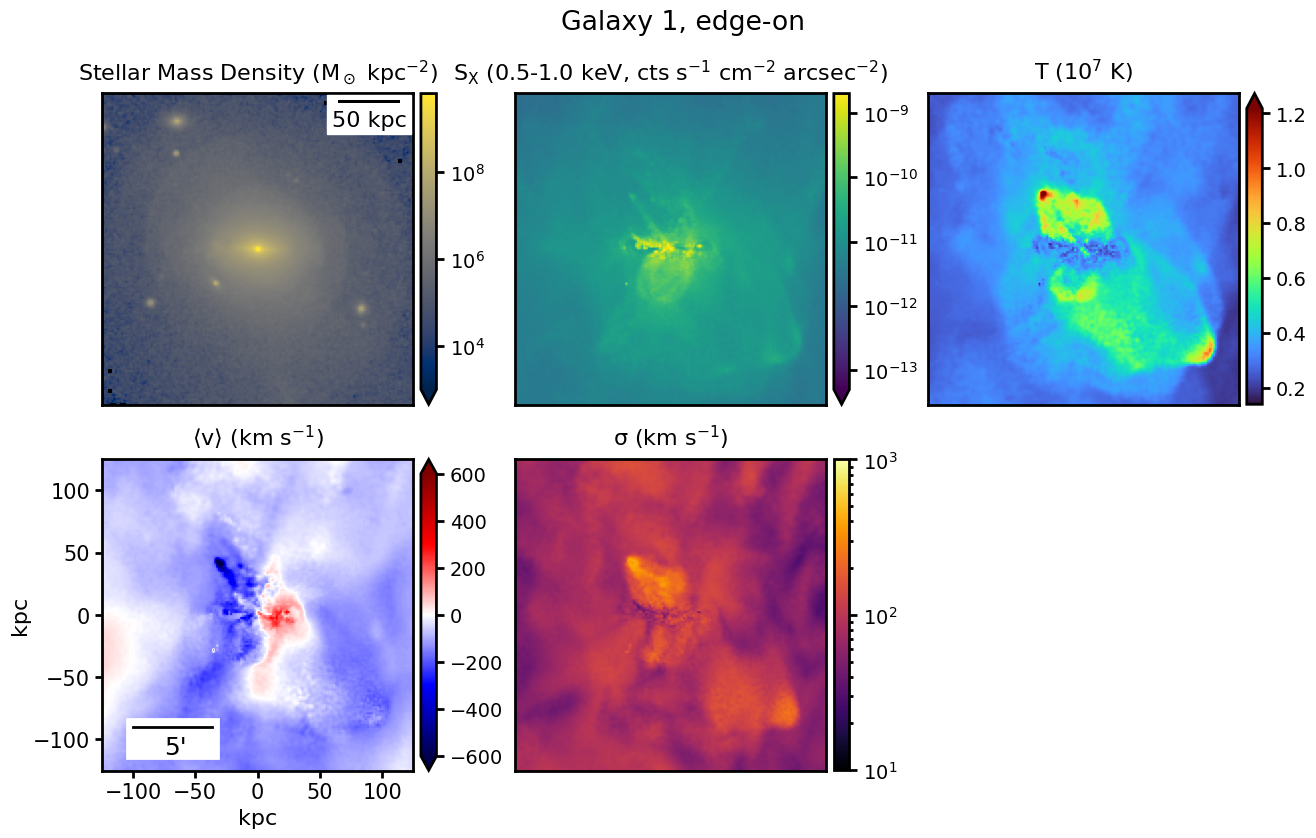}
\includegraphics[width=0.92\textwidth]{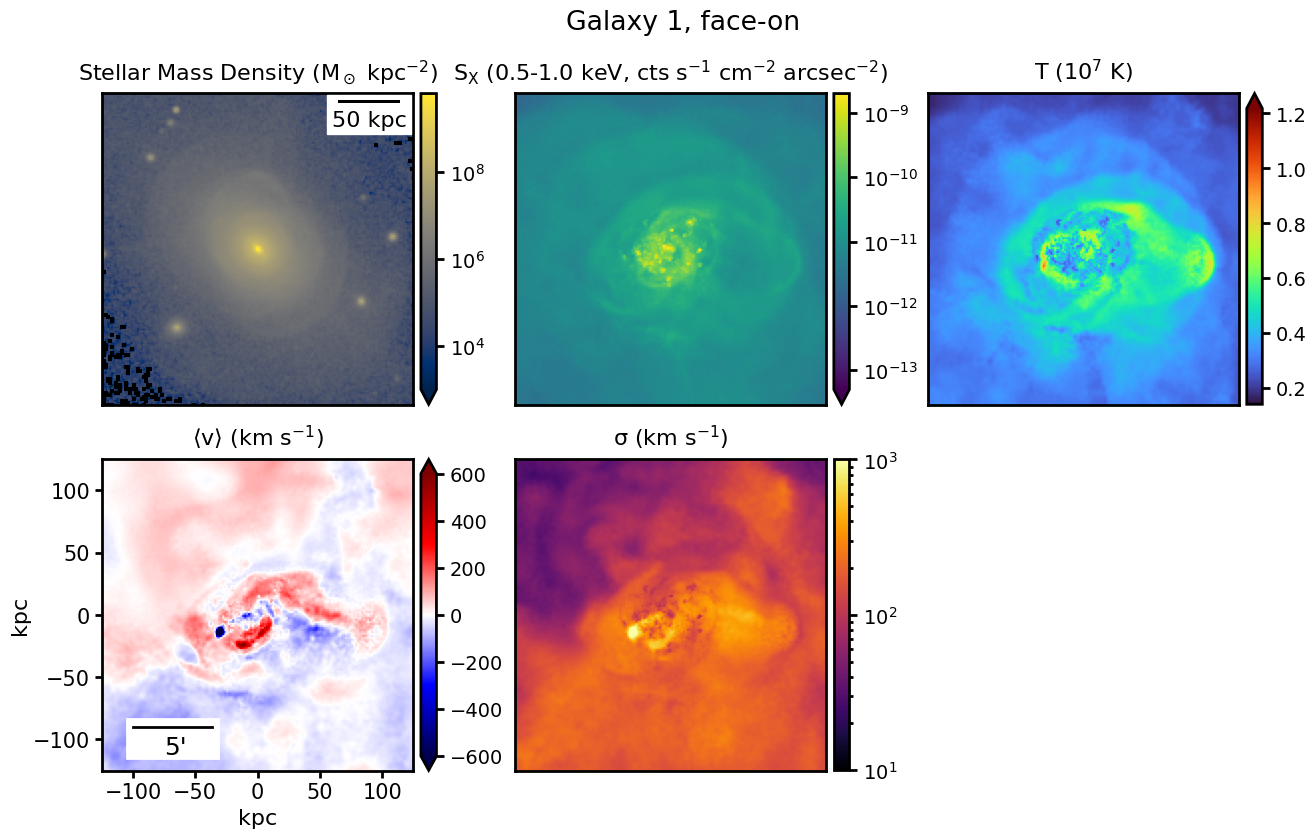}
\caption{Projections of various quantities from galaxy 1, viewed edge-on (upper panels) and face-on (lower panels) with respect to the galactic plane. For each set, top row from left: Stellar mass density, X-ray SB in the 0.5-1.0~keV band, emission-weighted gas temperature. Bottom row from left: emission-weighted gas mean line-of-sight velocity, emission-weighted gas line-of-sight velocity dispersion. Each panel is 250~kpc on a side, or $\sim$20' for the given redshift and cosmology. The properties for each galaxy can be found in Table \ref{tab:tab_gals}. Cavity-shaped regions are evident in SB, which have high temperature, and which are associated with fast mean velocities and velocity dispersions near the center. The hot gas is also rotating, as seen in the edge-on projection.\label{fig:gal1_proj}}
\end{figure*}
    
\begin{figure*}[!hp]
\centering
\includegraphics[width=0.92\textwidth]{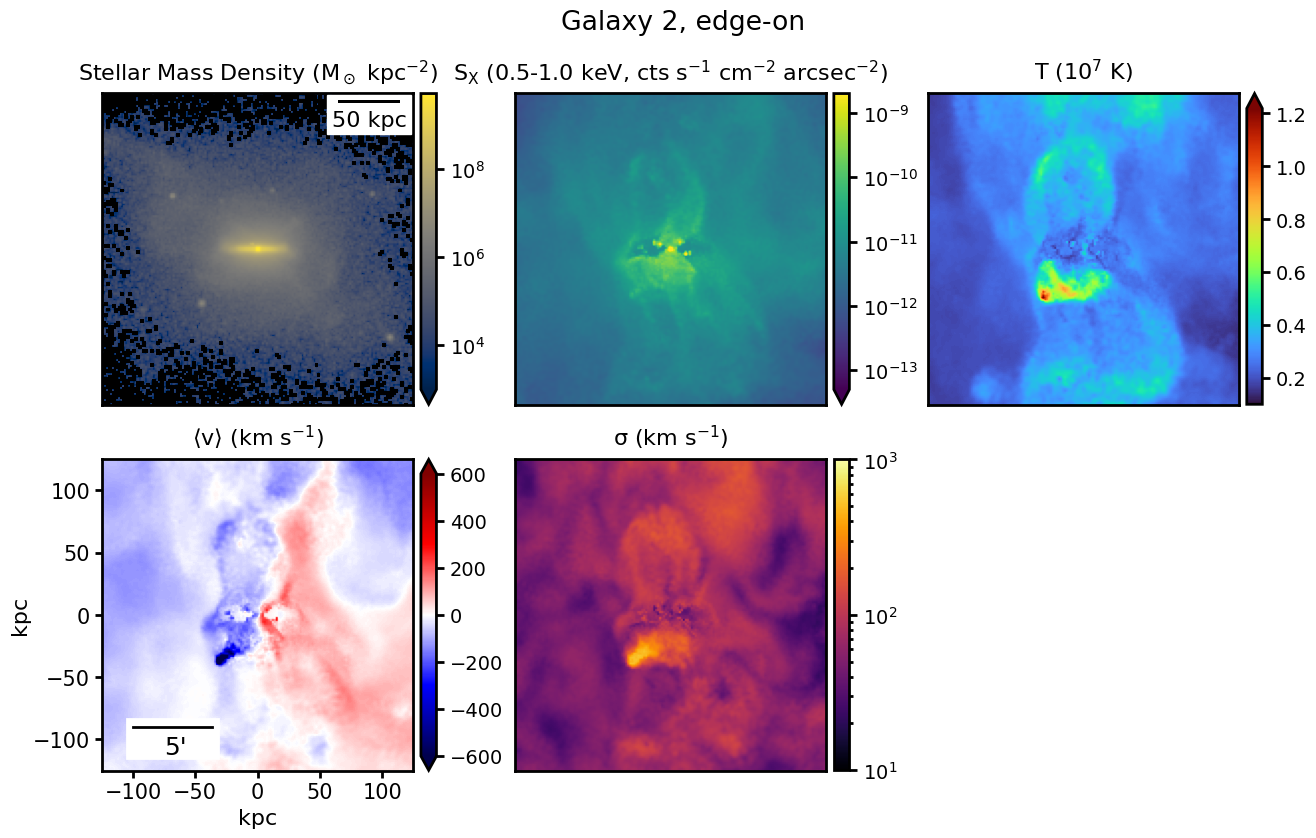}
\includegraphics[width=0.92\textwidth]{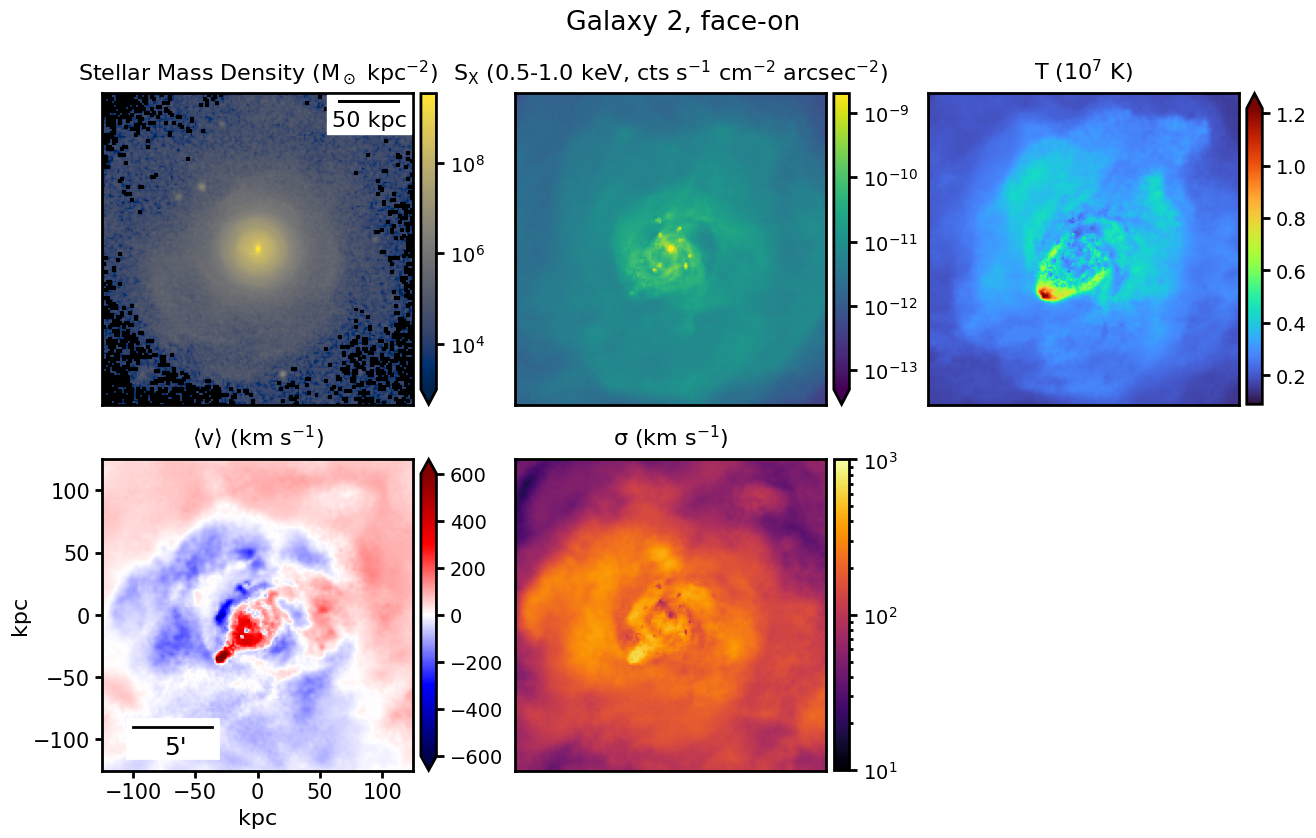}
\caption{Projections of various quantities from galaxy 2, viewed edge-on (upper panels) and face-on (lower panels) with respect to the plane of the galactic disk. Panel descriptions are the same as in Figure \ref{fig:gal1_proj}. Each panel is 250~kpc on a side, or $\sim$20' for the given redshift and cosmology. The same general features of hot outflows and rotation that are present in galaxy 1 are seen clearly in these maps.\label{fig:gal2_proj}}
\end{figure*}

\begin{figure*}[!hp]
\centering
\includegraphics[width=0.92\textwidth]{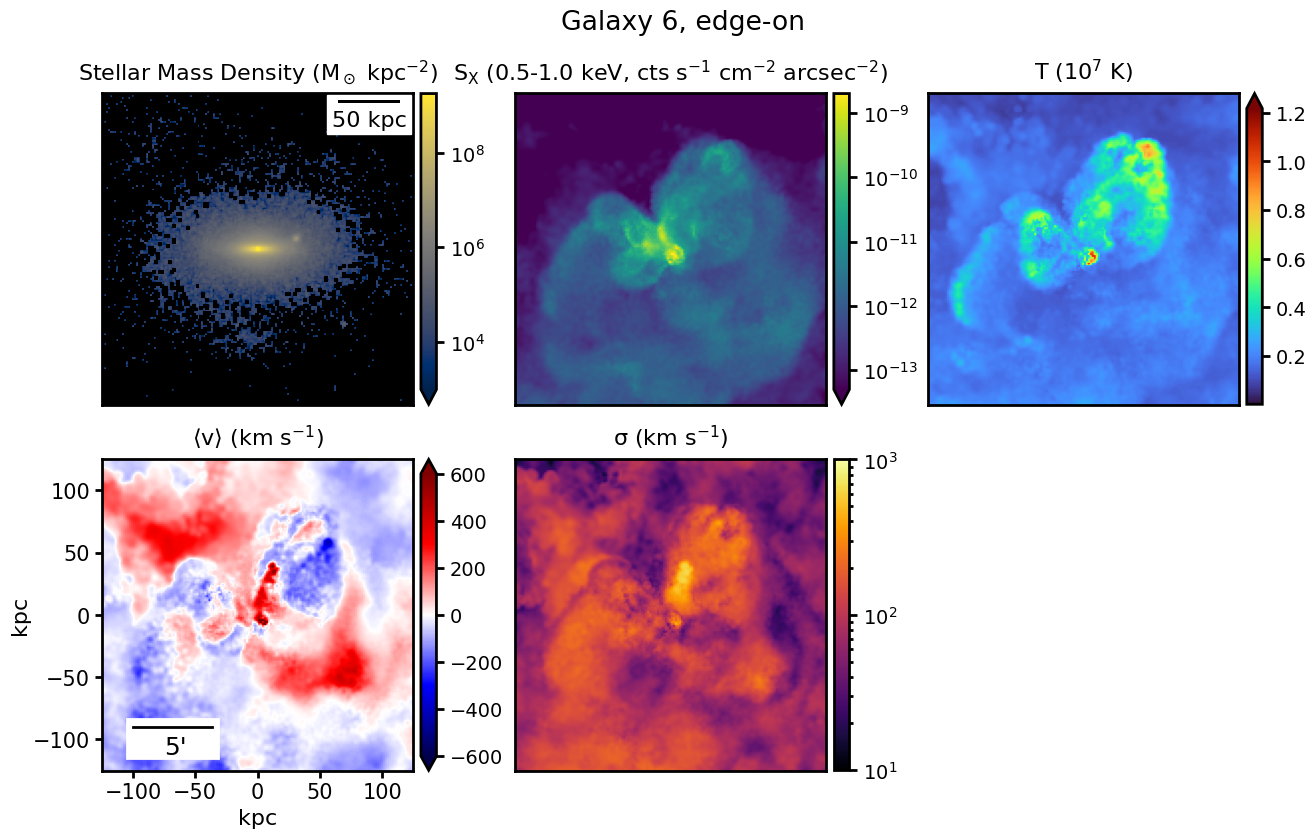}
\includegraphics[width=0.92\textwidth]{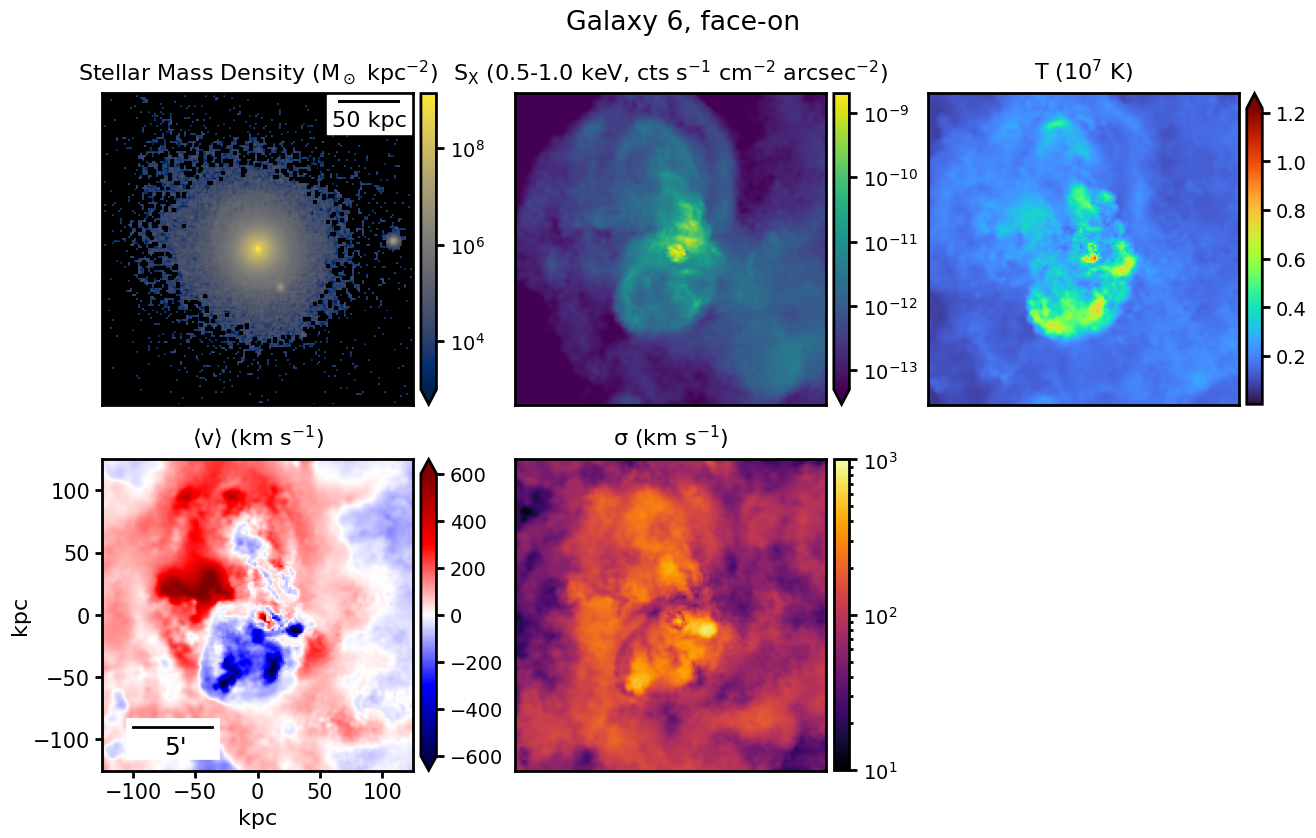}
\caption{Projections of various quantities from galaxy 6, viewed edge-on (upper panels) and face-on (lower panels) with respect to the plane of the galactic disk. Panel descriptions are the same as in Figure \ref{fig:gal1_proj}. Each panel is 250~kpc on a side, or $\sim$20' for the given redshift and cosmology. Unlike galaxies 1 and 2, this galaxy has no obvious pattern of rotation in the velocity map, and the AGN outflows are on the same side of the galactic disk.\label{fig:gal6_proj}}
\end{figure*}

\texttt{SOXS} also includes events from background and foreground models\footnote{More details about the background
models in \texttt{SOXS} can be found at \url{http://hea-www.cfa.harvard.edu/soxs/users_guide/background.html}.}. For the
non-X-ray particle background (NXB), a constant value of 4 counts~s$^{-1}$~keV$^{-1}$~deg$^{-2}$ is assumed. For the
cosmic X-ray background (CXB), we include resolved point sources with numbers and fluxes determined by a
$\log{N}-\log{S}$ distribution from \citet{Lehmer2012}. For the Galactic foreground, we first assume a model with two APEC
\citep{Smith2001} components: one for the ``hot halo'' with $kT$ = 0.225~keV ($T \sim 2.6 \times 10^6$~K), and another for the ``Local Hot Bubble'' with $kT$ = 0.099~keV ($T \sim 1.1 \times 10^6$~K), taken from \citet[][their Table 3]{McCammon2002}. We then add to this model another APEC component for the hot halo with $kT = 0.7$~keV ($T \approx 8.1 \times 10^6$~K) and a normalization parameter roughly 0.12$\times$ that of the first hot halo component, suggested by \textit{Halosat} observations \citep{Bluem2022}.\footnote{Evidence for a such a hot component was also found in \textit{eROSITA} observations by \citet{Ponti2022}.} All three components assume solar abundances. The CXB and the two hot halo components of the astrophysical foreground use the same absorption model and value for $N_H$ as the source emission for simplicity (see Appendix \ref{sec:AppendixA} for an investigation of varying the abundance and hydrogen column density of the foreground). Each galaxy and the included background and foreground emission is exposed for 1~Ms by the \texttt{SOXS} instrument simulator. 

The astrophysical background model detailed above does not include contribution of the line emission coming from the
heliospheric solar wind charge exchange \citep[SWCX; see ][for a recent review]{Kuntz2019}. This background component
originates from the interaction of the ionized particles of the Solar wind with the flow of neutral ISM through the
heliosphere. It is time-variable and much harder to predict and model, but one might expect it affecting mostly the
O~VII triplet, especially its forbidden component, and higher energy lines as well but to a smaller degree \citep[cf. a
recent measurement by][made with \textit{eROSITA} operating at the L2 point, so most closely corresponding to conditions
for the future \textit{Athena} and \textit{LEM} missions]{Ponti2022}. To the zeroth order, the presence of this
component would correspond to a moderate enhancement of the Galactic O~VII line emission, which should not significantly
affect the results presented here.

\section{Results}\label{sec:results}

\subsection{Maps of Projected Quantities from the Simulations}\label{sec:projected}

For three of the disk galaxies, we make projected maps of several quantities along the line of sight, which are shown in
Figures \ref{fig:gal1_proj}-\ref{fig:gal6_proj}. Each galaxy is projected along ``edge-on'' and ``face-on'' sight lines,
defined with respect to the stellar disk. The top-left panel of each sight line of these figures shows projected stellar
mass density. The stellar streams observed in both galaxies at large radii indicate past or ongoing merging activity
with satellites. 
                     
The other four panels in each figure show projected quantities associated with the X-ray emitting gas. The top-center
panels of each sight line of Figures \ref{fig:gal1_proj}-\ref{fig:gal6_proj} show X-ray SB in the 0.5-1.0~keV band,
which spans the prominent emission lines for the hot CGM as noted in Section \ref{sec:intro}. In the edge-on projections
(upper panels of Figures \ref{fig:gal1_proj}-\ref{fig:gal6_proj}), there are clear indications in the SB maps of AGN activity on either side of the galactic disk along its minor axis, including cavities surrounded by bright
rims, though for galaxy 6 the AGN outflows appear only above the stellar disk (in other edge-on projections of this
galaxy, these two cavities are projected along the same line of sight and make it appear that only one is present.) As
we show in Appendix \ref{sec:AppendixC}, this galaxy is very unlike the others as it has suffered a major recent
disruption of its ISM and CGM due to significant AGN feedback, which strongly affects its appearance. The face-on
projections of galaxies 1 and 2 (lower panels of Figures \ref{fig:gal1_proj} and \ref{fig:gal2_proj}) do not
show any such bimodality in their SB maps (as expected), but instead show roughly concentric edges along
various directions, likely generated from the expansion of the outflowing gas perpendicular to the disk axis or from
satellite merger activity. In the case of galaxy 6, the two bubbles on the same side of the galaxy in the
edge-on projection now appear on either side of the center in the face-on SB projection of galaxy 6 (lower SB panel of
Figures \ref{fig:gal6_proj}), resembling the edge-on cases in the other galaxies by pure chance.

Much stronger indications of the nature of the various X-ray features can be seen in the projected temperature and
velocity maps (which are weighted by the X-ray SB in the 0.5-1.0~keV band). The outflow regions shown in the SB
maps are clearly associated with regions of higher temperature (top-right panels of Figures
\ref{fig:gal1_proj}-\ref{fig:gal6_proj}). Most of the hot CGM has a projected temperature of $T \sim 3.5-4.6 \times
10^6$~K ($\sim 0.3-0.4$~keV), whereas the regions associated with the cavities in the SB maps range from $T
\sim 5.8-11.6 \times 10^6$~K ($\sim 0.5-1.0$~keV).
            
The bottom-left panels of Figures \ref{fig:gal1_proj}-\ref{fig:gal6_proj} show the line-of-sight bulk velocity. In the
edge-on projections of Figures \ref{fig:gal1_proj} and \ref{fig:gal2_proj} for galaxies 1 and 2, the mean velocity maps
show clear signs of rotation of the CGM in the inner $r \sim 50$~kpc for both galaxies. Rotation speeds measure up to
$\sim$200-300~km~s$^{-1}$. Outside of this radius, the bulk flows do not tend to follow a clear pattern of rotation, and
any measured velocities can be radial (whether inflowing or outflowing) or tangential. On the other hand, galaxy 6
(Figure \ref{fig:gal6_proj}) does not show a clear rotation pattern in the emission-weighted line-of-sight velocity map
in the edge-on projection--its velocity field appears highly turbulent and is correlated with the hot outflows
seen in the SB and temperature projections. The face-on projections of each galaxy show a complex pattern of mean
velocities in both directions near the center of the galaxy, ranging from -600---600 km~s$^{-1}$. For a
perfectly symmetric biconical outflow observed along the flow axis, the mean velocity would be zero. The complex pattern
of mean velocity that is instead observed is indicative of the turbulent and irregular nature of the outflows. At
larger projected radii, the mean velocities are smaller, at $\sim$-200---200~km~s$^{-1}$. In the face-on
projection of Galaxy 6, one of the outflow regions has a large line-of-sight velocity of $\sim$-400---600~km~s$^{-1}$.

The line-of-sight velocity dispersion maps are shown in the bottom-center panels of Figures
\ref{fig:gal1_proj}-\ref{fig:gal6_proj}. In the edge-on projections, velocity dispersions of $\sim$200-1000~km~s$^{-1}$
are seen primarily in the regions dominated by the outflows. These correspond primarily to the expansion of bubbles and
outflow regions along the sight line. In the face-on projections,
the whole inner $r \lesssim 100$~kpc region has projected velocity dispersions of $\sim$300-1000~km~s$^{-1}$, primarily
due to the directed hot outflows on either side of the disk. This too is very patchy, reflecting the complexity of the
outflows as seen in projection. Outside of these regions, the velocity dispersion is much smaller, around $\sigma \sim
100-200$~km~s$^{-1}$. We will see later (Section \ref{sec:velocity_profiles}) that the faster velocities (both mean and
dispersion) near the center are associated with outflows, while the slower velocities at larger projected radii are
associated with inflows.

\begin{figure*}[!t]
\centering
\includegraphics[width=0.47\textwidth]{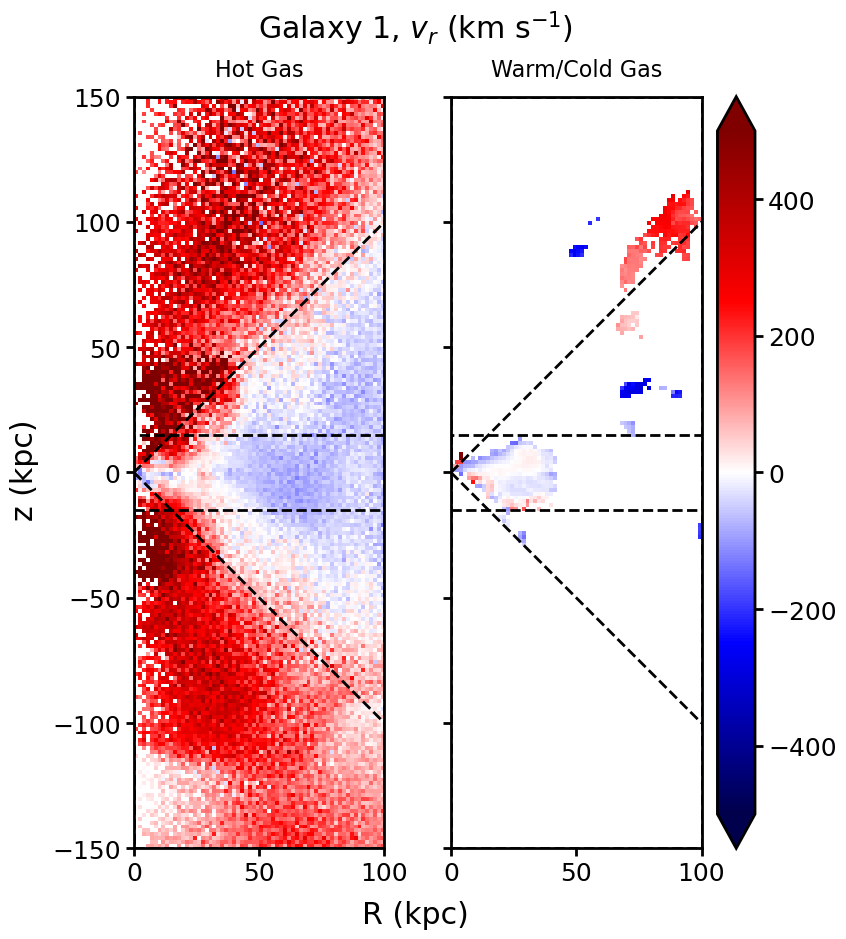}
\includegraphics[width=0.47\textwidth]{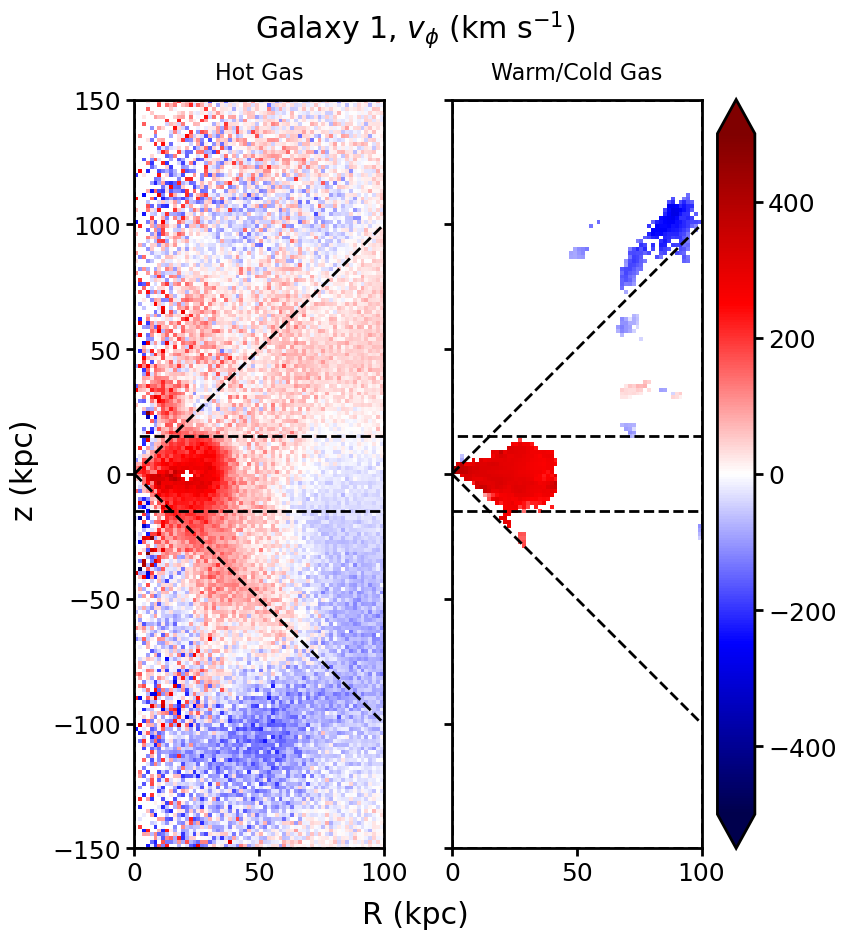}
\caption{Azimuthally averaged profiles in the radial ($R$) and vertical $z$ cylindrical coordinates of the spherical-$r$ (left) and cylindrical-$\phi$ (right) components of the gas velocity in hot (left sub-panels) and warm/cold (right sub-panels) gas phases for galaxy 1. The dashed horizontal lines indicate the region that is used to extract 1D cylindrical radial profiles in Section \ref{sec:1d_profiles}. Fast outflows in the hot gas perpendicular to the galactic plane are separated from slow inflows near the plane by 45$^\circ$ lines. The hot and warm/cold phases are co-rotating, though the former rotates more slowly.\label{fig:gal1_phase}}
\end{figure*}    

Projected maps in the face-on and edge-on directions for the other three galaxies (3, 4, and 5) are
presented in Figures \ref{fig:gal3_proj}, \ref{fig:gal4_proj}, and \ref{fig:gal5_proj} in
Appendix \ref{sec:appendixB}. In general, these maps show similar features in the different projections to galaxies 1, 2, and 6. 

\begin{figure*}
\centering
\includegraphics[width=0.47\textwidth]{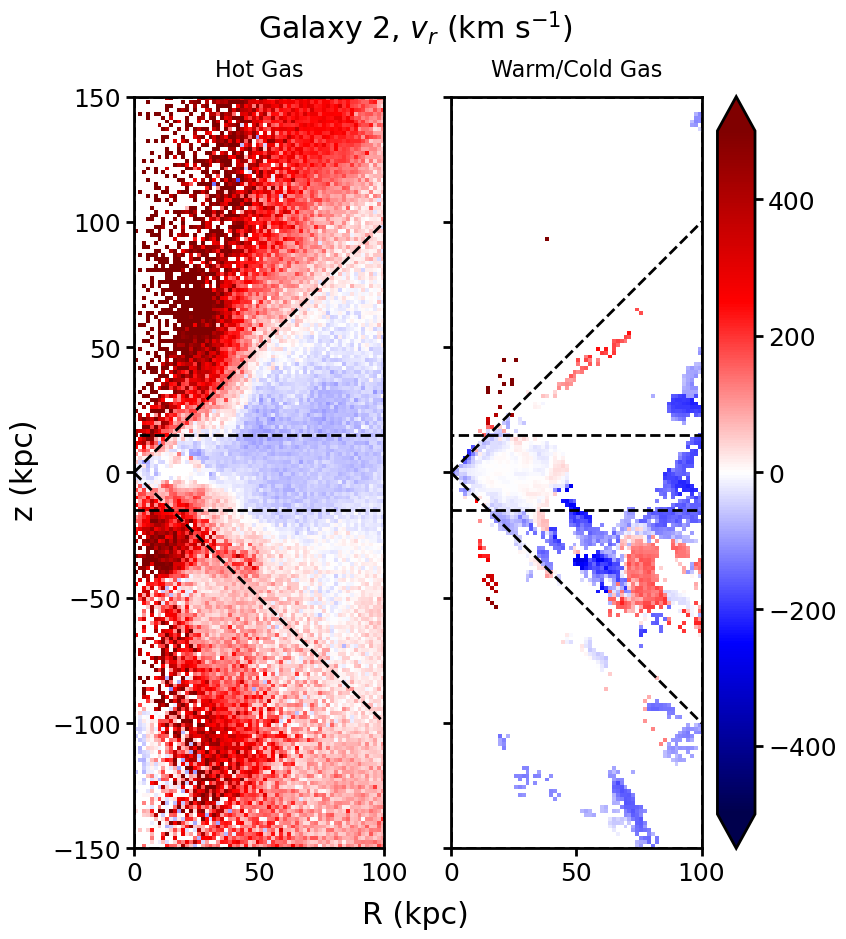}
\includegraphics[width=0.47\textwidth]{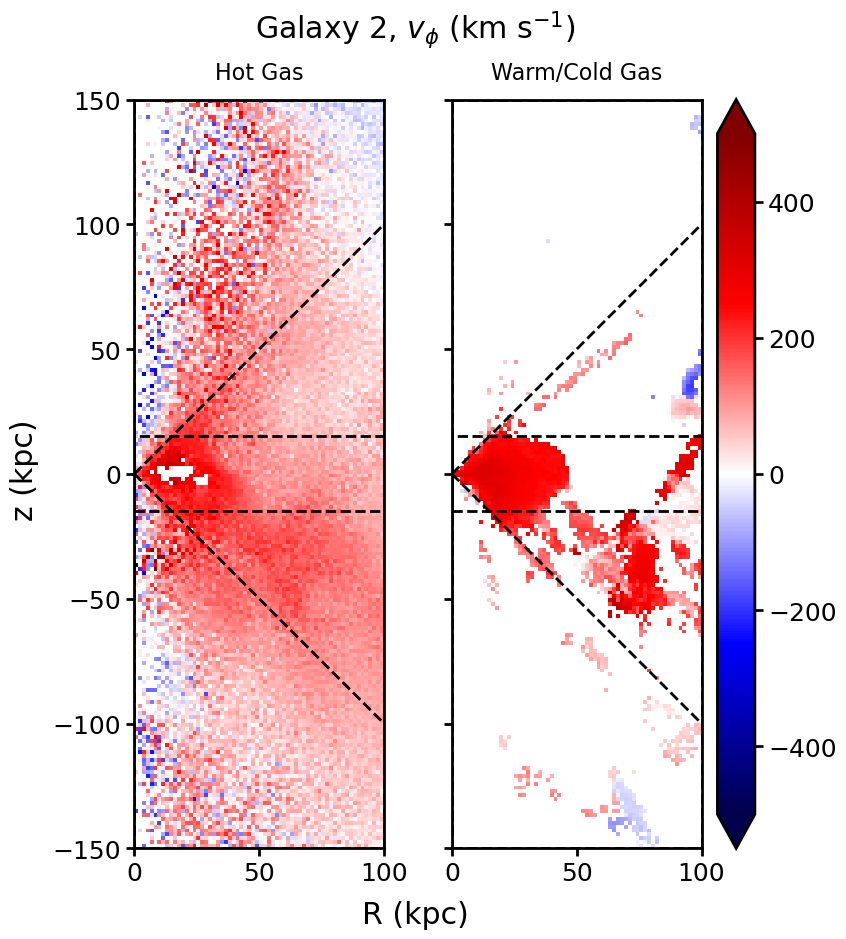}
\caption{Azimuthally averaged profiles in the radial ($R$) and vertical $z$ cylindrical coordinates of the spherical-$r$ (left) and cylindrical-$\phi$ (right) components of the gas velocity in hot (left sub-panels) and warm/cold (right sub-panels) gas phases for galaxy 2. The dashed horizontal and diagonal lines have the same meaning as in Figure \ref{fig:gal1_phase}. The clear pattern of outflows, inflows, and rotation is very similar to that in galaxy 1.\label{fig:gal2_phase}}
\end{figure*}    
    
\subsection{Velocity Profiles}\label{sec:velocity_profiles}

In this Section, we further examine the properties of the CGM velocity field, focusing on outflows, inflows, and
rotation. For this purpose, we adopt a cylindrical coordinate system where the vertical $z$-axis is perpendicular to the
disk, and the radial $R$ and angular $\phi$ directions define planes parallel to the disk. The origin of the coordinate
system is defined to be at the center of the galaxy, where in particular $z = 0$ defines the galactic plane. Note that
in the following we will also refer to the coordinate $r = \sqrt{R^2+z^2}$, which is the radial coordinate in the
spherical coordinate system, as it is easier to distinguish inflows and outflows in this coordinate.

\subsubsection{2D Velocity Profiles}\label{sec:2d_profiles}

\begin{figure*}
\centering
\includegraphics[width=0.47\textwidth]{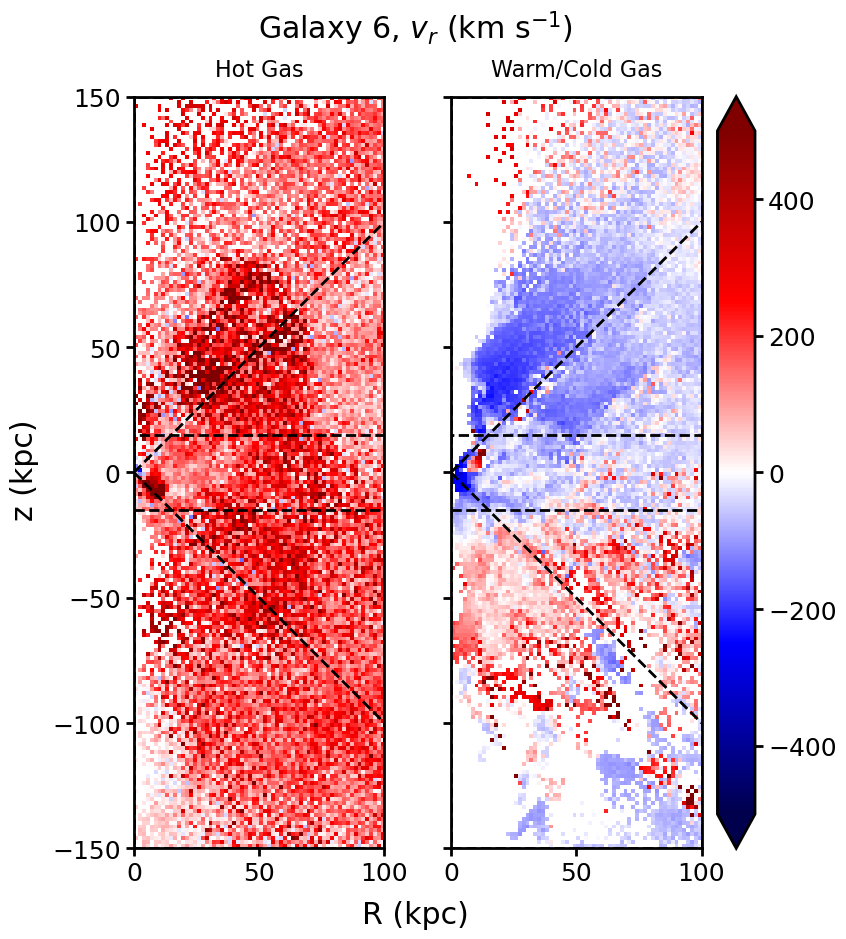}
\includegraphics[width=0.47\textwidth]{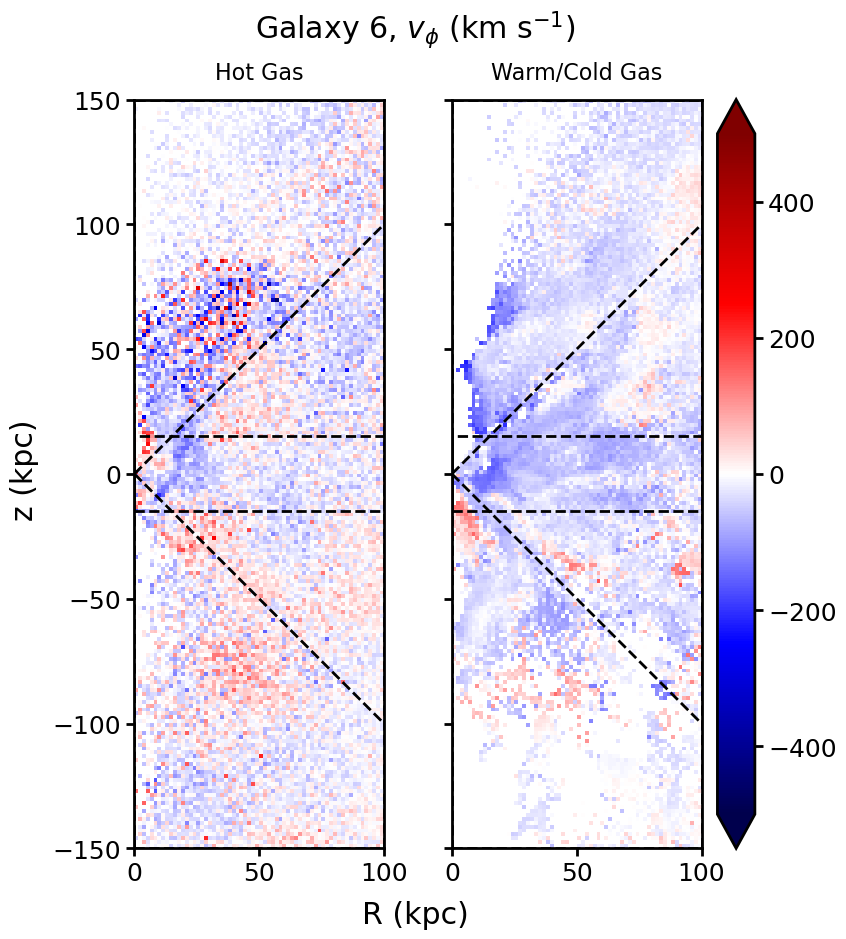}
\caption{Azimuthally averaged profiles in the radial ($R$) and vertical $z$ cylindrical coordinates of the spherical-$r$ and cylindrical-$\phi$ components of the gas velocity in hot (left sub-panels) and warm/cold (right sub-panels) gas phases for galaxy 6. The dashed horizontal and diagonal lines have the same meaning as in Figure \ref{fig:gal1_phase}. Unlike galaxies 1 and 2, the azimuthally averaged hot phase in galaxy 6 is all outflowing, and there is no coherent rotation. The warm/cold phase is counter-rotating to the stellar disk (see also Figure \ref{fig:gal6_profiles}).\label{fig:gal6_phase}}
\end{figure*}

Figures \ref{fig:gal1_phase}-\ref{fig:gal6_phase} show 2D profiles of the mass-weighted velocity in the spherical-$r$
direction (left sub-panels), representing inflows and outflows, and the cylindrical-$\phi$ direction (right sub-panels),
representing rotation and general tangential motions. The profiles are a function of $z$ and $R$ and are azimuthally
averaged over the $\phi$-direction, for both the ``hot'' ($T \geq 5 \times 10^5$~K) and ``warm/cold'' ($T < 5 \times
10^5$~K) gas phases in both galaxies.

The left panels of Figures \ref{fig:gal1_phase}-\ref{fig:gal6_phase} show the azimuthally averaged mass-weighted
spherical radial velocity $v_r$ for galaxies 1, 2, and 6. Galaxies 1 and 2 (Figures \ref{fig:gal1_phase} and
\ref{fig:gal2_phase}) display a straightforward geometry -- in conical regions above and below the disk plane at $z =
0$, aligned with the $z$-axis and with an opening angle of approximately 45$^\circ$, the hot phase (left sub-panels)
flows outward at speeds of $\sim$400-500~km~s$^{-1}$ or more. These regions are dominated by feedback. The hot gas flows
in this direction will be most easily observed in face-on disk galaxies, though the morphological features in X-ray SB
and temperature which accompany these flows will of course be observed most easily in edge-on disk galaxies. This basic
structure in SB and temperature in TNG50 galaxies was previously described in detail in \citet{Truong2021}. 

Outside of these conical regions, closer to the galactic plane, the hot phase is mostly slowly inflowing with a velocity
of $\sim$100~km~s$^{-1}$. In these two galaxies, there is not a significant amount of gas in the warm/cold phase, but it
is largely confined to the volume away from the conical hot outflow regions and is mostly inflowing at velocities of
$\sim$200-300~km~s$^{-1}$. Galaxy 6 (Figure \ref{fig:gal6_phase}), however, is quite different. Essentially all of the
hot phase is flowing radially outward at velocities of $\sim$300-500~km~s$^{-1}$, and the warm/cold phase, which makes
up a larger fraction of the mass of the CGM in this galaxy than galaxies 1 and 2, is mostly inflowing above the galactic
plane and mostly outflowing below it, with similar speeds. 
    
The right panels of Figures \ref{fig:gal1_phase}-\ref{fig:gal6_phase} show the azimuthally averaged mass-weighted
velocity in the $\phi$-direction. In galaxies 1 and 2, the hot phase (left sub-panels) shows coherent rotation of the
CGM within a cylindrical radius of at least $\sim$50~kpc and a height above the disk out to $\sim$75~kpc, though for
galaxy 2 it extends somewhat further out. The majority of the warm/cold phase (right sub-panels) rotates in a disk of
$\sim$50~kpc radius and $\sim$20-30~kpc thickness near the center, with other parts of the cold gas phase at large radii
largely co-rotating with the hot phase. The rotation of the gas will obviously be most easily observed in edge-on disk
galaxies. The hot phase in galaxy 6 shows essentially no coherent rotation and instead is moving mostly randomly in the azimuthal direction, and very slowly with speeds of $\sim$50~km~s$^{-1}$. The warm/cold phase is moving more coherently in the azimuthal direction, though in the opposite direction of rotation of the young stars (see Section \ref{sec:1d_profiles}), with speeds of $\sim$100-200~km~s$^{-1}$.

In summary, galaxies 1 and 2 are very similar, showing coherent hot outflows directed above and below the disk that push
hot gas outward in the $R$ and $z$-directions in conically-shaped regions on either side of the galaxy \citep[as shown
previously by][]{Nelson2019, Pillepich2021,Truong2021}, while hot gas flows slowly inward closer to the galactic plane.
Both galaxies also show coherent rotation of both phases, albeit at slightly different velocities (this will be explored
more in Section \ref{sec:1d_profiles}). Galaxy 6 does not have any coherent rotation of the hot phase, and does not have
any significant amount of gas which is inflowing. This is consistent with the disturbed appearance of the velocity field
in the maps in Figure \ref{fig:gal6_proj} in Section \ref{sec:projected}. Phase plots for the other three galaxies in
the sample (3, 4, and 5) are shown in Figures \ref{fig:gal3_phase}, \ref{fig:gal4_phase}, and \ref{fig:gal5_phase} in
Appendix \ref{sec:appendixB}. All three of these galaxies appear very similar to galaxies 1 and 2 in terms of outflows, inflows, and rotation, though Galaxy 5 (Figure \ref{fig:gal5_phase}) does have a slightly more complicated velocity structure in general. This galaxy has a stellar mass which is near the threshold for which the formation of a stable virial shock is expected, and thus may be more susceptible to denser and colder accreting streams of gas which can penetrate the hot and diffuse phase of the CGM \citep{Birnboim2003,Keres2005,Dekel2006}.
        
\subsubsection{1D Velocity Profiles}\label{sec:1d_profiles}

\begin{figure*}
\centering
\includegraphics[width=0.92\textwidth]{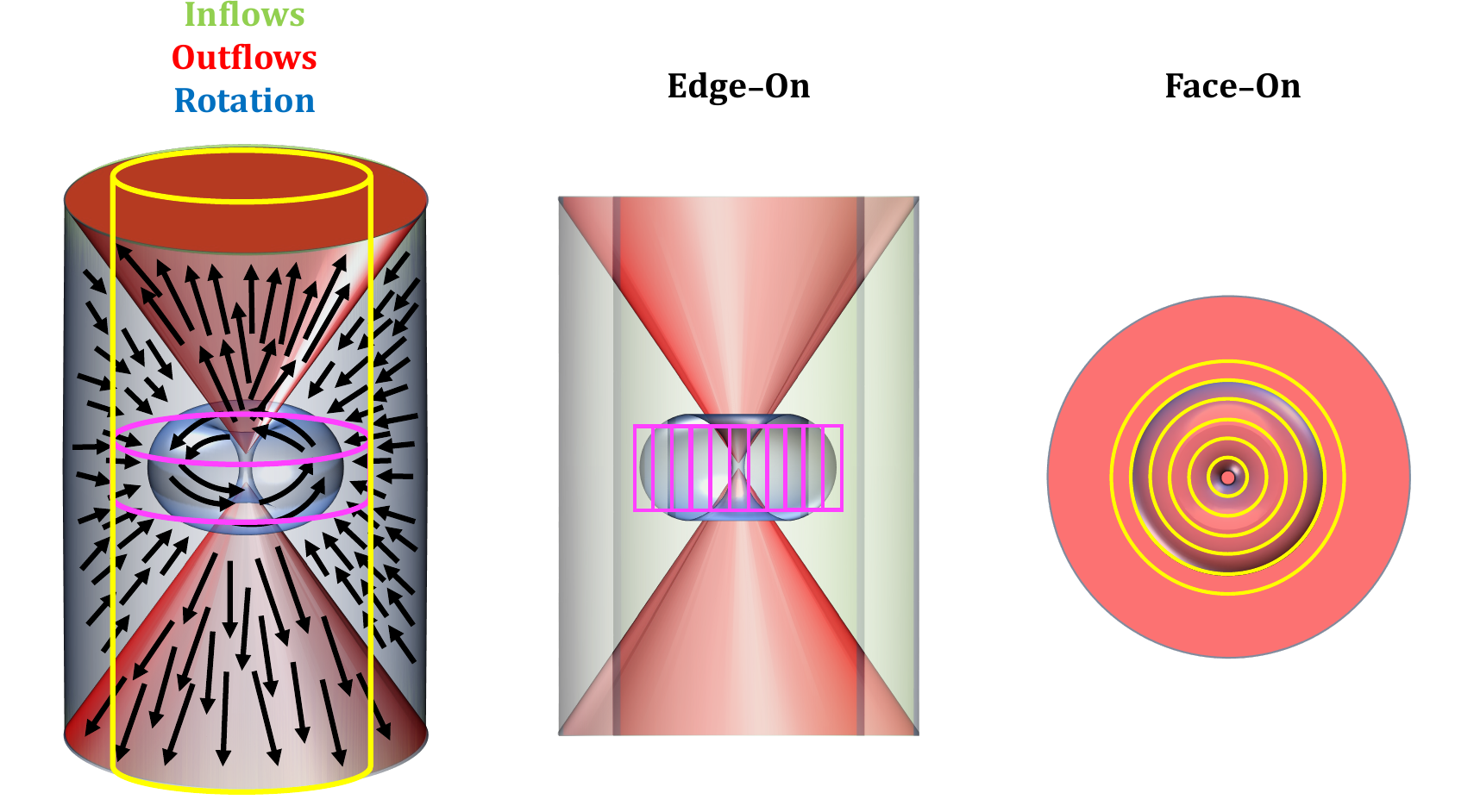}
\caption{Schematic of the general properties of the galaxies in our sample with a simple geometry. The left image shows the three regions of inflow (light green), outflow (light red), and rotation (light blue), with vectors indicating the directions of flow in the different regions. Two cylindrical regions (magenta and yellow) which would be used to analyze the X-ray emission in two different projections are also shown. The center image shows how this would appear in the edge-on projection, with the small magenta cylinder appearing as a rectangle within which the rotation curve can be measured. The right image shows how the larger yellow cylinder would appear in the face-on projection.\label{fig:schematic}}
\end{figure*}

How are the properties of the velocity field seen in the 2D profiles in the previous Section and the properties of the velocity field in projection seen in the maps in Section \ref{sec:projected} connected? To illustrate this, and to motivate the discussion in this and the following Sections, in Figure \ref{fig:schematic} we show a schematic representation of the major components of the velocity field for the galaxies in our sample (1-5) which have clearly distinguished regions of inflow (light green), outflow (light red), and rotation (light blue), shown in the left image with vectors indicating the directions of flow in the different regions. Given the cylindrical geometry, and the fact that X-ray spectra must be extracted over regions large enough to contain a statistically significant number of X-ray counts, two natural choices for regions to analyze the X-ray emission in cylindrical radial profiles in two different projections are also shown. The small magenta cylinder would be a logical choice for studying the radial profile of the rotation curve of the galaxy using line shifts in the edge-on projection, in which it would appear as a rectangular region (center image), where the smaller inset rectangles represent the radial bins and the regions from which spectra would be extracted. In addition to rotation in the inner hot CGM, in these regions the radial inflows in the outer regions would also have velocity components along the sight line, fully aligned with it at the very center and decreasing with projected radius. Assuming cylindrical symmetry, the radial inflow would produce a small line broadening, strongest near the center. In the face-on projection (right image), the larger yellow cylinder would represent measuring radial profiles in a set of circular annuli, which will probe the fast outflows near the center but also the slower inflows in the outer regions. Again assuming symmetry, both the outflows and the inflows would produce line broadening.

Motivated by these considerations, in this Section we produce azimuthally and height-averaged 1D profiles of the first
two moments of the velocity field along the three coordinate directions $R$, $\phi$, and $z$ in the cylindrical
coordinate system, as a function of cylindrical radius $R$ from the center of the galaxy. We also compare the profiles
of the gas velocities to those from young stars (defined for our purposes here to have ages $<$ 5~Gyr).
In what follows, for the $R$ and $\phi$ directions we extract 1D velocity profiles for the gas and young stars
(which will be viewed in edge-on projections) from a thin cylinder with a half-height of 30~kpc and radius of 100~kpc
(represented by the magenta cylinder in Figure \ref{fig:schematic} and shown with dashed lines in Figures
\ref{fig:gal1_phase}-\ref{fig:gal2_phase}). For the $z$-direction, we extract 1D velocity profiles (which will be viewed
in the face-on projection) from a cylinder of the same radius but a half-height of 1000~kpc, which corresponds to a long
cylinder with the axis projected along the line of sight (corresponding to the yellow cylinder in Figure
\ref{fig:schematic}).

\begin{figure*}
\centering
\includegraphics[width=0.82\textwidth]{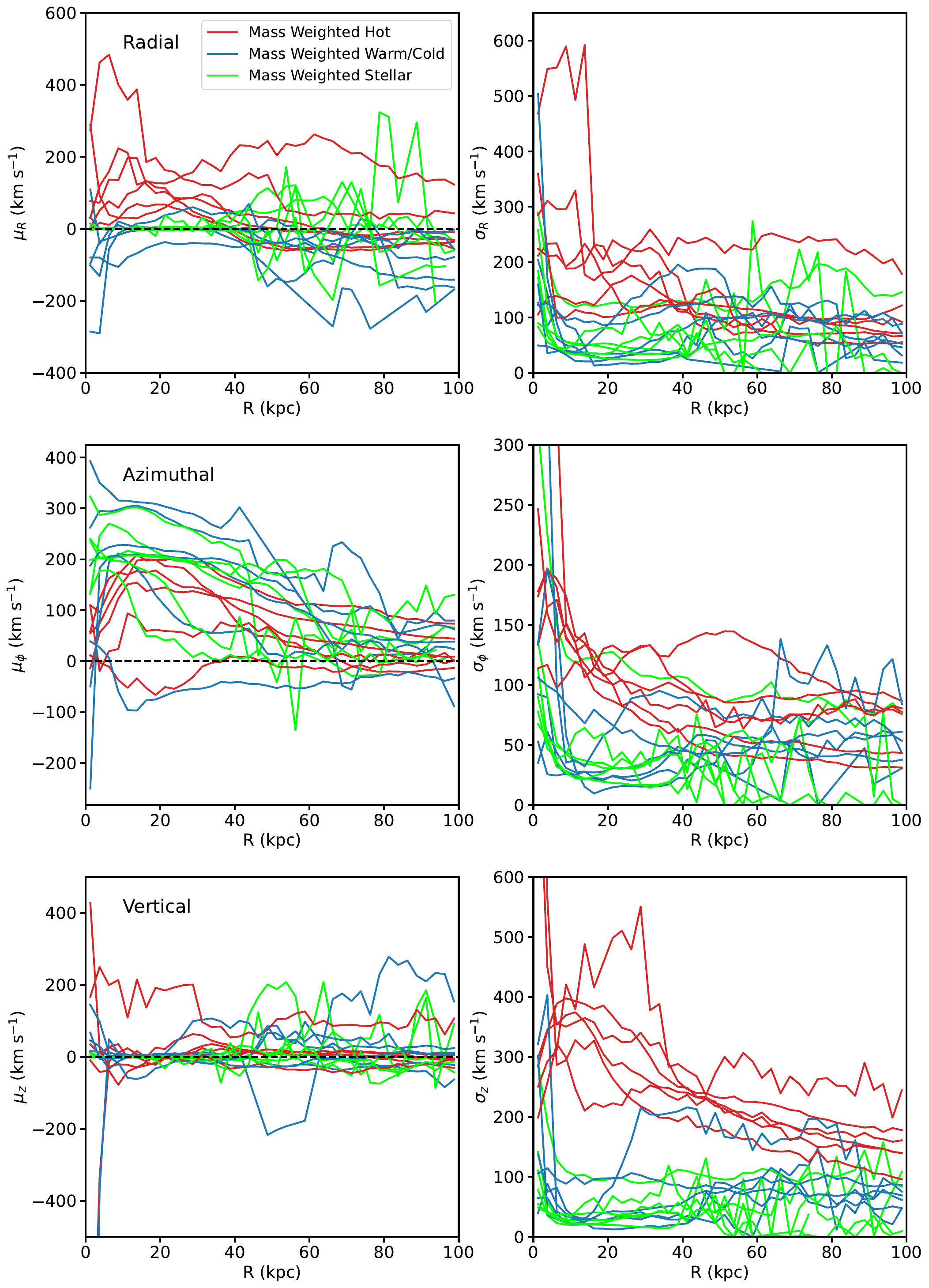}
\caption{Azimuthally and height-averaged mass-weighted radial profiles of the velocity of the gas and young stars for galaxies 1-6. The top, middle, and bottom panels show profiles of the $R$, $\phi$, and $z$-components of the velocity, respectively. Left panels show the mean velocity for each galaxy and right panels show the velocity dispersion. With two exceptions, the hot gas is outflowing near the center, inflowing at large cylindrical radii near the disk, and co-rotating with the stars and warm/cold gas. Warm/cold gas is mostly slowly inflowing at all radii. The hot gas also exhibits more dispersion in its motion in each direction than the warm/cold gas.\label{fig:all_profiles}}
\end{figure*}
    
We first show mass-weighted velocity profiles of the young stars (green), hot gas (red), and warm/cold gas
(blue) for all six galaxies in Figure \ref{fig:all_profiles}. The top panels of Figure
\ref{fig:all_profiles} show the mean velocity (left) and the velocity dispersion (right) in the
$R$-direction. In this direction, a complication from the geometrical considerations discussed above
immediately arises. For the galaxies with coherent hot outflows, the top and bottom of the thin
cylinder used to extract the profiles (30~kpc away from the galactic plane) intersects with the
boundary of this outflow at a radius of roughly $\sim$30-50~kpc (see Figures
\ref{fig:gal1_phase}-\ref{fig:gal2_phase}). Within this radius, the hot phase is outflowing with an
average velocity $\mu_{R} \sim +50-200$~km~s$^{-1}$. Outside of this region, the hot gas is
either outflowing with a similar velocity, or inflowing with $\mu_R \sim -50-100$~km~s$^{-1}$,
depending on whether this phase has the simple outflow/inflow structure, which is the case for
galaxies 1-5. The warm/cool phase gas, regardless of radius, is mostly inflowing with an
average velocity of $\mu_R \sim -100-300$~km~s$^{-1}$. The velocity dispersion in the radial
direction is typically larger for the hot phase, with $\sigma_R \sim 100-500$~km~s$^{-1}$, than the
warm/cold gas, which has $\sigma_R \sim 50-100$~km~s$^{-1}$. By symmetry, the mean radial profile of the young stars is near zero. The radial velocity dispersion of the young stars is low, $\lesssim~100$~km~s$^{-1}$.
 
\begin{figure*}
\centering
\includegraphics[width=0.82\textwidth]{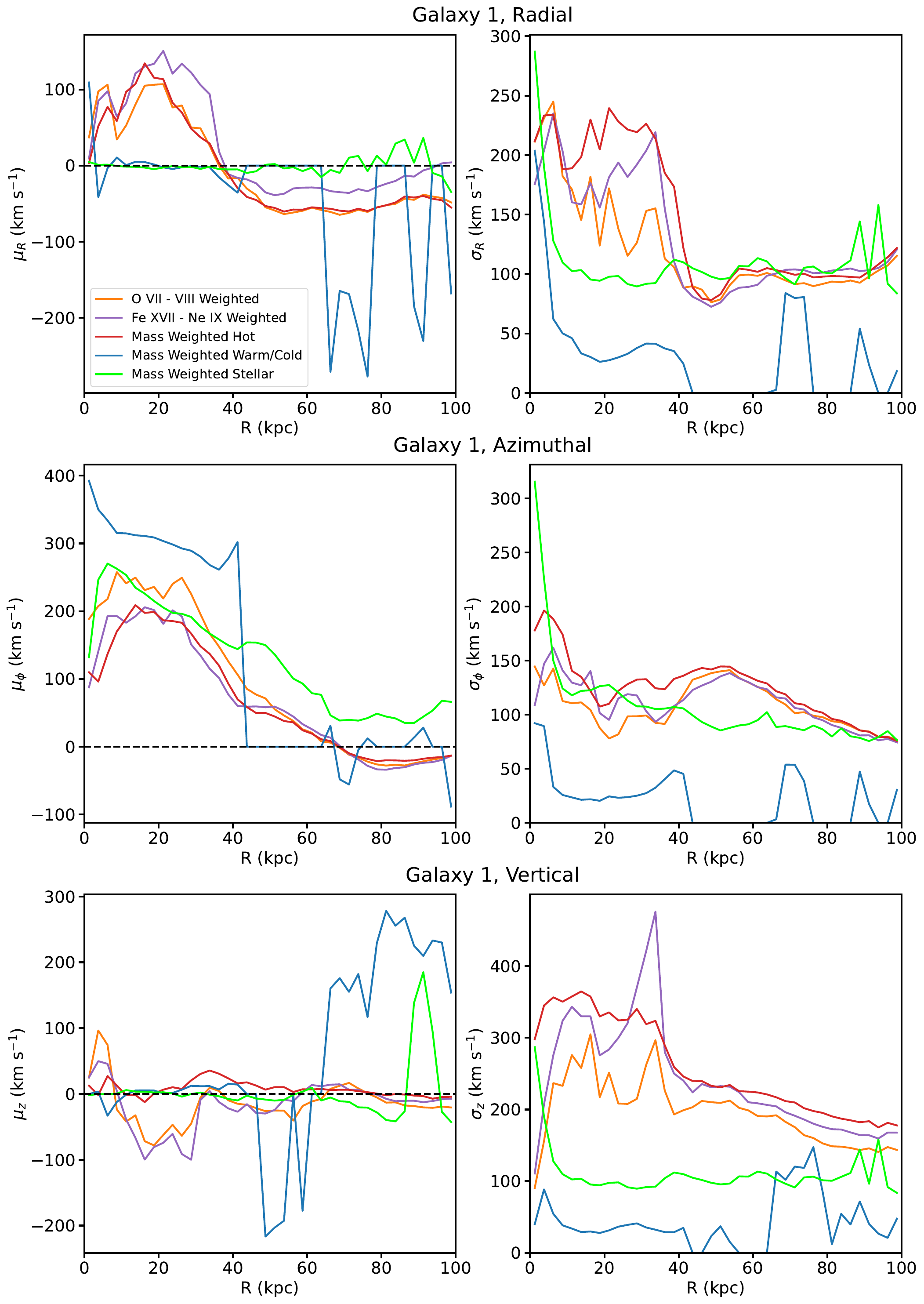}
\caption{Azimuthally and height-averaged mass-weighted and emission-weighted radial profiles of the gas and stellar velocity for galaxy 1. The top panels show profiles of the $R$-component of the velocity, the middle panels show profiles of the $\phi$-component, and the bottom panels show the $z$-component. Left panels show the mean velocity for each galaxy and right panels show the velocity dispersion. The same patterns of outflows, inflows, and rotation are seen as in the 2D profile in Figure \ref{fig:gal1_phase}. The emission-weighted velocity profiles are similar to the mass-weighted profiles, but higher-energy lines (Fe~XVII and Ne~IX) track hotter and faster phases of gas.\label{fig:gal1_profiles}}
\end{figure*}

\begin{figure*}
\centering
\includegraphics[width=0.82\textwidth]{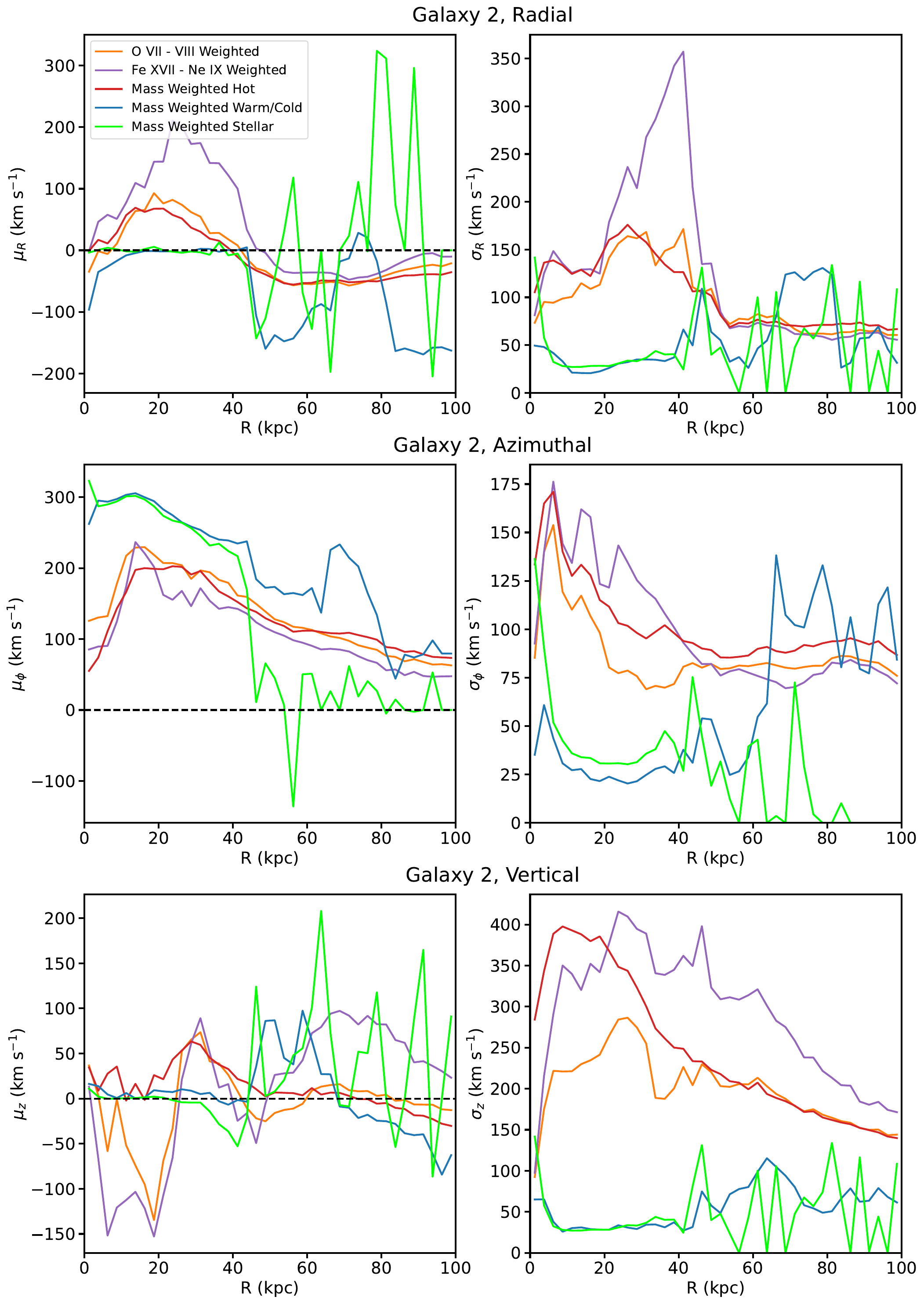}
\caption{Azimuthally and height-averaged mass-weighted and emission-weighted radial profiles of the gas and stellar velocity for galaxy 2. The description of the panels are the same as in Figure \ref{fig:gal1_profiles}. The general shape of the profiles is very similar to that of galaxy 1.\label{fig:gal2_profiles}}
\end{figure*}

\begin{figure*}
\centering
\includegraphics[width=0.82\textwidth]{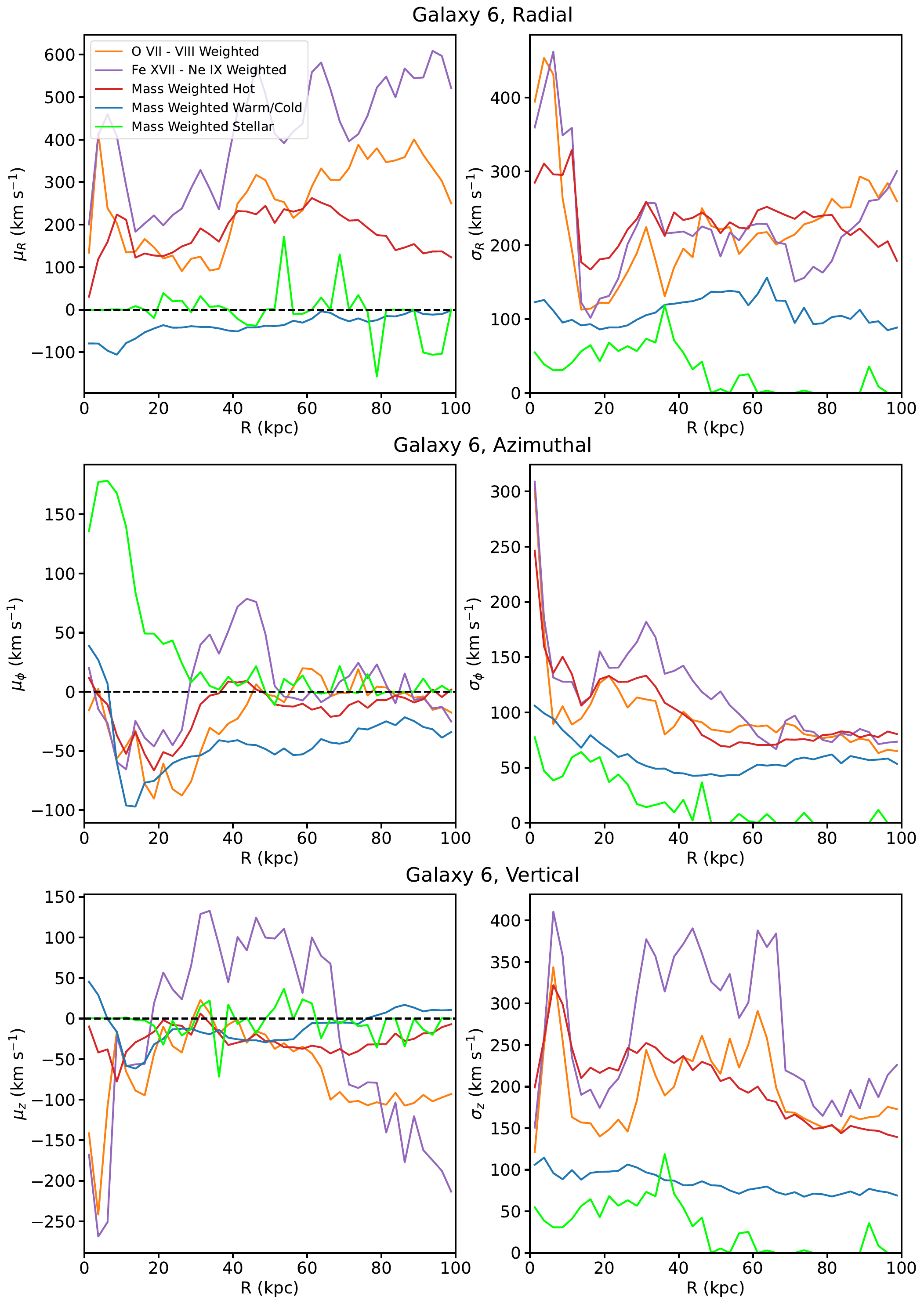}
\caption{Azimuthally and height-averaged mass-weighted and emission-weighted radial profiles of the gas and stellar velocity for galaxy 6. The description of the panels are the same as in Figure \ref{fig:gal1_profiles}. Unlike galaxies 1 and 2, this galaxy exhibits much faster outflows over a larger radial range, and the gas is counter-rotating to the stars.\label{fig:gal6_profiles}}
\end{figure*}
  
\begin{figure*}
\centering
\includegraphics[width=0.82\textwidth]{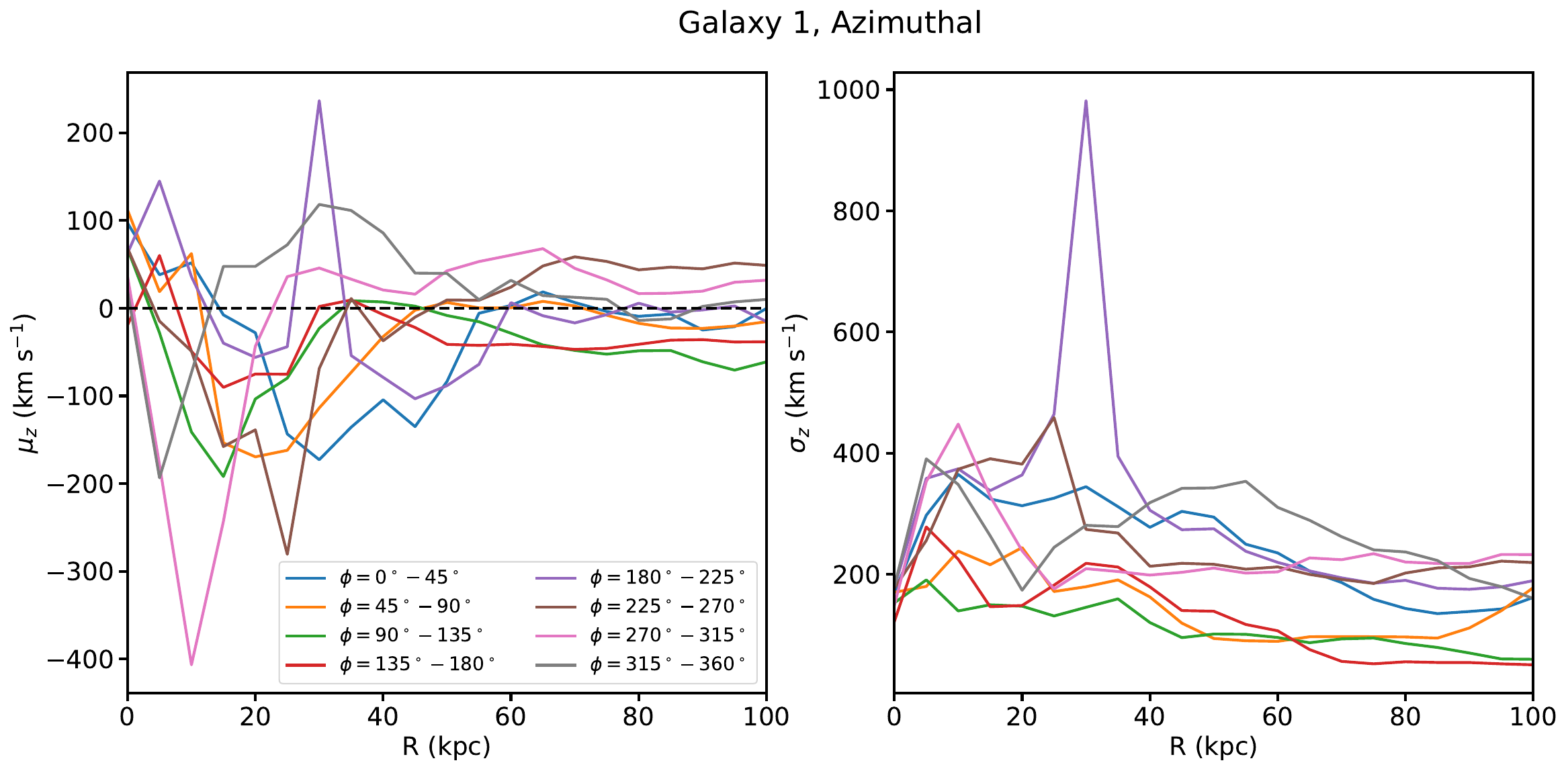}
\caption{Height-averaged emission-weighted (Fe~XVII-Ne~IX band) radial profiles of the hot gas velocity in the $z$-direction, separated into 45$^\circ$ sectors in the azimuthal direction around the profile. There is a considerable amount of scatter in the mean velocity profiles, which will at least partially contribute to the velocity dispersion measured in the full annulus.\label{fig:theta_profiles}}
\end{figure*}

The middle panels of Figure \ref{fig:all_profiles} show the same profiles for the $\phi$-component of the velocity. The
middle-left panel shows the mean $\phi$-velocity--essentially the rotation curves of the different phases. Though there
is a clear spread, in general the rotational speed of the cold gas is faster than the hot gas within a radius of
$\sim$50-kpc, with the former rotating at $\sim$200-400~km~s$^{-1}$ and the latter rotating at $\sim$50-200~km~s$^{-1}$.
The young stellar disks are rotating at $\sim$150-300~km~s$^{-1}$. All of the warm/cold phase
curves indicate rotation of this phase in the same direction as the stars--with the exception of one, which is galaxy 6
and will be discussed below. The middle-right panel shows that the $\phi$-velocity dispersions within $\sim$50~kpc for
the hot gas are slightly higher than the cold gas--$\sim$100-150~km~s$^{-1}$ versus $\sim$50-100~km~s$^{-1}$,
respectively. The lack of complete rotational support for the hot CGM in these TNG50 galaxies is consistent with
previous results from other works \citep{Oppenheimer2018,Huscher2021,Hafen2022}. We also note the fact that there is
more angular momentum in the warm/cold gas than the stars, in agreement with previous studies
\citep[e.g.][]{Oppenheimer2018} and shown to be common to simulations of galaxy formation by \citet{Stewart2017},
arising at least in part from cold, high-angular-momentum streams of infalling gas. The velocity dispersion of
the young stars in this direction is low, $\lesssim~100$~km~s$^{-1}$.
                                          
The bottom panels show the same profiles in the $z$-direction, which would be seen if a galaxy were
viewed face-on. The mean $z$-velocity profile hews very closely to zero for nearly all of the
profiles. This is expected if the outflows are nearly equal and opposite on either side of the disk
in galaxies observed face-on. Exceptions to this are most prominent in the very center ($r \lesssim
10-20$~kpc), where the volumes of the radial annuli are small enough that the average can dominated
by a few cells with high velocity (see also the bottom-left panels of Figures
\ref{fig:gal1_proj}-\ref{fig:gal6_proj} in Section \ref{sec:projected}, which show large velocity
shifts near the center). Deviations from zero velocity mean are more pronounced in the warm/cold
phase, which is sometimes dominated by large and coherent parcels of gas (see the phase plots in
Section \ref{sec:2d_profiles}). The $z$-velocity dispersion profiles show a clear separation
between the hot phase and the warm/cold phase--the former has velocity dispersions within
$\sim$40~kpc of $\sim$300-500~km~s$^{-1}$, and the latter has very low dispersions of
$\lesssim$100~km~s$^{-1}$, with one outlier curve with a dispersion of $\sim$200~km~s$^{-1}$ (galaxy
5) over almost the entire radial range. The high dispersions in the hot phase come from the
oppositely directed outflows on either side of the galaxy. By symmetry, the mean profile in the $z$-direction of the young stars is near zero. The $z$-velocity dispersion of the young stars is low, $\lesssim~100$~km~s$^{-1}$.
    
We now briefly look in more detail at the 1D velocity profiles of the individual galaxies. These are
shown for galaxies 1, 2, and 6 in Figures \ref{fig:gal1_profiles}-\ref{fig:gal6_profiles}. Here, we
also show the velocity profiles for the hot gas weighted by the X-ray emission in specific
source-frame bands around prominent emission lines in the CGM: O~VII and O~VIII (0.558-0.656~keV
band, orange lines), and Fe~XVII and Ne~IX (0.723-0.924~keV band, purple lines). These weightings
are significant since they correspond more closely to what X-ray microcalorimeter instruments will
be able to measure. The arrangement of the panels in Figures
\ref{fig:gal1_profiles}-\ref{fig:gal6_profiles} is the same as in Figure \ref{fig:all_profiles}. 
    
In the top panels of each figure, we show the first and second moments of the $R$-component of the
velocity. For galaxies 1 and 2 (Figures \ref{fig:gal1_profiles} and \ref{fig:gal2_profiles}), the hot gas is outflowing within $\sim$50~kpc with a velocity up to $\sim$100-200~km~s$^{-1}$, depending on the weighting used. For example, in galaxy 2 the hotter gas probed by the higher-energy emission lines of Fe~XVIII and Ne~IX is moving faster than both the cooler hot phase probed by the lower-energy O emission lines, as well as the gas probed by the mass weighting. At these inner radii, this is indicative of the hot, outflowing gas from the central SMBH. The warm/cold phase in this region has essentially zero radial velocity. Beyond this radius in these two galaxies, the hot gas is inflowing at $\sim$-50-100~km~s$^{-1}$, with slightly slower speeds for the phase weighted by the Fe~VIII-Ne~IX emission. Parcels of warm/cold gas are also inflowing at these radii with higher velocities near $\sim$150-300~km~s$^{-1}$. Galaxy 6 is very different in that the hot gas is strongly outflowing at all radii with velocities of $\sim$100-400~km~s$^{-1}$ depending on the weighting. The cold phase is inflowing at all radii, especially near the center, with velocity up to $\sim$100~km~s$^{-1}$.

In the $\phi$-direction (middle panels of Figures \ref{fig:gal1_profiles}-\ref{fig:gal6_profiles}),
the mean and dispersion of the velocity between the different weightings for the gas are all more
similar for galaxies 1 and 2. This is expected, since the azimuthal direction is least affected by the hot outflow. Galaxy 6 is once again seen to be quite different from the other two--both its warm/cold and hot phases have a mean velocity in the opposite direction of the rotation curve of the young stars. This is discussed in more detail in Appendix \ref{sec:AppendixC}.
    
The bottom panels of each figure shows the moments of the $z$-component of velocity. We note again
(as seen in Figure \ref{fig:all_profiles}) that in this projection the mass-weighted mean velocities
of both the hot and warm/cold phases are close to zero (as expected), and that the mass-weighted
velocity dispersion is higher for the hot phase. In the emission-weighted profiles, the absolute
value of the mean $z$-velocity can be significant in places, up to $\sim$150~km~s$^{-1}$. Similar to the $R$-component, the velocity dispersion in the $z$-direction weighted by the Fe~XVIII
and Ne~IX lines can be noticeably higher than that weighted by the O lines. Similar trends between the profiles are seen in the other three galaxies (3, 4, and 5), as seen in
Figures \ref{fig:gal3_profiles}, \ref{fig:gal4_profiles}, and \ref{fig:gal5_profiles} in Appendix
\ref{sec:appendixB}.

As already noted, azimuthally averaged profiles such as these are motivated not only by the geometry
but also by the number of X-ray counts available for an observation. This immediately introduces a
complicating factor--the mean and the standard deviation of the velocity in such a large region may
either arise from the velocity distribution along the sight line or from the velocity distribution
across the sky plane within the region. To check for this effect, we plot profiles of the velocity
in the $z$-direction for Galaxy 1 in the face-on projection in 8 different azimuthal sectors of
width 45$^\circ$ each in Figure \ref{fig:theta_profiles} (to be compared to the top panels of Figure
\ref{fig:gal1_profiles}.) This figure clearly shows that there is an effect on both the measured
velocity mean and dispersion from the azimuthal averaging (which can also be predicted from the
bottom panels of Figure \ref{fig:gal1_proj}). This should be taken into consideration when
interpreting the results of the azimuthally averaged profiles, and mitigated by splitting into
subregions if there are enough counts to do so.

\begin{figure*}[!t]
\centering
\includegraphics[width=0.92\textwidth]{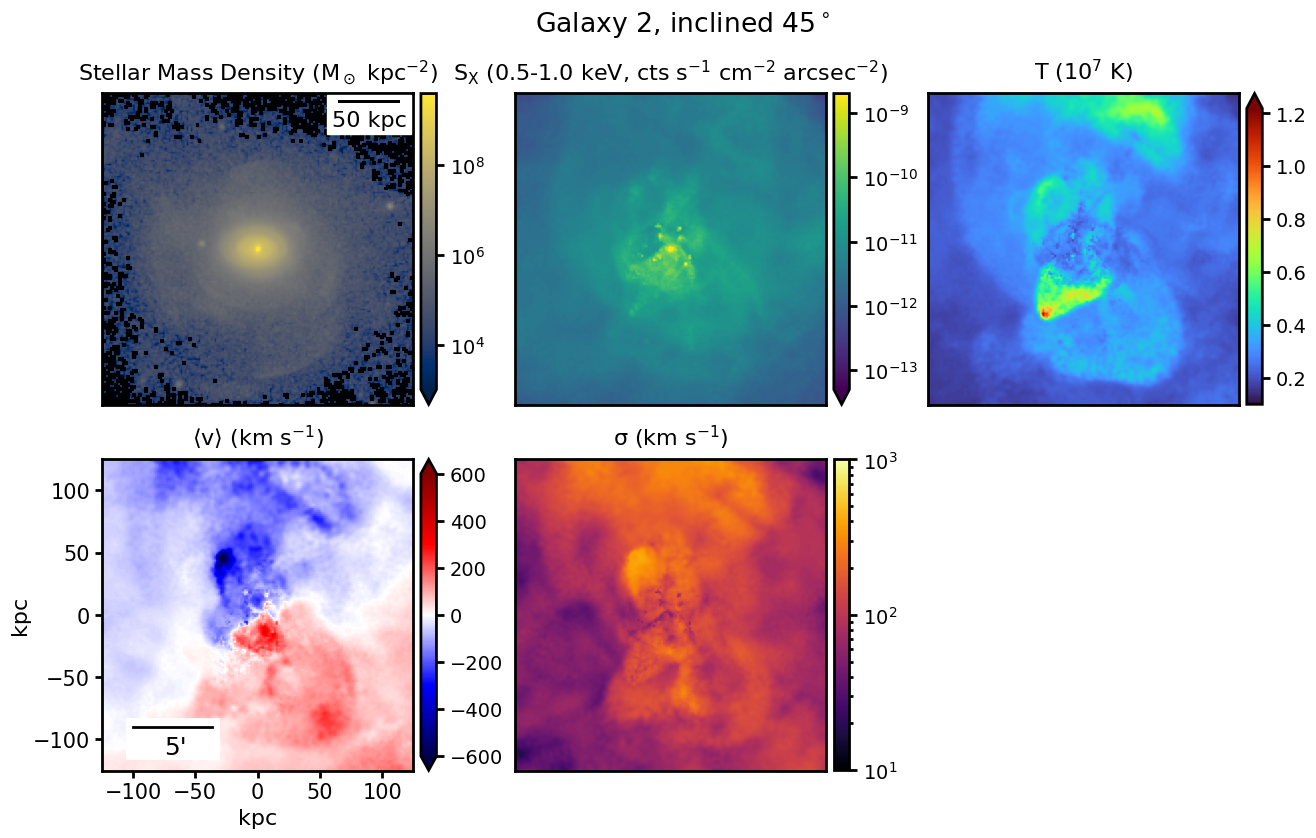}
\caption{Projections of various quantities from galaxy 2, viewed at a 45$^{\circ}$-angle with respect to the plane of the galactic disk. Panel descriptions are the same as in Figure \ref{fig:gal1_proj}. Each panel is 250~kpc on a side, or $\sim$20' for the given redshift and cosmology. Similar to the edge-on maps, observable signatures of the hot outflows appear in both SB and temperature above and below the galactic center, but in the mean velocity map there is a clear superposition of both rotation and outflow velocities that contributes to the overall pattern in the map.\label{fig:mid_maps}}
\end{figure*}
    
\begin{figure*}[!t]
\centering
\includegraphics[width=0.95\textwidth]{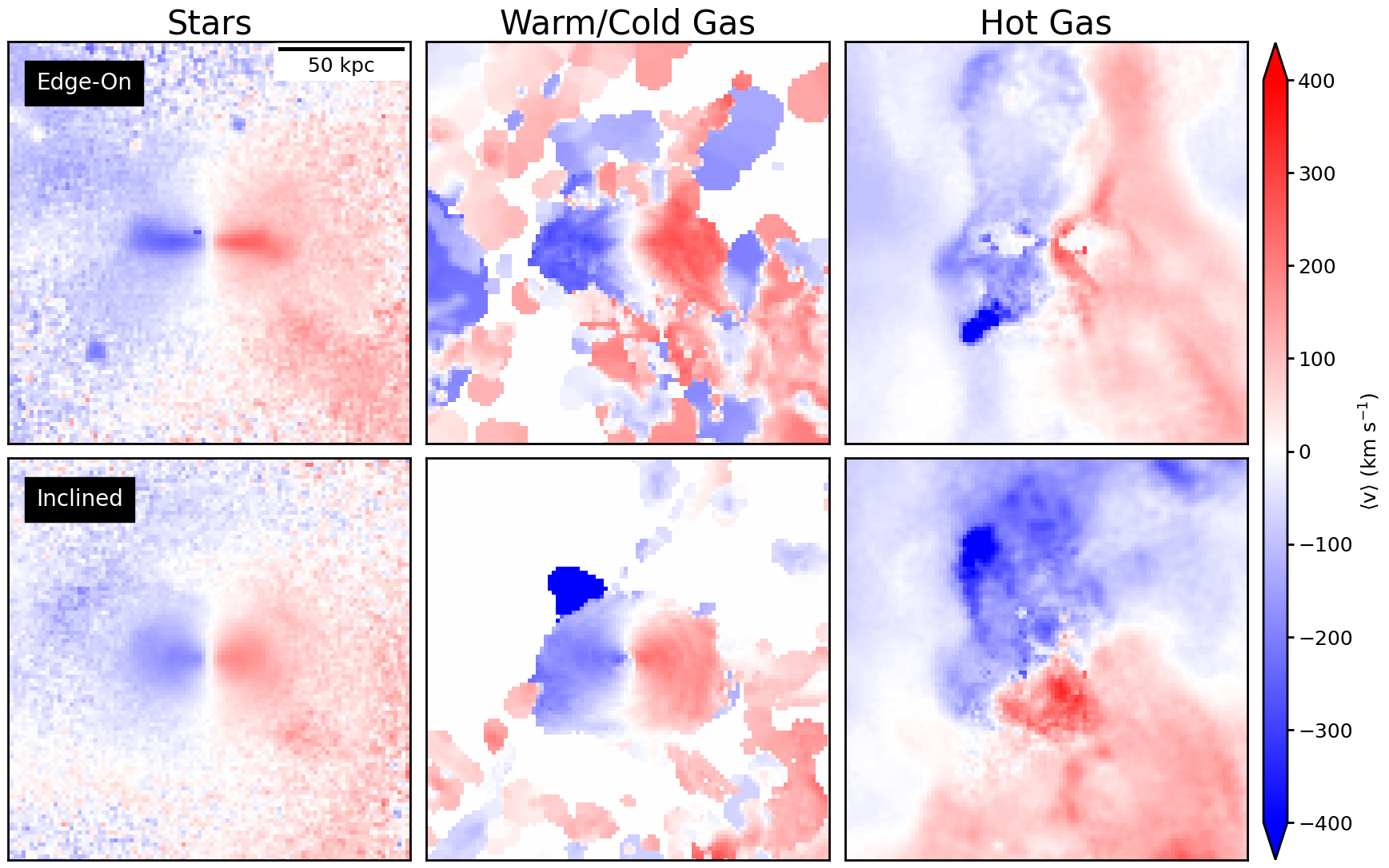}
\caption{Line-of-sight velocity maps in the edge-on (top panels) and inclined (bottom panels) projections. Stellar velocities (left) and warm/cold gas (middle) are weighted by mass, and the hot gas is weighted by X-ray emission in the 0.5-1~keV band. For both the stars and the warm/cold gas, the central regions are dominated by rotation in both the edge-on and inclined projections (with the motions becoming more random at larger distances), whereas the hot gas exhibits rotation in the edge-on projection and a superposition of outflow and rotation velocities together in the inclined projection.\label{fig:edge_vs_45}}
\end{figure*}
                        
\subsection{Off-Axis Projections}\label{sec:off_axis}

Of course, most galaxies will not be inclined either perfectly edge-on or face-on to our sight line. In off-axis
projections, components of the velocity field from both the rotating CGM and the hot outflows will be observable
together. In the two projections we have examined so far, the outflow velocities could not be easily distinguished via line shifts, either because they were mainly out of the sight line in the edge-on case, or these line shifts were largely canceled out by the approximately biconical symmetry. In the case of an
off-axis projection, these outflow velocities could be measured, and if the inclination angle can be constrained from
the stellar disk the total outflow velocity may be estimated. 
        
Figure \ref{fig:mid_maps} shows maps of the same quantities as shown in Section \ref{sec:projected},
except along a sight line 45$^\circ$ away from both the edge-on and face-on projections, for
galaxy 2, to give an example. The most intriguing of these images are the projected mean
velocity maps (bottom-left panels for each galaxy). The outflow velocities on either side of the
galaxy are clearly seen, and can be spatially matched with features in X-ray SB (top-middle panel
for each galaxy) showing outflows and cavities. The pattern of the velocity field in the map is also
twisted from a purely vertical dipole, showing the effect of both CGM rotation and outflows in the
same projection.

\begin{figure*}[!t]
\centering
\includegraphics[width=0.47\textwidth]{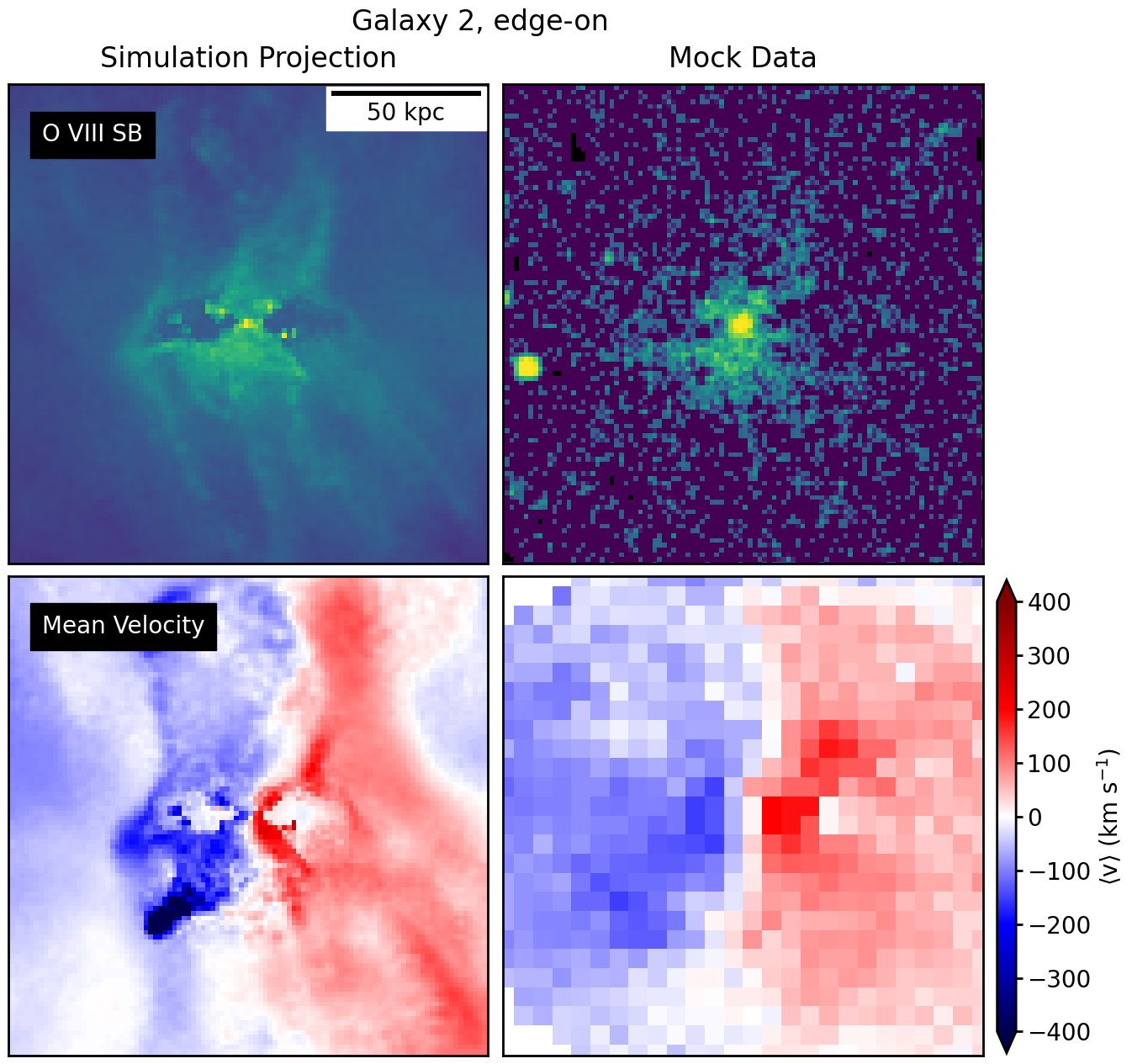}
\includegraphics[width=0.47\textwidth]{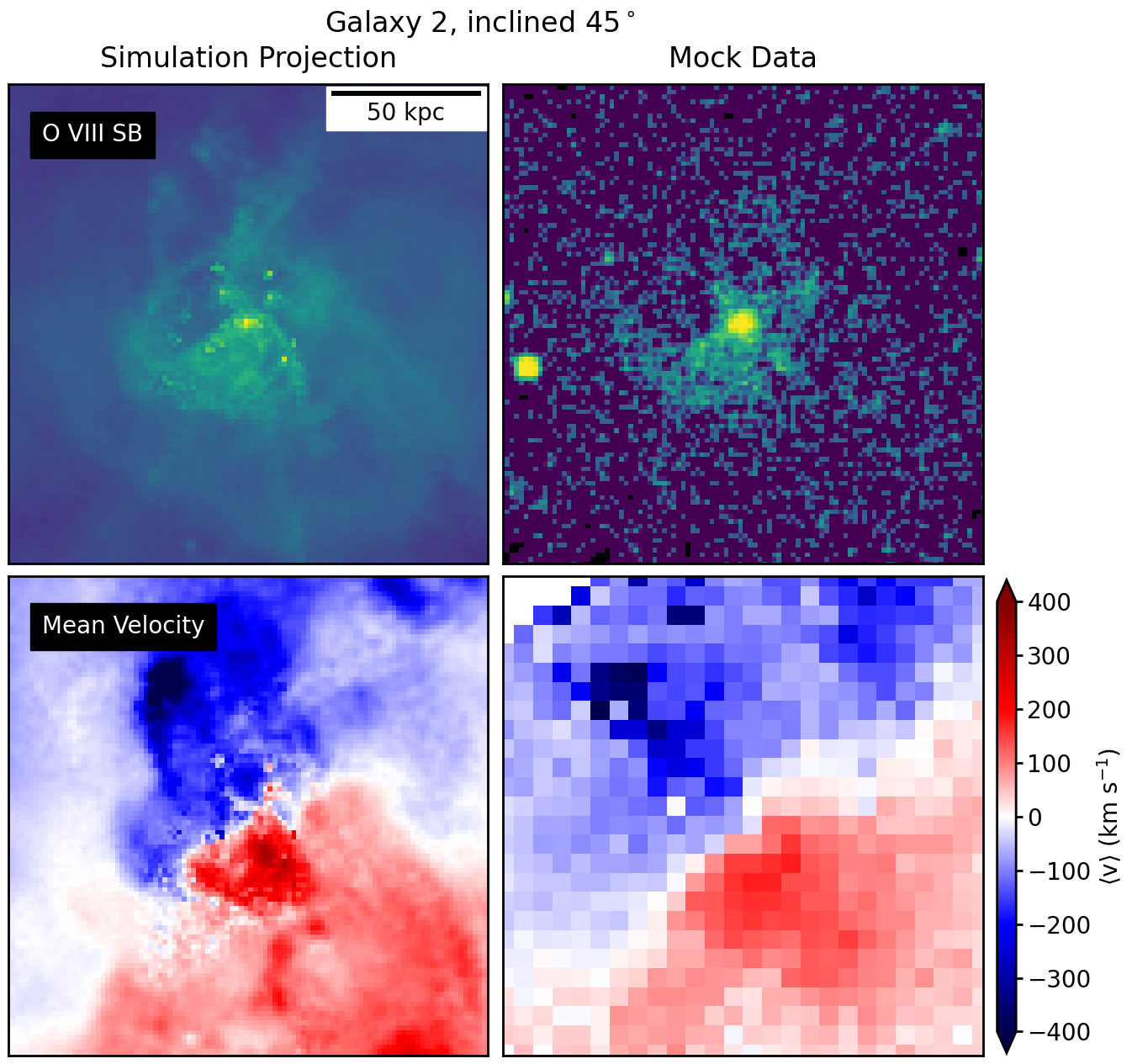}
\caption{Projected O~VIII emission line SB and velocity maps from the simulations and the mock observations, in the edge-on (left panels) and inclined by 45$^\circ$ (right panels) directions. In each direction, the left subpanels show quantites projected directly from the simulations, and the right subpanels show the X-ray counts map (top, with background included and brightest CXB sources removed) and the velocity map (bottom) obtained from spectral fitting as detailed in Section \ref{sec:mocks_velmaps}. The fits to the mock observations recover the mean velocity field map to a significant degree, with typical uncertainties on the velocity of $\pm 20-30$~km~s$^{-1}$.\label{fig:sim_and_mock}}
\end{figure*}
    
This last effect is particularly interesting, as it reveals how the different baryonic phases can
have very different kinematic properties. Figure \ref{fig:edge_vs_45} shows line-of-sight mean
velocity maps for the stars and warm/cold and hot gas phases, for both the edge-on and inclined
projections. In the edge-on projection, the stars and both gas phases show a common rotation pattern
in the inner $\sim$30~kpc region of the galaxy, though they may rotate at slightly different speeds
as noted in Section \ref{sec:velocity_profiles}. In the inclined projection, though the stars and
warm/cold gas show the same axis of rotation as before, the velocity pattern of the CGM is tiltled
with respect to both due to the hot, directed outflow signature combined with the rotation signature. This is an intriguing prediction which can only be tested with an X-ray microcalorimeter.
                                
\subsection{Mock X-ray Observations}\label{sec:mocks_results}

In this Section, we produce synthetic X-ray observations of galaxies 1 and 2 using the procedure
described in Section \ref{sec:xrays}, using a model with instrument characteristics similar to
\textit{LEM} \citep{LEMWhitePaper}. In the spectral analysis that follows,
the brightest $\sim$50-100~CXB point sources have been identified using \texttt{wavdetect}
\citep{Freeman2002} and removed.

\subsubsection{Velocity Maps from Spectral Fitting}\label{sec:mocks_velmaps}

If statistics permit, X-ray IFUs will be most useful in producing maps of projected quantities from
model fits to spectra. To demonstrate this, we carry out such a procedure on two of our model event
files to produce maps of line-of-sight mean velocity. This analysis is carried out using the CIAO
\citep{Ciao2006} and Sherpa packages \citep{Burke2020}. 
        
We first extract a spectrum from a region by removing all emission within a radius of 12.5' from the galaxy center. We fit this spectrum to a combined model for the MW foreground (\texttt{apec + TBabs*(bapec+bapec)}, where \texttt{TBabs} is the same foreground absorption model described in Section \ref{sec:xrays}, and \texttt{bapec} is an APEC CIE model with thermal line broadening), one power-law component for the (unresolved) CXB, and another for the NXB. 

With the background determined, we proceed to produce the velocity map. We then bin the counts
images in the O~VIIf\footnote{Only the O~VII forbidden line is sufficiently redshifted away from the
MW foreground lines at $z = 0.01$ to be used for this purpose.}, O~VIII, and Fe~XVII lines (defined
by narrow bands around the line centroids at $z = 0.01$ with width 3~eV) into 30" pixels (twice the
size of the pixels in the simulated instrument). Each of these larger pixels is the center of a
circular region where the radius is expanded until it reaches a SNR of 7, where the maximum allowed
radius of each circle is 4.5' (18 pixels). Spectra are extracted from these regions, and grouped so
that there is at least one count in each energy bin of the spectra. 
    
Each circular region is then fit to a single-component \texttt{TBabs*bapec} model for the source,
with the parameters for the model components corresponding to the MW foreground and the NXB frozen to the values obtained from the background-only fit (rescaled by area), and the normalization
of the CXB component free to vary to account for the variable CXB contributions in each localized
circular region. We fit each spectrum in 8~eV bands around the O~VII, O~VIII, and Fe~XVII lines. For
the source model, the temperature, redshift, line width, and normalization parameters are free to
vary. The hydrogen column density for foreground galactic absorption is fixed to $N_H = 1.8 \times 10^{20}$~cm$^{-2}$, and the abundance parameter is fixed to $Z = 0.3~Z_\odot$, assuming \citet{angr89} relative abundances. These parameters are fixed given the narrow spectral bands we use in the fits; neither of them will be well-constrained by the fit and the measurement of the line shift is not sensitive to their value in any case. We fit by minimizing the Cash statistic \cite{Cash1979}. Once a best fit for a region is found, we refine it by running a Markov Chain Monte Carlo (MCMC) analysis with 2500 steps.

Figure \ref{fig:sim_and_mock} shows the result of this procedure on the observation of galaxy 2 with
the sight line facing edge-on (left subpanels) and inclined 45$^\circ$ to the plane of the galactic
disk (right subpanels). The top subpanels show maps of SB in the O~VIII line, with the idealized SB
map projected from the simulation in the top-left subpanels and the counts map from the mock
observation in the top-right subpanels. The bottom subpanels show the line-of-sight mean velocity,
computed from the simulation by weighting by the emission in the 0.5-1~keV band (bottom-left
subpanels), and produced from the fitted line centroid as described above (bottom-right subpanels).
Typical uncertainties on the mean velocity from the fits are $\pm 20-30$~km~s$^{-1}$. There is
remarkable agreement between the simulated and fitted velocity maps, with the model fits reproducing
the overall shape of the velocity distribution as well as the magnitude of the velocity in either
direction (see also \citet{Truong2024} for similar analyses in TNG-Cluster; \citet{Nelson2024}). 
The absolute values of the most extreme values of the idealized map are slightly
underestimated due to the fact that they appear in small regions with faint emission that do not
contribute greatly to the spectra in their respective circular regions. The reproduction of the
general features in the map demonstrates that an X-ray IFU with $\sim$1~eV spectral resolution will
be able to map the velocity field of the CGM to sufficient detail to observe the effects of rotation
and outflows.
    
\subsubsection{Velocity Distributions in Regions from Spectral Fitting}\label{sec:mocks_profiles}

\begin{figure*}[!t]
\centering
\includegraphics[width=0.85\textwidth]{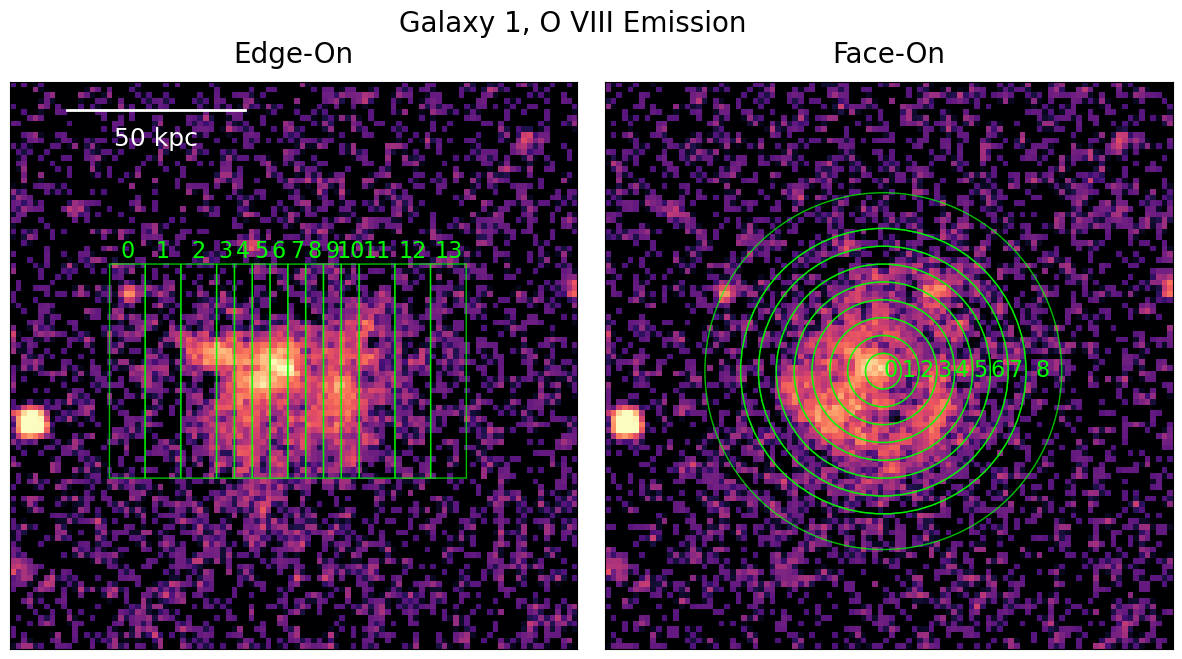}
\caption{Mock X-ray images of galaxy 1 in the edge-on (left panel) and face-on (right panel) projections. The images consist of source, background AGN and galaxy point sources, and MW foreground counts in the 0.645-0.649~keV band, which bounds the redshifted O~VIII line doublet at z = 0.01. Regions shown are those from which spectra are extracted for model fitting in Section \ref{sec:mocks_profiles}. The regions in both panels are labeled with numbers which are referenced in Figures \ref{fig:gal1_mock_edge}-\ref{fig:fit_phase_space}.\label{fig:mock_image}}
\end{figure*}

Figure \ref{fig:mock_image} shows a 1~Ms exposure of galaxy 1 in the face-on and edge-on
projections, where the plotted events have been restricted to the 0.646-0.649~keV band, which bounds
the redshifted O~VIII line at z = 0.01. As noted in Section \ref{sec:xrays}, all backgrounds are
included in this image. Also overlaid on the two panels in Figure \ref{fig:mock_image} are numbered
regions from which spectra are extracted for fitting to emission models for the analysis in Section
\ref{sec:mocks_profiles}. In the edge-on image (left panel), the regions are made of rectangles so
that the velocity profile of the CGM can be measured across the disk. In the face-on image, the
regions are made of annuli, reflecting the approximate cylindrical symmetry along this sight line. 

We extract spectra from the regions shown in Figure \ref{fig:mock_image} and fit them using \texttt{XSPEC}
\citep{Arnaud1996}. In each region, we model the CGM emission using a single \texttt{TBabs*bapec} component, where the
hydrogen column density for foreground galactic absorption and the metallicity parameters are fixed as above, and the
temperature, redshift, velocity broadening, and normalization parameters are free to vary. In this model, the
velocity distribution function is modeled as a single Gaussian. For the galactic foreground emission, we assume the
model given in Section \ref{sec:xrays}, holding all parameters fixed except an overall constant normalization which is
free to vary. A power-law component is included to model the CXB, with its photon index and normalization parameters
free to vary. Finally, the normalization of the constant particle background component is also free to vary. We fit
within the 0.64-0.83~keV band (covering the O~VIII and Fe~XVII lines), and use the Cash \citep{Cash1979} statistic for
minimization. 

\begin{figure*}[!t]
\centering
\includegraphics[width=0.95\textwidth]{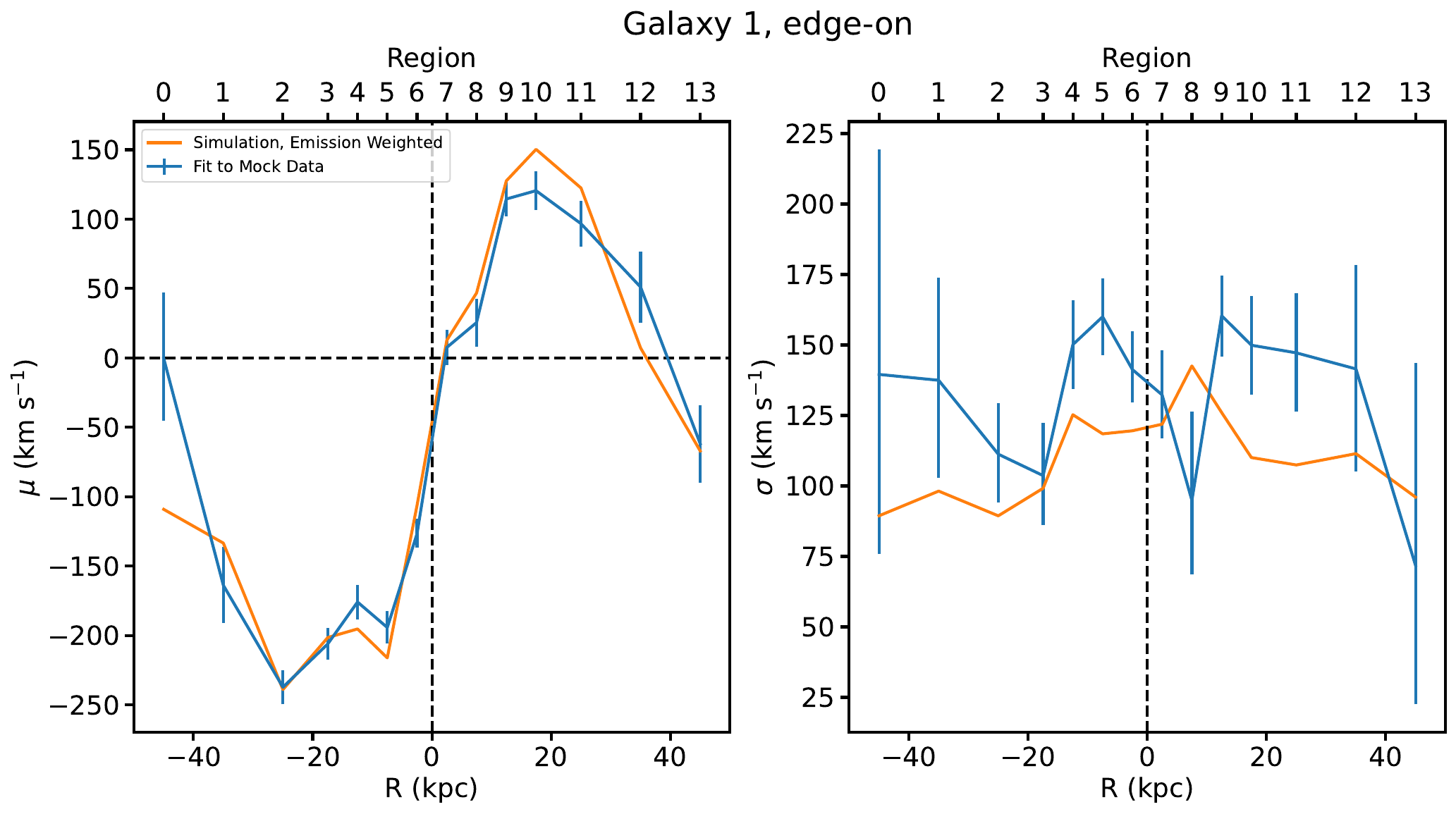}
\caption{Profiles of mean velocity / line shift (left) and velocity broadening / line width (right). Blue lines show the result from spectral fits to the numbered box regions in the edge-on projection (left panel of Figure \ref{fig:mock_image}). Orange lines show the projected emission-weighted values directly from the simulation. The rotation curve of the hot CGM is clearly recovered in the line shift measurements, whereas line broadening measures oppositely directed inflows near the galactic center and dispersion in the tangential velocity at larger radii.\label{fig:gal1_mock_edge}}
\end{figure*}

The result is shown in Figure \ref{fig:gal1_mock_edge}, where the blue lines show the mean (left panel) and standard deviation (right panel) of the velocity as determined from the spectral fitting, and the orange lines show the same quantities projected directly from the simulation weighted by the X-ray emission in the 0.5-1.0~keV band. The shape of the mean velocity measurements clearly shows the rotation curve, and is in excellent agreement with the simulation projection and broad agreement with the azimuthally averaged curve of the same quantity in the middle-left panel of Figure \ref{fig:gal1_profiles}. The measured velocity dispersion is in somewhat less agreement with the simulation projection, though the uncertainties are large. This quantity is more difficult to constrain than the mean velocity, and the assumption of a single Gaussian distribution may not be the best model for the underlying velocity distribution. Despite these facts, the disagreement between the velocity broadening parameter from the simulation prediction and that from the fitted model is modest, always within $\sim$50~km~s$^{-1}$ across the profile. For the reasons discussed in Section \ref{sec:1d_profiles}, this quantity will be dominated by the oppositely directed radial ($R$) inflows near the center of the galaxy, while at larger projected radius the contribution of differences along the azimuthal ($\phi$) direction will become more important. The numbers measured here are consistent with those from the 1D radial profiles in the top-right and middle-right panels of Figure \ref{fig:gal1_profiles}.
                               
For the face-on projection, the story is somewhat different. In this case, where we project along
the $z$-axis of the cylinder, there is a complex distribution of outflows and inflows with different
temperatures and velocities. As we have already seen (Figure \ref{fig:theta_profiles}), azimuthally
averaging within an annular region also combines different phases in a non-trivial way. We find for
many of the annular regions shown in Figure \ref{fig:mock_image} that a single thermal emission
model component does not adequately represent the observed emission from the CGM within them. To
this end, we have fit these 9 regions with multiple components to attempt to capture multiple
temperature and velocity components in the hot gas from both the outflows and the inflows, which we
would predict to be observed especially along this sight line, from the results of Sections
\ref{sec:projected} and \ref{sec:velocity_profiles}. For these fits, we have successively added
additional \texttt{bapec} components until no more components were statistically required. In order
to make sure we can also correctly detect and characterize weaker emission components, we used the
whole energy range (0.3-2~keV) and kept background parameters free to vary in this part of the
analysis. This kind of deep, multi-component analysis will only be possible with
microcalorimeter-quality data.
 
\begin{figure*}[!t]
\centering
\includegraphics[width=0.95\textwidth]{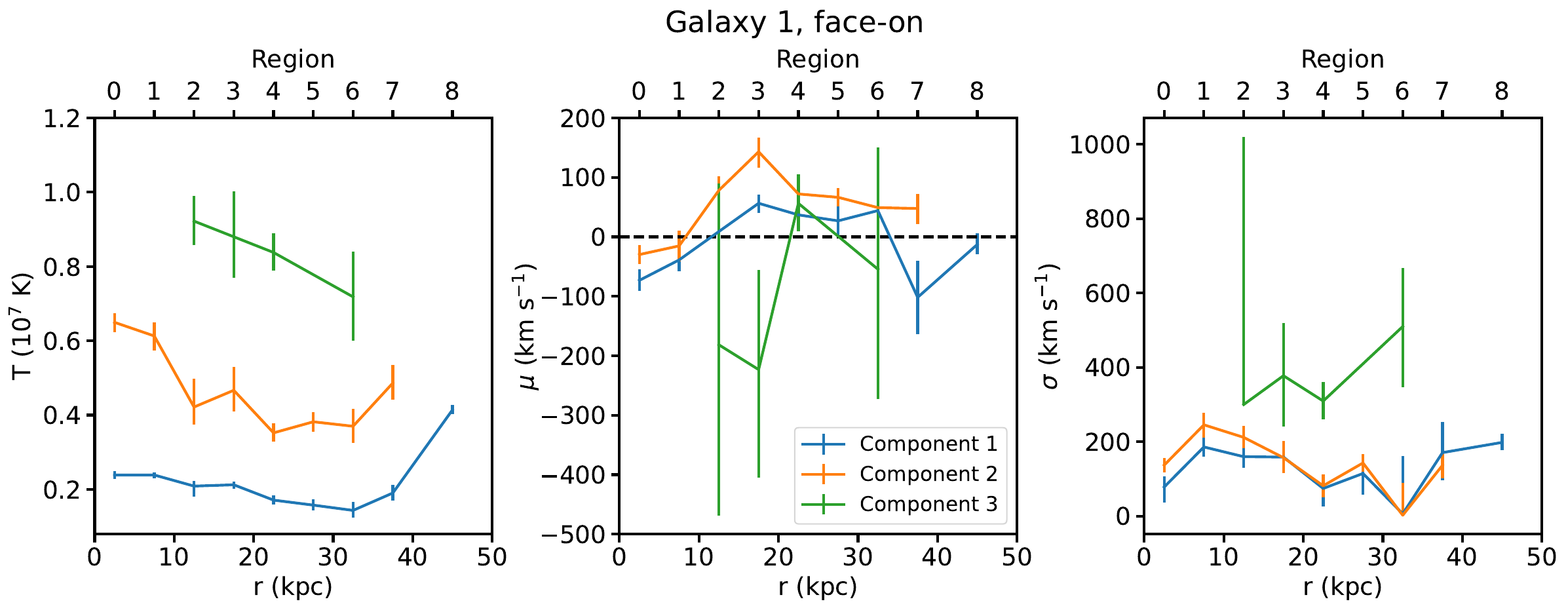}
\caption{Results of 1, 2, and 3 \texttt{bapec} component fits to spectra extracted from the numbered face-on regions shown in the right panel of Figure \ref{fig:mock_image}. Left panel: Temperature. Center panel: Mean velocity / line shift. Right panel: Velocity dispersion / line width. The multi-component fits are able to clearly recover distinct phases of gas, with colder/hotter temperatures corresponding to slower/faster velocities.\label{fig:gal1_mock_face}}
\end{figure*}

\begin{figure*}[!t]
\centering
\includegraphics[width=0.94\textwidth]{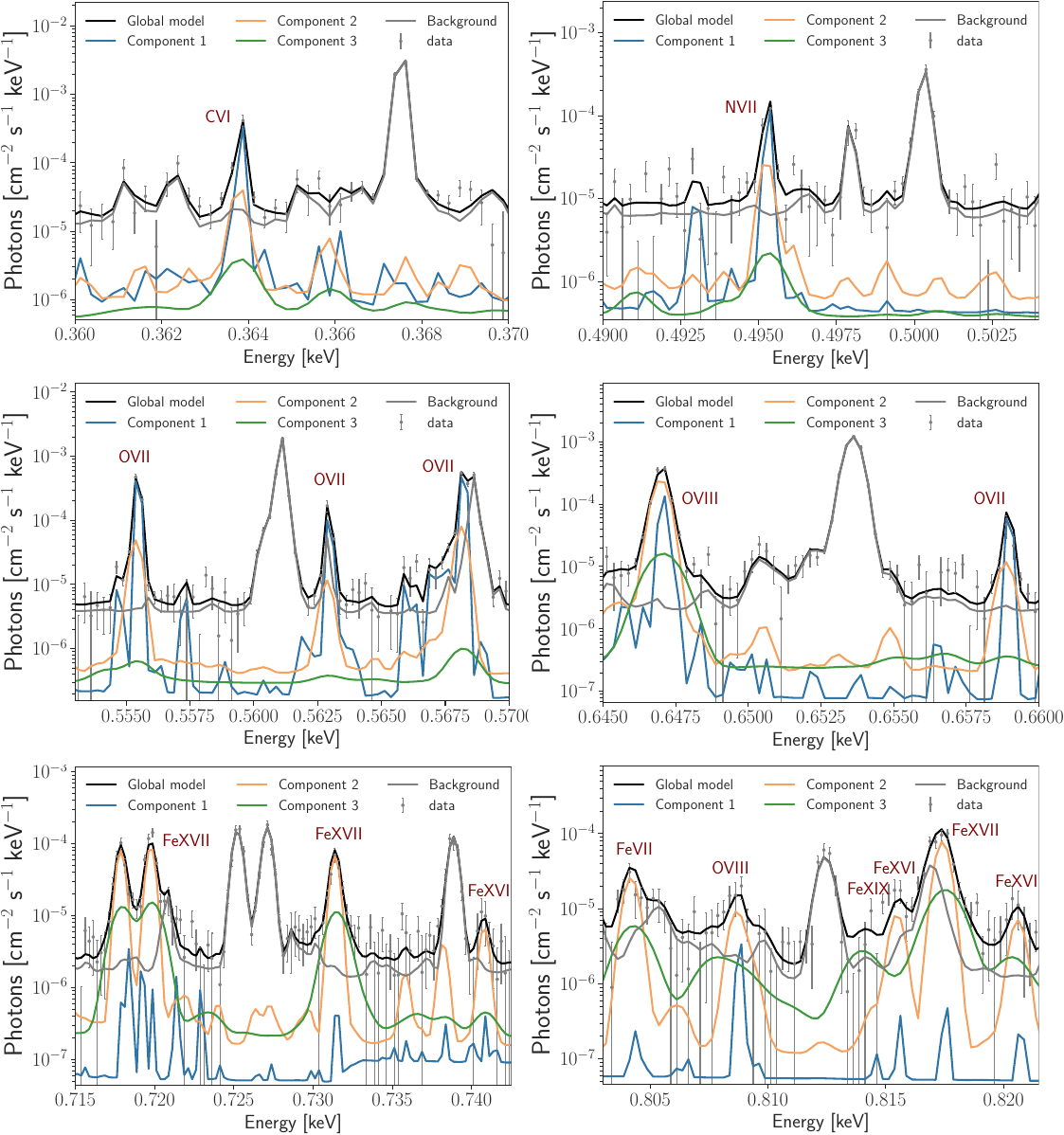}
\caption{Spectrum and best-fit model of Region 4 from the right panel of Figure \ref{fig:mock_image}, shown in selected energy bands around key CGM emission lines. The full multicomponent model is shown in black. We show the three kinematic components in colors matching Figure \ref{fig:gal1_mock_face}, the data are the grey points, and the background is shown in grey. Microcalorimeter resolution allows us to clearly separate the low surface brightness emission from the CGM from the very bright Milky Way lines (which is impossible with CCD detectors). We also highlight the diversity of the ions that contribute to the signal depending on the component temperature. Please note that only narrow energy bands are shown, while the whole spectrum spans 0.3-2~keV and consists of 12,000 data points.\label{fig:spectra}}
\end{figure*}

The results are shown in Figure \ref{fig:gal1_mock_face}, for which all regions are numbered for
reference back to Figures \ref{fig:mock_image}. The left panel shows the fitted gas temperature for
each region, for fits with 1, 2, and/or 3 \texttt{bapec} components. Not all of the annular regions
were well-fit by a second or a third component---only regions where a new component was
statistically required are shown. It can be seen that there are three distinct gas temperatures
recovered by the fits (left panel)---the dominant Component 1 is at $T \sim 2.3 \times 10^6$~K,
($\sim 0.2$~keV), with Component 2 at $T \sim 4.6-5.8 \times 10^6$~K ($\sim 0.4-0.5$~keV), and
Component 3 at $T \sim 7.0-9.3 \times 10^6$~K ($\sim 0.6-0.8$~keV). The mean velocities for these
three components are shown in the center panel, with Components 1 and 2 averaging around zero
velocity with a range of $\sim -100-100$~km~s$^{-1}$. Component 3, which is the hottest gas, has
mean velocities as high as $\sim -200$~km~s$^{-1}$ near a radius of $\sim$10-20~kpc, but these
values are also more uncertain. The right panel shows the velocity dispersion for each component,
which is $\sim$100-200~km~s$^{-1}$ for Components 1 and 2, and $\sim$400~km~s$^{-1}$ for Component
3, but again these latter values are more uncertain. The lower values of the mean and dispersion at
lower temperature (Components 1 and 2) are consistent with the fact that the slower inflows are
cooler, and the higher values for both of these quantities of Component 3 are consistent with the
fact that this component is associated with the hot outflowing gas that we have seen in the previous
sections, especially Figure \ref{fig:gal1_profiles}, which showed the same for the velocity profiles
weighted by the higher energy band, which is more sensitive to hotter gas. The spectra for Region 4
in the 0.3-2~keV band, with the best-fit model overlaid with three components, are shown in Figure
\ref{fig:spectra}.
    
\begin{figure*}[!t]
\centering
\includegraphics[width=0.95\textwidth]{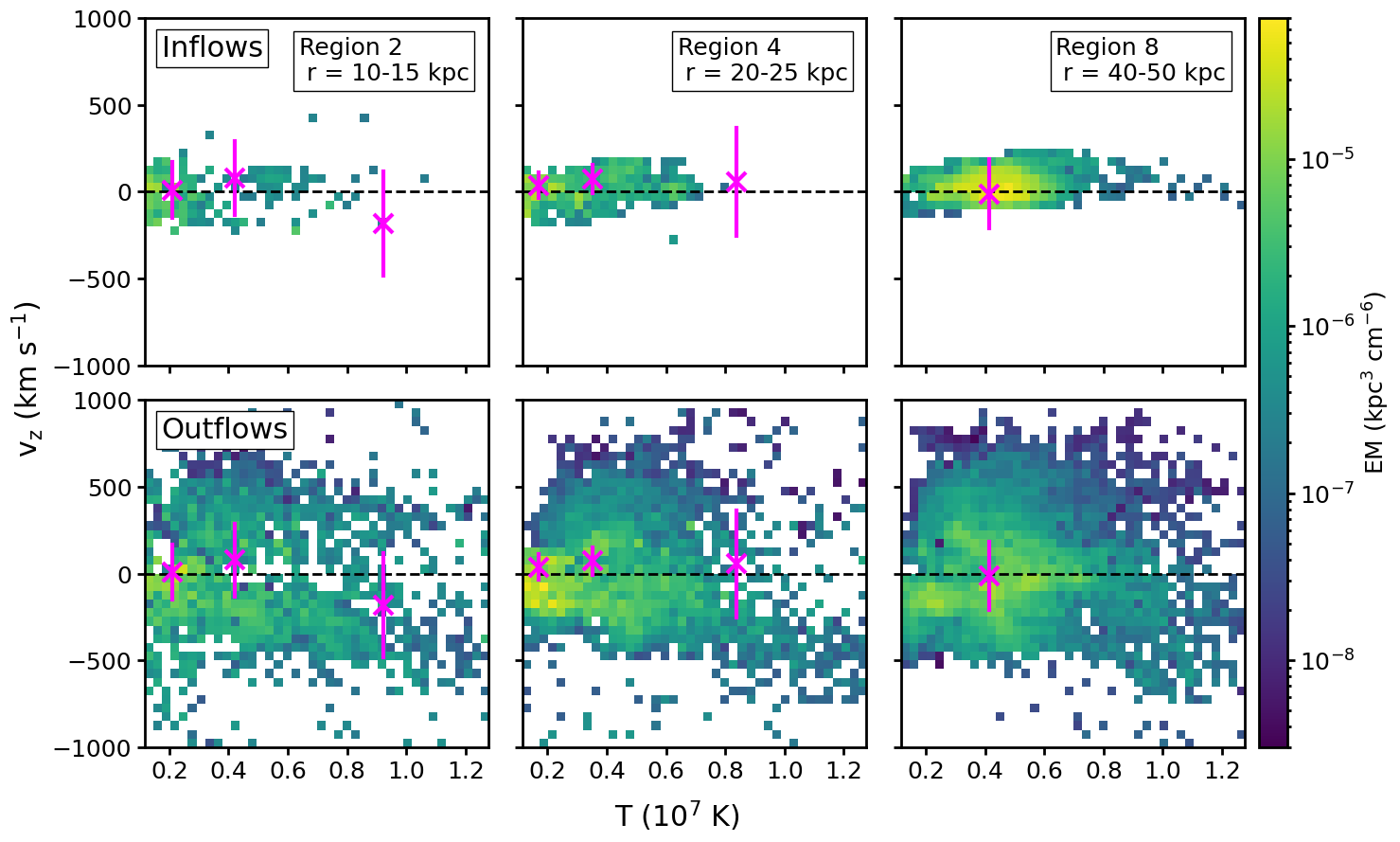}
\caption{Phase space of temperature vs. line-of-sight velocity for 3 of the cylindrical annuli corresponding to the numbered face-on regions shown in the right panel of Figure \ref{fig:mock_image}. The top three panels correspond to cells which are inflowing, and the bottom three panels indicate cells which are outflowing. The colormap indicates summed emission measure at each point. Magenta points indicate the position of the fitted temperature and mean velocity from the \texttt{bapec} components shown in Figure \ref{fig:gal1_mock_face}, whereas the vertical error bars on each point indicate the velocity dispersion. In general, the results from the multiple component fits are consistent with the simulation, though projection effects will make it difficult to distinguish between inflows and outflows.\label{fig:fit_phase_space}}
\end{figure*}

To further examine the consistency of the fitted values with the data from the simulation, we show in Figure \ref{fig:fit_phase_space} the phase space of temperature vs. line-of-sight velocity for  3 of the cylindrical annuli corresponding to the numbered face-on regions shown in the right panel of Figure \ref{fig:mock_image}. The top panels only show gas which is inflowing with $v_r < 0$ and the bottom panels only show gas which is outflowing with $v_r > 0$ (see also Section \ref{sec:2d_profiles}). The colormap indicates summed emission measure at each point, with yellow indicating the highest values. For each region, the general trend is for the phase space to be most concentrated at temperatures of $T \sim 3-6 \times 10^6$~K ($\sim 0.25-0.5$~keV), where the spread of velocities is also the lowest. As we move to higher temperatures, the spread of velocities increases, though the phase space is less populated in these regions. 

Overplotted on each phase space panel are magenta points indicating the position of the fitted temperature and mean velocity from the \texttt{bapec} components shown in Figure \ref{fig:gal1_mock_face}, whereas the vertical error bars on each point indicate the velocity dispersion. For all of the panels, the coldest magenta point (Component 1 from Figure \ref{fig:gal1_mock_face}) is usually consistent with the highest emission measures in the region. We can also see from the top row of panels that this component is also consistent with inflowing gas, especially Region 8, which is at large radius and the gas with the highest emission measure should be near the disk and thus inflowing. However, there is also outflowing gas in these regions in projection (bottom row of panels) at the same velocity/temperature phase, so the distinction is not always clear-cut. Where present, Components 2 and 3 also appear generally consistent with the data, though once again there are large uncertainties. These temperatures, especially Component 3, are hotter and are more consistent with the outflowing gas. This particular analysis relied on combinations of single temperature models, but it is likely that better models accounting for the temperature and velocity distributions in a more general way will need to be developed to properly model calorimeter-quality data in the future. 
    
\section{Summary}\label{sec:summary}

The hot, X-ray-emitting phase of the CGM has so far eluded detailed study, even for nearby galaxies,
due to the brightness of the MW's own CGM and the lack of X-ray instruments with sufficient spectral
resolution to distinguish between the emission lines of the latter and the former. In disk galaxies
with mass greater than or equal to the MW, the hot phase will be dominant. In the coming decades,
determining the properties of this phase of the CGM will be crucial to the further development of
our understanding the processes of galaxy formation and evolution. If our own galaxy is any
indication, many of these galaxies may be expected to possess hot outflows (such as evidenced by the
cavities seen in the \textit{eROSITA} all-sky survey) and rotational motions in their CGM. Only
X-ray IFUs will be able to map the velocity field of these galaxies to observe these processes in
action. 
    
Using a small selection of disk galaxies from the TNG50 simulation which were already shown to have cavities, we have shown that the CGM of such galaxies can exhibit velocities representing gas
outflows, inflows, and rotation that can be mapped by microcalorimeter instruments of the future. Our main conclusions are as follows:

\begin{itemize}

\item Most of the TNG50 galaxies examined in this sample (1-5) have a simple geometrical structure in the hot phase of the CGM, comprised of oppositely directed hot and fast outflows in conical regions on either side of the disk, rotation in the inner hot CGM (within $\sim$50~kpc), and slower inflows of hot gas ($\sim100-150$~km~s$^{-1}$) close to the galactic plane at larger radii. When viewed exactly edge-on, line shift maps exhibit the rotation curve clearly, with velocities of $\sim100-200$~km~s$^{-1}$. Hot outflows can also be seen edge-on in line shift ($\sim400-600$~km~s$^{-1}$) if they are not launched exactly in the plane of the sky, and the expansion of the outflow along the sight line can be seen in line broadening measurements ($\sim200-400$~km~s$^{-1}$). When viewed exactly face-on, line shift maps of the hot CGM of the same galaxies show a turbulent line-of-sight velocity structure with mean velocities of $\sim \pm 200-500$~km~s$^{-1}$, and velocity dispersions of $\sim 300-1000$~km~s$^{-1}$ within $\sim$50-100~kpc of the galactic center, and $\sim$100-200~km~s$^{-1}$ at larger radii.

\item The hot CGM of these galaxies rotates in the same direction as the stellar disk and has a similar rotation speed ($\sim100-200$~km~s$^{-1}$), but is slower than the colder CGM and ISM ($\sim200-400$~km~s$^{-1}$). Conversely, the velocity dispersion in the azimuthal direction of the hot phase is greater ($\sigma_\phi \sim$100-150~km~s$^{-1}$) than the warm/cold phase ($\sigma_\phi \sim $50-100~km~s$^{-1}$). Outside of the rotation and outflow regions and closer to the disk, both phases of gas are inflowing, the hot phase at $\sim100-150$~km~s$^{-1}$ and the warm/cold phase at $\sim100-300$~km~s$^{-1}$.

\item Our lowest-mass galaxy (6) does not have either the pattern of outflows on opposite side of the disk, or coherent rotation of the hot CGM in the same direction as the ISM and stars. Instead, the hot CGM of this galaxy has a more chaotic and turbulent velocity field, and at the current epoch the outflows appear on only one side of the stellar disk. A closer inspection of this galaxy reveals that it does not have a cold gas disk co-rotating with the stars either, which at one point did exist but was recently completely disrupted by AGN activity (see Appendix \ref{sec:AppendixC}).

\item If viewed face-on, the mean line-of-sight velocity in most of these galaxies azimuthally averages out to values near zero (assuming as we did in this work that the velocities are measured with respect to the center-of-mass frame). The velocity dispersion in this direction averages out to $\sigma \sim 200-500$~km~s$^{-1}$ within $\sim$50-100~kpc and $\sigma \sim 100-200$~km~s$^{-1}$ at larger radii. This structure is indicative of a complex pattern of flows that nevertheless when averaged over the azimuthal direction is composed of conically-shaped outflows away from the disk near the center and inflows at larger projected radii. We find that the velocity dispersion that is obtained is sensitive to which emission lines are used, since these probe different gas phases from each other and from the mass-weighted average.

\item For X-ray observations, using regions larger than the angular resolution of the detector will often be necessary to obtain sufficient counts to measure the line centroid shift and broadening. This will not only measure velocity differences along the sight line, but also across the sky plane within the region, which also contributes to the measured centroid shift and broadening. We find that these contributions can be as significant to the overall measurement. To separate out these effects, splitting up the regions as finely as count rate statistics allow may be necessary.

\item When our galaxies are viewed at an angle inclined away from the disk, signs of both rotation and hot outflows are observed, the latter of which will be especially prominent in line shift measurements ($\sim \pm 200-500$~km~s$^{-1}$) in regions of X-ray SB which show evidence of cavities and bubbles. The combination of these effects produce a velocity pattern in our simulated galaxies that is distinct from the stellar and ISM velocity patterns, as the velocity fields of the latter two are dominated by rotation. 

\item We produced mock X-ray microcalorimeter observations of galaxy 2 and used a spectral fitting technique to produce maps of the mean velocity field along two sight lines; edge-on and tilted 45$^\circ$ to the rotation axis of the stellar disk. In both cases we are able to reproduce the features of the mean velocity field of the simulation to high-accuracy, enabling us to determine the properties of the rotation curve and the hot outflows. 

\item We produced similar mock observations of galaxy 1 along the edge-on and face-on sightlines. We then selected regions in each projection to measure the first two moments of the velocity field by extracting spectra from these regions and fitting thermal emission models to them. We find that in the edge-on projection that the mean and standard deviation of the velocity are well-fit by a single thermal emission model, enabling us to measure the rotation curve of the CGM from line shifts and estimate the inflow velocity using line widths. In the face-on direction, the different phases of the gas with different velocities appear in projection along the sight line, and thus we require multiple thermal emission components to reproduce their properties. We find that the lower-temperature hot phase is consistent with lower velocity dispersions, and the higher-temperature gas is consistent with higher velocity dispersions. The former may be consistent with inflows (especially at large projected radii), whereas the latter is consistent with outflows, but projection effects make unambiguous identification of these two different phases difficult.

\end{itemize}
    
Our results show that future microcalorimeter observations of the hot CGM of galaxies will be able to measure the temperature and velocity fields of the gas, and determine if the hot CGM has the main structures we identified in this work: directed inflows, outflows, and rotation, or if is dominated by a chaotic and turbulent flow. Detecting hot outflows and measuring their velocities will help determine the mass and energy fluxes of these outflows, and thus their impact on the evolution of the galaxy and its environment. Measuring the rotation and inflow velocities of the hot CGM, especially in comparison to measurements of the cooler phase in the UV, will help determine how gas accretes from the CGM onto the galaxy itself in its different temperature phases and drives its evolution. 

This analysis could be extended in a number of ways. The spectral analysis of the mock observations in this work merely scratched the surface of what is possible. The closest analog to studies of the hot halos of galaxies are their more massive counterparts in groups and clusters of galaxies in the intragroup and intracluster media, which are much brighter in X-rays and hence easier to study. In the era of \textit{Chandra}, \textit{XMM-Newton}, \textit{Suzaku}, \textit{NuSTAR}, and now \textit{eROSITA}, spectral analysis of these extended sources has been largely limited to the $\sim$100~eV resolution of the imaging instruments on these telescopes. This prevents analysis of the velocity field in groups and clusters, and limits the ability to distinguish between different gas phases of different temperatures and compositions. This latter issue has not been a major limitation for most studies of the hot gas in groups and clusters, since for most applications it is well-approximated by a single-temperature phase over relevant spatial regions. However, this is not the case for the CGM, and to characterize it adequately we will need microcalorimeter instruments that can resolve the velocity field and the different gas phases. Extensions of the work presented here should focus on improvements to the process of extracting and fitting spectra to decompose the emission into these multiple thermodynamic and kinematic components, which will likely require more sophisticated statistical methods than have been required for spectra with CCD-like energy resolution, and/or machine learning techniques. 

We chose to focus on systems with active AGN feedback in this work because of the strong temperature and kinematic signatures they are expected to produce in future X-ray microcalorimeter observations of the CGM, and to determine if the velocity patterns they produce can be disentangled from other effects such as rotation and turbulence in projection. By focusing on simulated galaxies with such centrally-driven outflows, we are necessarily unable to make predictions for what the velocity field of the same galaxies would look like when feedback is not active, less powerful, the outflows are more isotropic, or more distributed throughout the disk. These considerations are reserved for future studies.

Another important factor in driving the velocity fields of the CGM that we did not explore is the effect of mergers. Given our sample selection criteria (see Section \ref{sec:tng50}), at late times our galaxies are relatively isolated with a lack of major mergers. Nevertheless, galaxies are also shaped by minor mergers with smaller galaxies. These may have an effect on the X-ray emitting CGM in a number of ways. Stellar and supernova feedback from satellite galaxies will heat the CGM locally and drive small outflows. ISM or CGM stripped from orbiting satellites which pass through dense regions of the CGM can drive turbulence and bulk flows in their wake that will distort the simple picture presented in this work of feedback, rotation, and inflows. Small satellites are present in the projected stellar mass maps of all 6 galaxies in our small sample (Figures \ref{fig:gal1_proj}-\ref{fig:gal3_proj}, \ref{fig:gal4_proj}, \ref{fig:gal5_proj}, and \ref{fig:gal6_proj}); the maps of Galaxy 1 in particular show hints of some flows that may have been originated in such an encounter with a smaller galaxy. Future work should analyze systems with minor mergers in detail and make observational predictions for possible correlations between stellar streams and stripped gas from these satellites.

We have also only used disk galaxies from the TNG simulations, which prescribe particular modes of AGN and stellar feedback. It would be instructive to perform similar analyses on other simulated galaxies, including from cosmological simulations such as EAGLE \citep{Schaye2015,Crain2015}, SIMBA \citep{Dave2019}, FIRE \citep{Hopkins2018, Hopkins2023}, Magneticum \citep{Biffi2013}, as well as simulations with much higher resolution available in the CGM, such as FOGGIE \citep{Peeples2019} and GIBLE \citep{Ramesh2023d}, and idealized simulations \citep{Fielding2017a,Schneider2018a,Stern2023}. These models have different prescriptions for feedback and different spatial resolutions, which will likely have significant impacts on the observables discussed here. For example, if feedback manifests itself as a more gentle and/or distributed heating of the gas (either from AGN or stellar feedback), we would not expect to see the hot, fast outflows associated with cavities above and below the stellar disk. This would be observable in surface brightness and temperature measurements, but would also be observable in the velocity maps of galaxies not perfectly oriented edge-on (Figure \ref{fig:sim_and_mock}), or perhaps in edge-on galaxies if the outflow has a significant diagonal component. Looking down along the outflows in more face-on galaxies (Section \ref{sec:mocks_profiles}), the spectrum may be better represented by a single cooler temperature ($T \sim 2 \times 10^6$~K) and slower velocity ($\sigma \sim 200$~km~s$^{-1}$) component (Figure \ref{fig:gal1_mock_face}), instead of the multi-phase structure that we observed in our simulations. Alternatively, the kinetic feedback may be more efficient, driving faster and hotter outflows, and may expel a significant fraction of the hot CGM from the halo entirely, making it difficult to observe. We have recently undertaken a project to produce these observables from some of these different simulations for comparison with the results from this work. Since all of the simulations listed above have exhibited some success in reproducing the stellar properties of the galaxy population, making these distinguishing observations in the X-ray will be crucial to narrowing down the feedback prescriptions that are most successful in reproducing the observed properties of galaxies and their halos overall.

\begin{acknowledgments}

    We thank the anonymous referee for their reading of the paper and their valuable comments, which have improved the manuscript greatly. JAZ, AB, PEJN, and RPK are funded by the Chandra X-ray Center, which is operated by the Smithsonian Astrophysical Observatory for and on behalf of NASA under contract NAS8-03060. JS was supported by the Israel Science Foundation (grant No. 2584/21). DN acknowledges funding from the Deutsche Forschungsgemeinschaft (DFG) through an Emmy Noether Research Group (grant number NE 2441/1-1). IK acknowledges support by the COMPLEX project from the European Research Council (ERC) under the European Union's Horizon 2020 research and innovation program grant agreement ERC-2019-AdG 882679.

This material is based upon work supported by NASA under award number 80GSFC21M0002.

The TNG50 simulation was run with compute time granted by the Gauss Centre for Supercomputing (GCS) under Large-Scale Projects GCS-DWAR on the GCS share of the supercomputer Hazel Hen at the High Performance Computing Center Stuttgart (HLRS).

\end{acknowledgments}

\software{
    \texttt{AstroPy} \citep{2013A&A...558A..33A,2018AJ....156..123A,Astropy2022},  
    \texttt{CIAO} \citep{Ciao2006},
    \texttt{emcee} \citep{emcee},
    \texttt{Matplotlib} \citep{Hunter2007},
    \texttt{NumPy} \citep{Harris2020},
    \texttt{pyXSIM} \citep{pyxsim2016},
    \texttt{schwimmbad} \citep{schwimmbad},
    \texttt{Sherpa} \citep{Burke2020},
    \texttt{SOXS} \citep{soxs2023},
    \texttt{XSPEC} \citep{Arnaud1996},
    \texttt{yt} \citep{Turk2011}
}

\bibliography{ms}{}
\bibliographystyle{aasjournal}

\appendix

\section{Effects of Varying Background Model Parameters}\label{sec:AppendixA}

As noted in Section \ref{sec:xrays}, for the galactic foreground model we have assumed a sum of three \texttt{APEC} components, one for the Local Hot Bubble and two for the MW's own hot CGM. The latter two components have foreground absorption (\texttt{TBabs} model) applied with a hydrogen column density of $N_H = 1.8 \times 10^{20}$~cm$^{-2}$ \citep[][]{McCammon2002}. This same absorption model and column density value is applied to the CXB sources and the CGM photons from our mock galaxies. Needless to say, the hydrogen column density and the abundance of the MW CGM will vary across the sky \citep[e.g.][]{Wang2020,Das2019b,Das2021,Ponti2023}. It is instructive to examine what the effect of a larger column density or a different MW CGM abundance would have on our results.

\begin{figure*}[!t]
    \centering
    \includegraphics[width=0.95\textwidth]{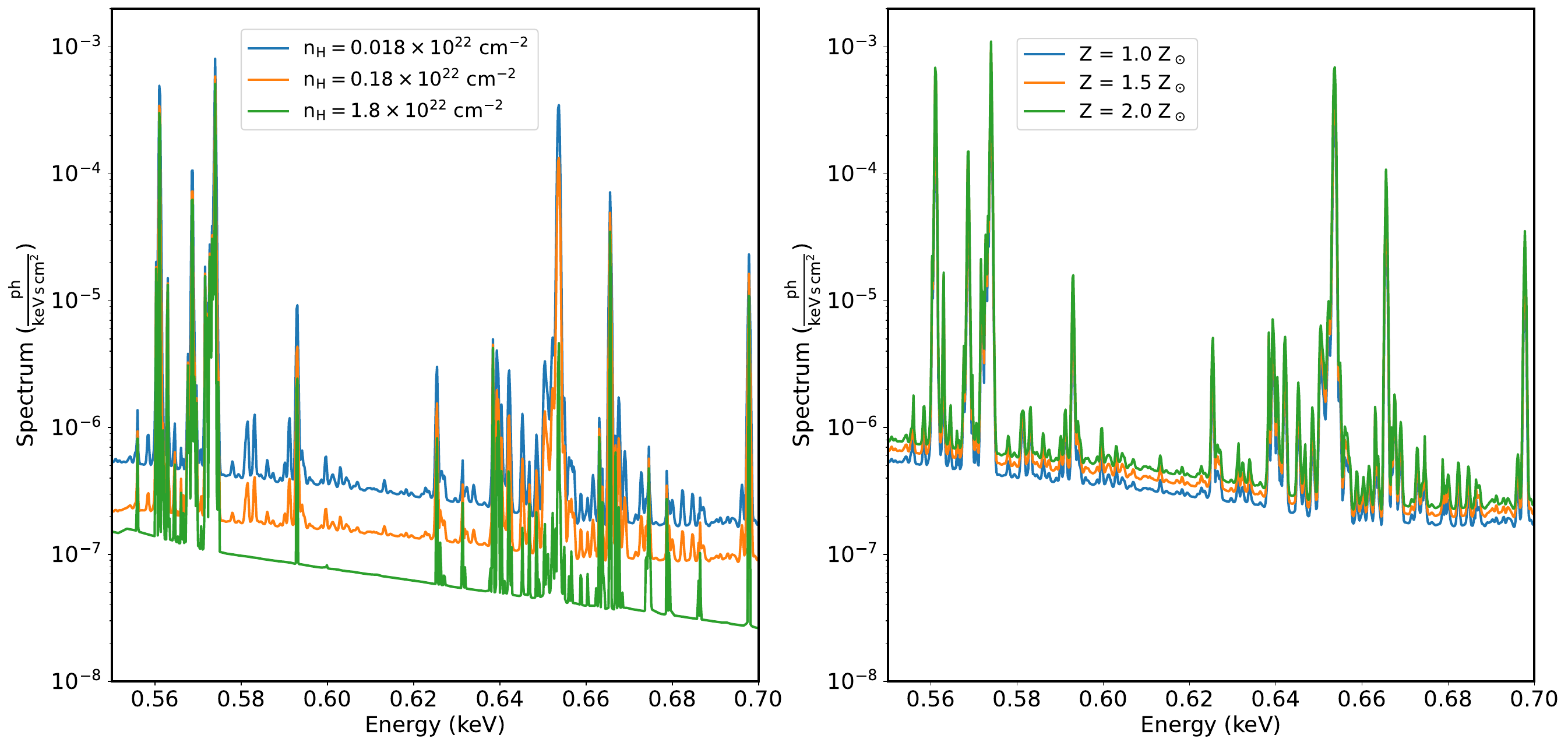}
    \caption{The effect of varying model parameters on the foreground spectrum in the energy range of 0.55-0.7~keV where most of the CGM emission lines of interest are located. Left panel: the effect of varying the hydrogen column density. Right panel: the effect of varying the abundance in the MW CGM components.}\label{fig:vary_frgnd}
\end{figure*}

In the left panel of Figure \ref{fig:vary_frgnd}, we show the effect of increasing $N_H$ by one and two orders of magnitude over our default value on the spectrum of our foreground model in the energy range of 0.55-0.7~keV, where most of our source emission lines are located. If we increase the column density for the foreground absoprtion, then the emissivity from the two MW CGM components and the unresolved CXB will decrease, however the distant galaxy's CGM source will also decrease in brightness. The Local Hot Bubble component, which does not suffer from foreground absorption in our model, will become dominant for $N_H \sim 1.8 \times 10^{22}$~cm$^{-2}$. At this point, the distant galaxy's lines will be significantly more difficult to see amidst the foreground continuum. In reality, the best observations of the CGM of distant galaxies will be possible at higher galactic latitudes, away from the larger column densities near the MW disk.

The right panel of Figure \ref{fig:vary_frgnd} shows the effect of varying the MW CGM abundance. Since our technique for observing the CGM of distant galaxies relies on them being cosmologically redshifted, an increased MW CGM abundance will not significantly affect the analysis. However, to the extent that an increased MW abundance increases the value of the foreground continuum, this could pose a problem for analyzing fainter regions of the CGM in our galaxies.

\section{Plots for Galaxies 3, 4, and 5}\label{sec:appendixB}

This appendix shows the main figures for galaxies 3, 4, and 5, mentioned previously in Section \ref{sec:results}. These galaxies are lower-mass and not as bright in X-rays as galaxies 1 and 2 (though galaxy 6 is the lowest-mass galaxy in our sample). In general, galaxies 3, 4, and 5 are very similar to 1 and 2 in the sense that they have the same general pattern of inflows, outflows, and rotation, though galaxy 5 has a slightly more complicated velocity distribution in the hot phase (see Figure \ref{fig:gal5_phase}. This lower-mass galaxy is closer to the mass threshold below which virial shocks are not stable and gas inflows may be more filamentary and anisotropic \citep{Keres2005}.

\begin{figure*}[!hp]
\centering
\includegraphics[width=0.92\textwidth]{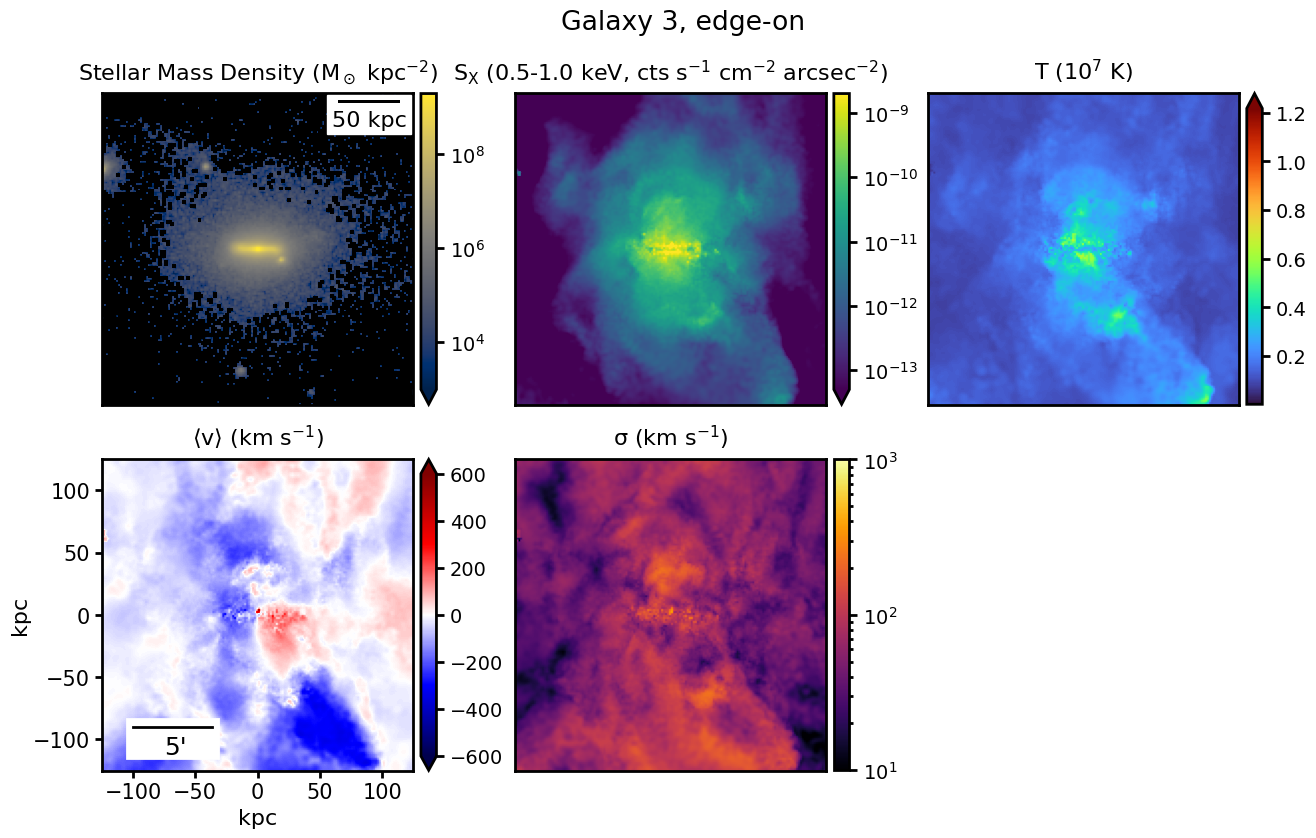}
\includegraphics[width=0.92\textwidth]{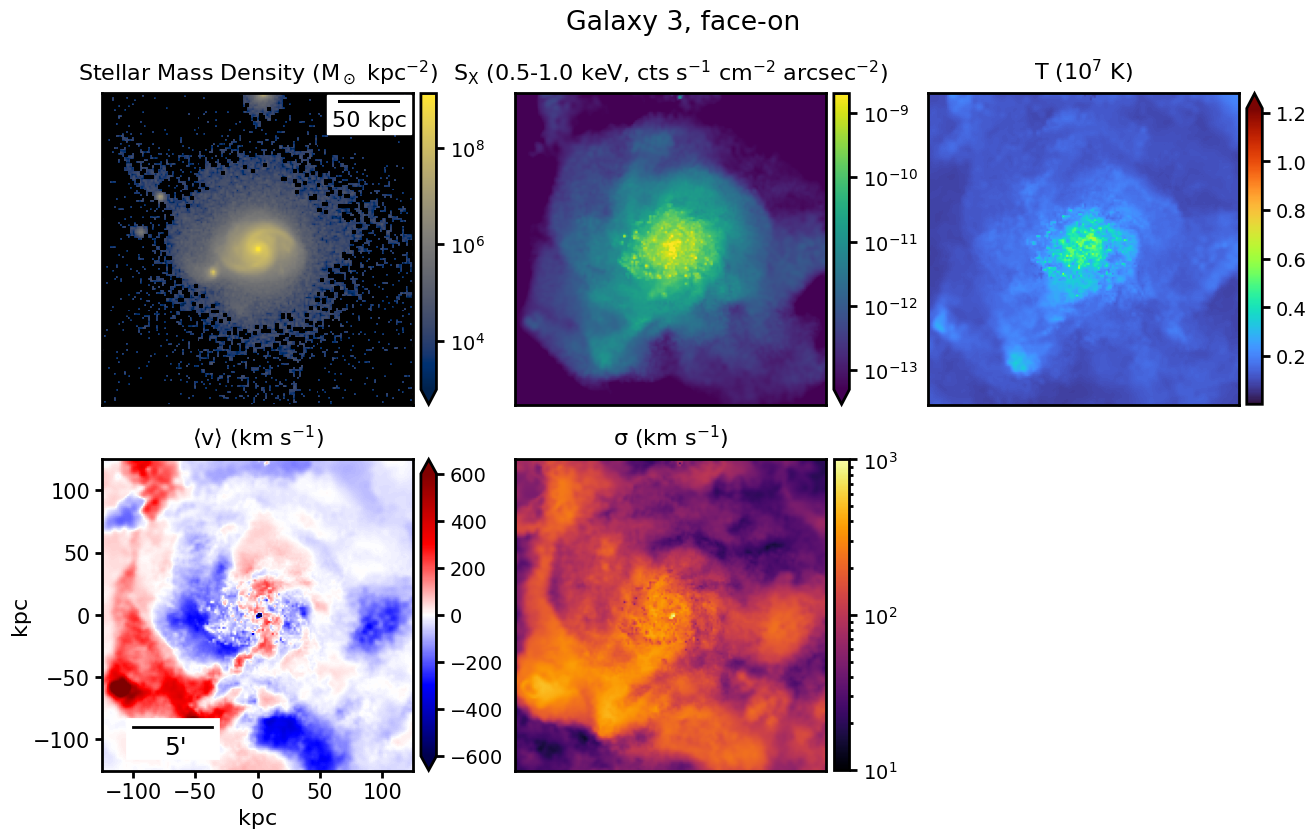}
\caption{Projections of various quantities from galaxy 3, viewed edge-on (upper panels) and face-on (lower panels) with respect to the plane of the galactic disk. Panel descriptions are the same as in Figure \ref{fig:gal1_proj}. Each panel is 250~kpc on a side, or $\sim$20' for the given redshift and cosmology. There is a spiral galaxy just to the north of the main galaxy in the face-on panels, which for some lines of sight in the disk plane of the latter appears to be very close in projection but is in fact much further away.}\label{fig:gal3_proj}
\end{figure*}
    
\begin{figure*}
\centering
\includegraphics[width=0.47\textwidth]{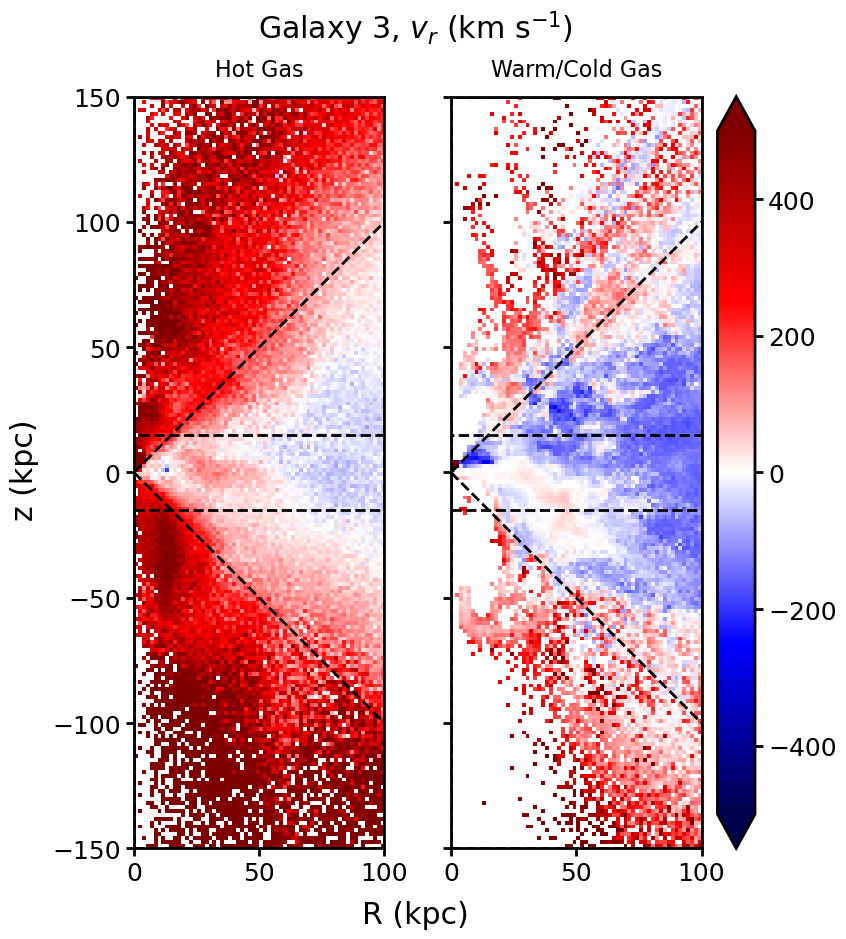}
\includegraphics[width=0.47\textwidth]{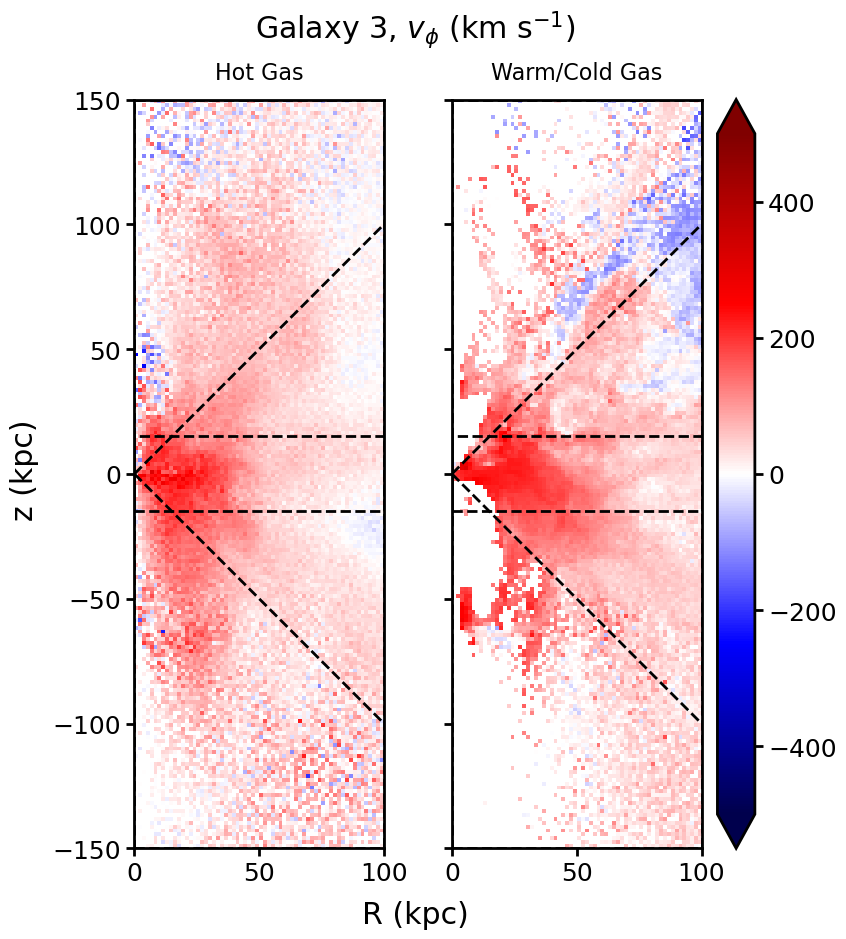}
\caption{Azimuthally averaged mass-weighted profiles in the radial ($R$) and vertical $z$ cylindrical coordinates of the spherical-$r$ and cylindrical-$\phi$ components of the gas velocity in hot (left sub-panels) and warm/cold (right sub-panels) gas phases for galaxy 3. The dashed horizontal lines indicate the region that is used to extract 1D cylindrical radial profiles in Section \ref{sec:1d_profiles}. 45$^\circ$ lines show the approximate boundary between the hot outflow and the slow inflow for the hot gas.\label{fig:gal3_phase}}
\end{figure*}    
    
\begin{figure*}
\centering
\includegraphics[width=0.82\textwidth]{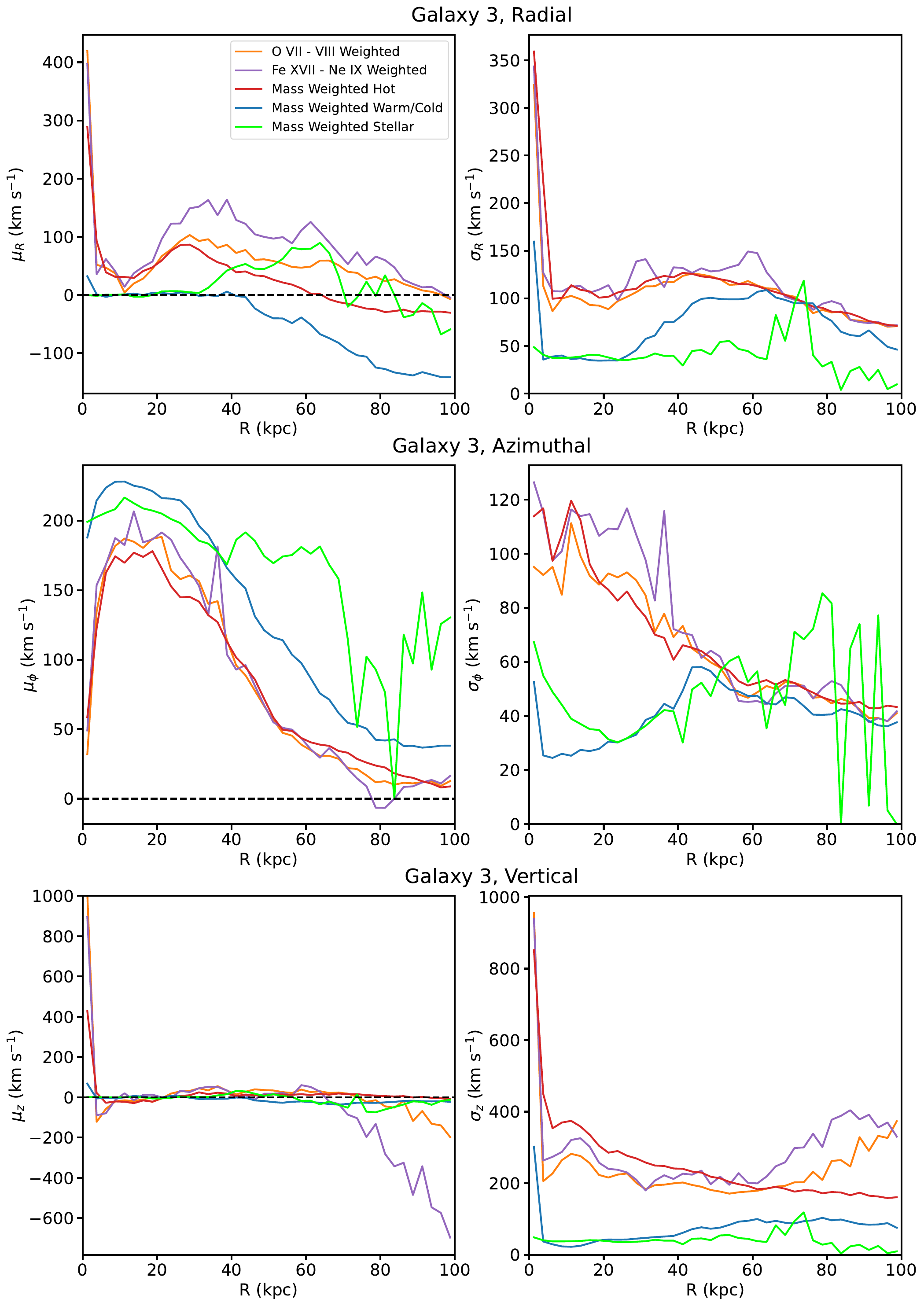}
\caption{Azimuthally and height-averaged mass-weighted and emission-weighted radial profiles of the gas and stellar velocity for galaxy 3. The description of the panels are the same as in Figure \ref{fig:gal1_profiles}.\label{fig:gal3_profiles}}
\end{figure*}

\begin{figure*}[!hp]
\centering
\includegraphics[width=0.92\textwidth]{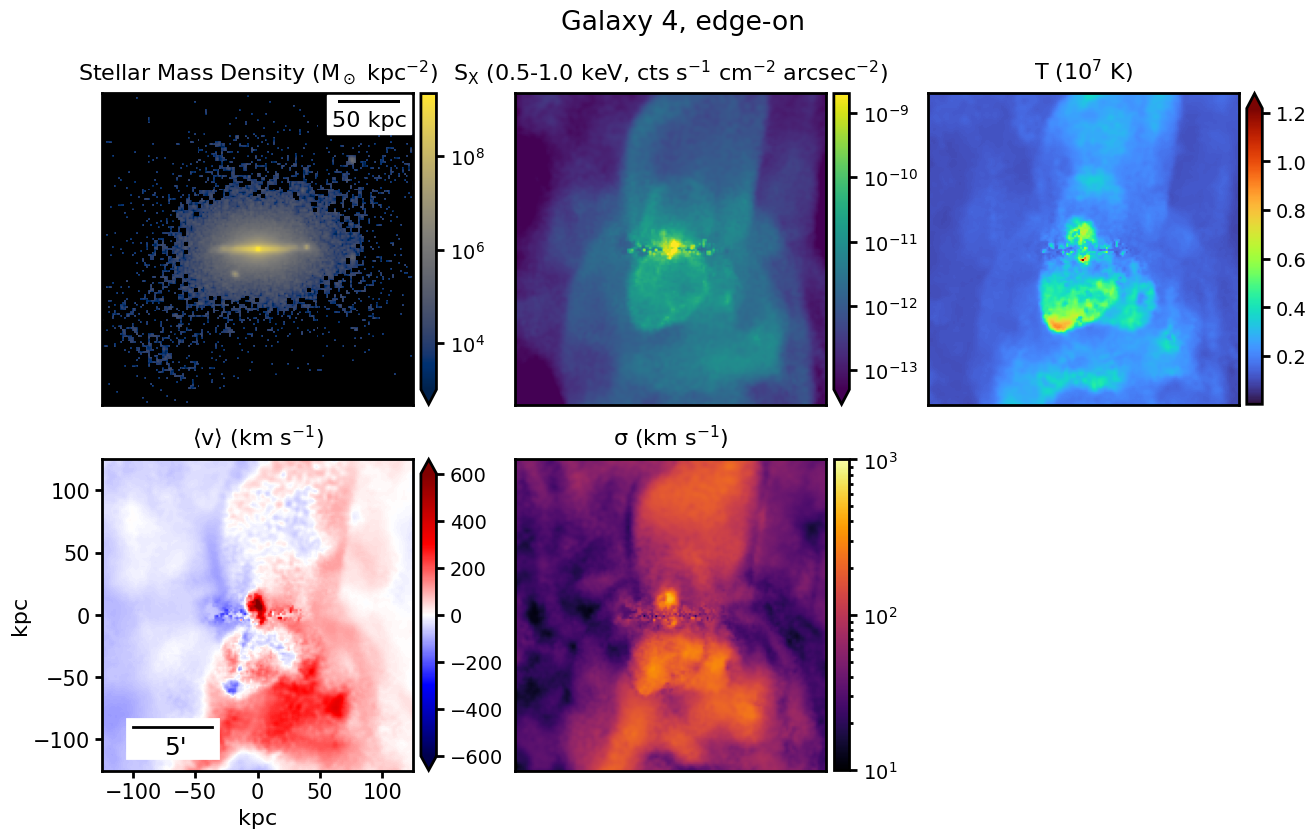}
\includegraphics[width=0.92\textwidth]{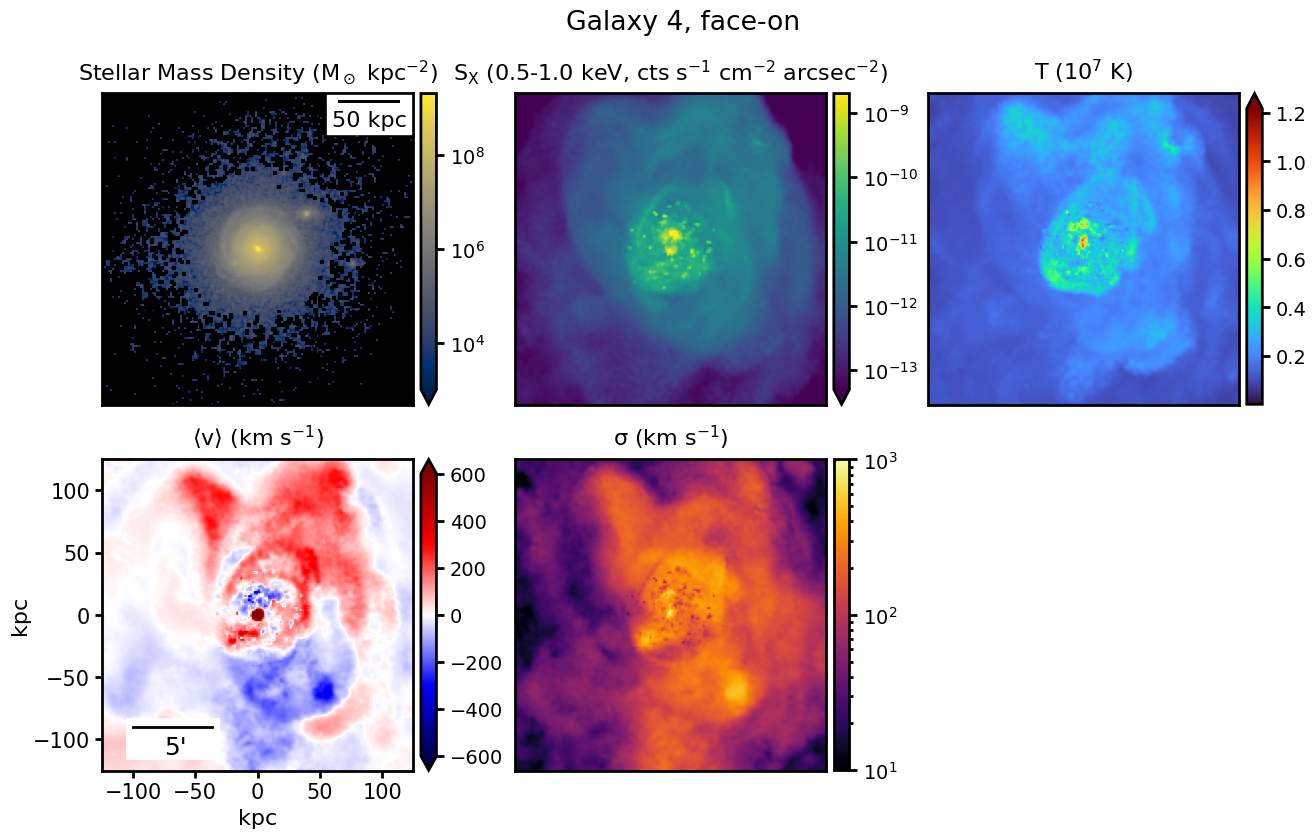}
\caption{Projections of various quantities from galaxy 4, viewed edge-on (upper panels) and face-on (lower panels) with respect to the plane of the galactic disk. Panel descriptions are the same as in Figure \ref{fig:gal1_proj}. Each panel is 250~kpc on a side, or $\sim$20' for the given redshift and cosmology.\label{fig:gal4_proj}}
\end{figure*}
        
\begin{figure*}
\centering
\includegraphics[width=0.47\textwidth]{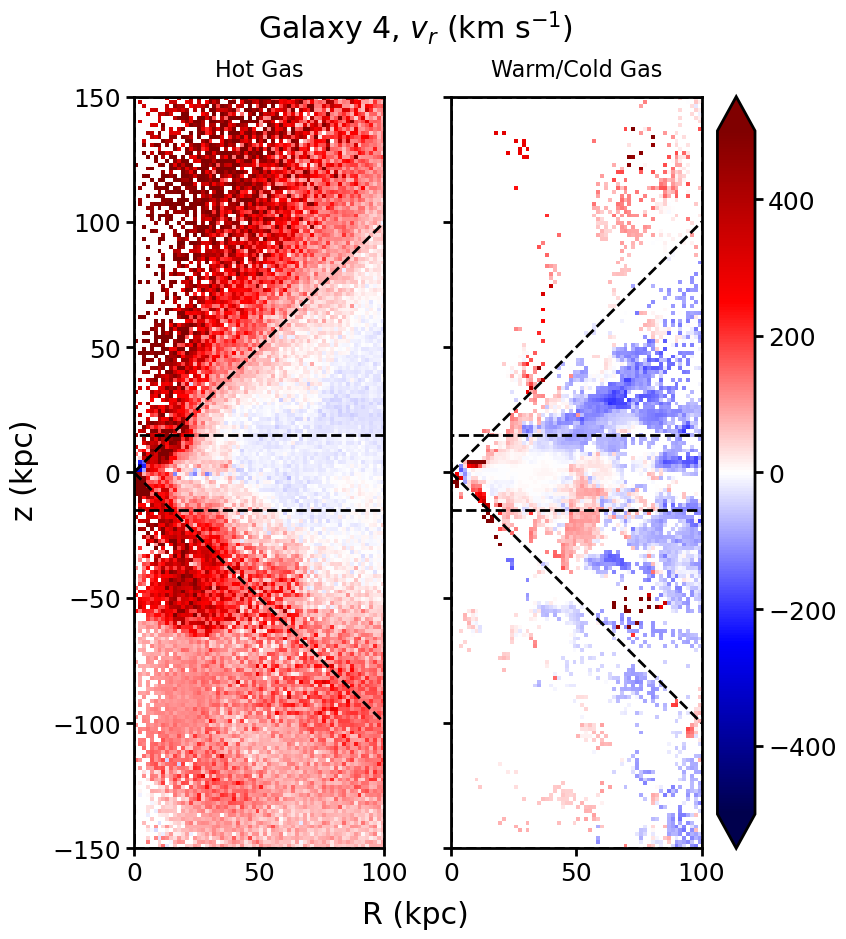}
\includegraphics[width=0.47\textwidth]{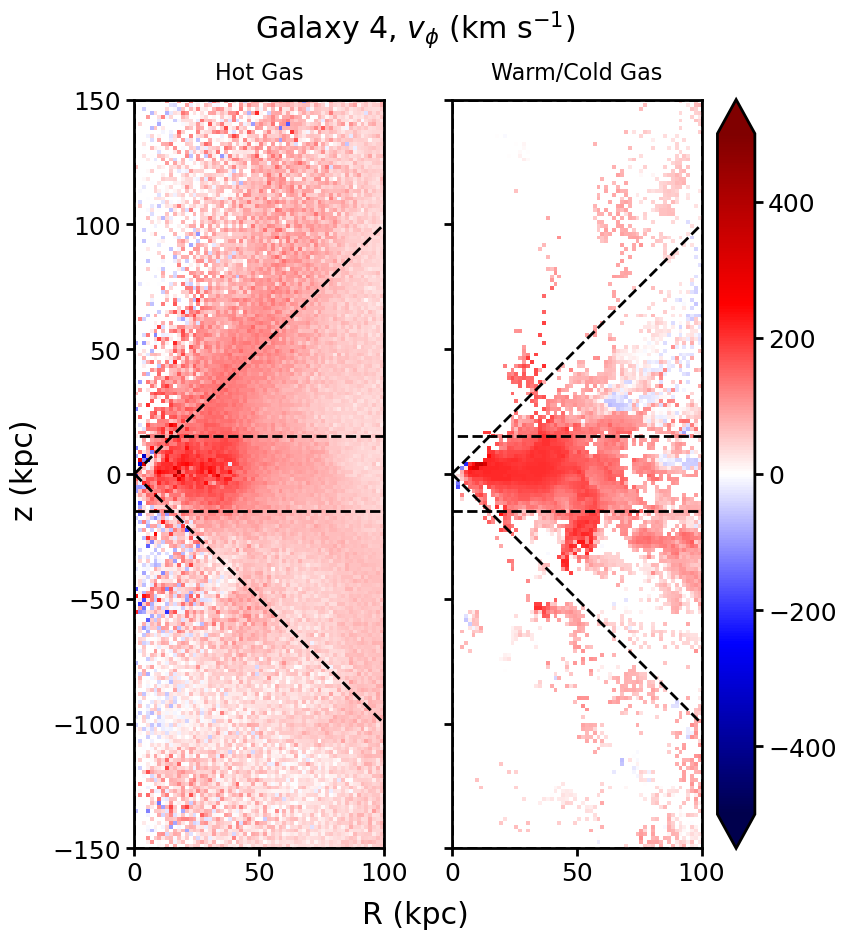}
\caption{Azimuthally averaged mass-weighted profiles in the radial ($R$) and vertical $z$ cylindrical coordinates of the spherical-$r$ and cylindrical-$\phi$ components of the gas velocity in hot (left sub-panels) and warm/cold (right sub-panels) gas phases for galaxy 4. The dashed horizontal lines indicate the region that is used to extract 1D cylindrical radial profiles in Section \ref{sec:1d_profiles}. 45$^\circ$ lines show the approximate boundary between the hot outflow and the slow inflow for the hot gas.\label{fig:gal4_phase}}
\end{figure*}    

\begin{figure*}
\centering
\includegraphics[width=0.82\textwidth]{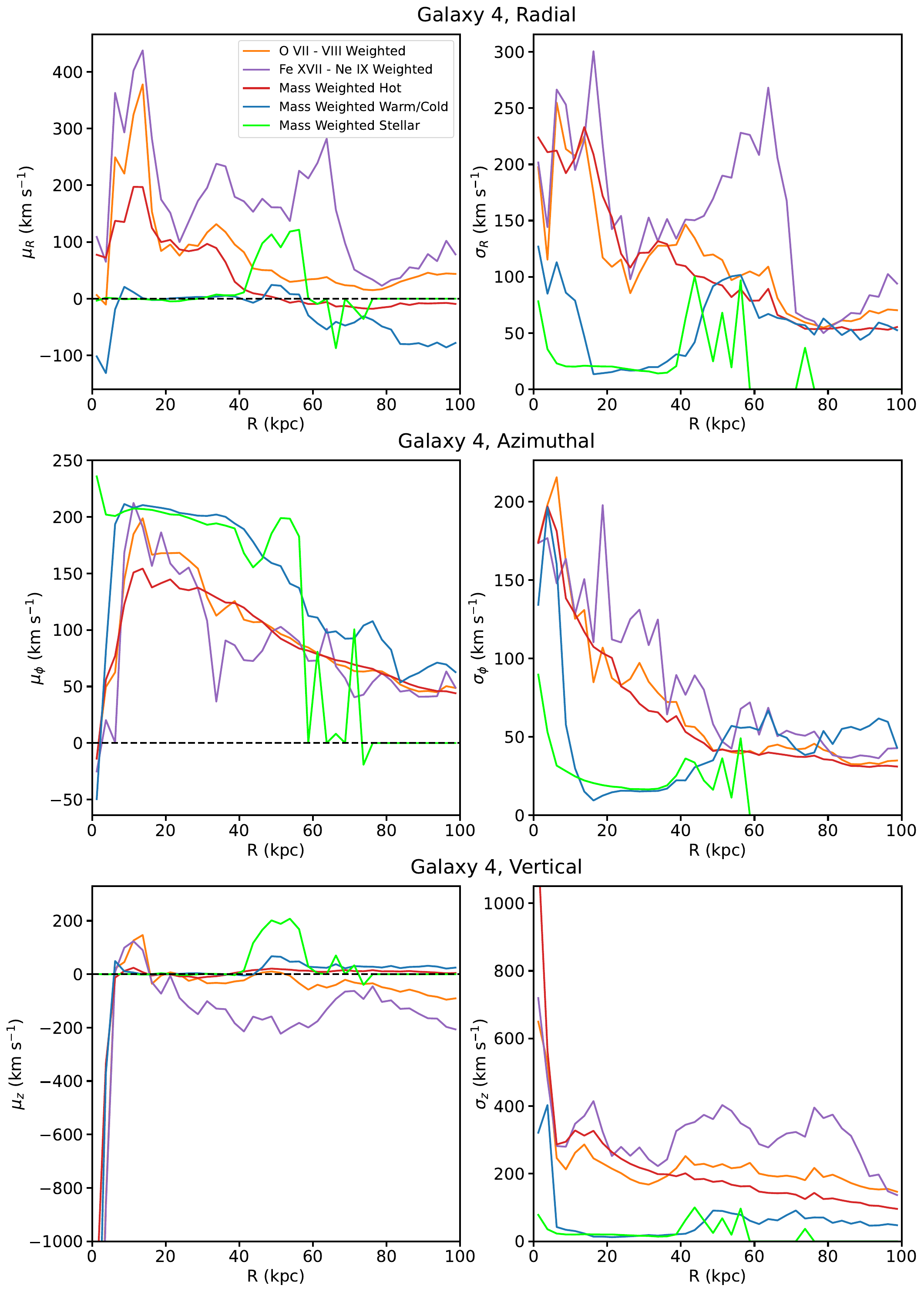}
\caption{Azimuthally and height-averaged mass-weighted and emission-weighted radial profiles of the gas and stellar velocity for galaxy 4. The description of the panels are the same as in Figure\ref{fig:gal1_profiles}.\label{fig:gal4_profiles}}
\end{figure*}
    
\begin{figure*}[!hp]
\centering
\includegraphics[width=0.92\textwidth]{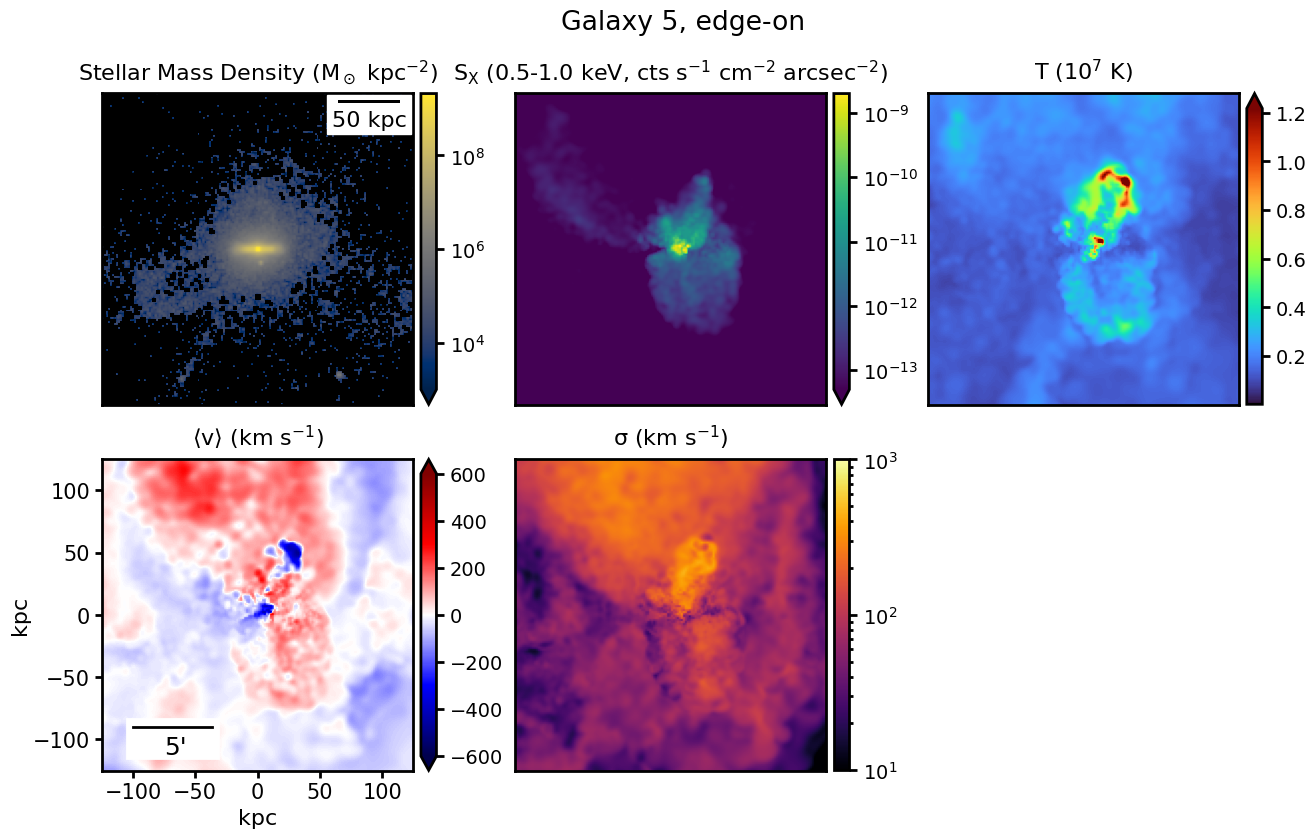}
\includegraphics[width=0.92\textwidth]{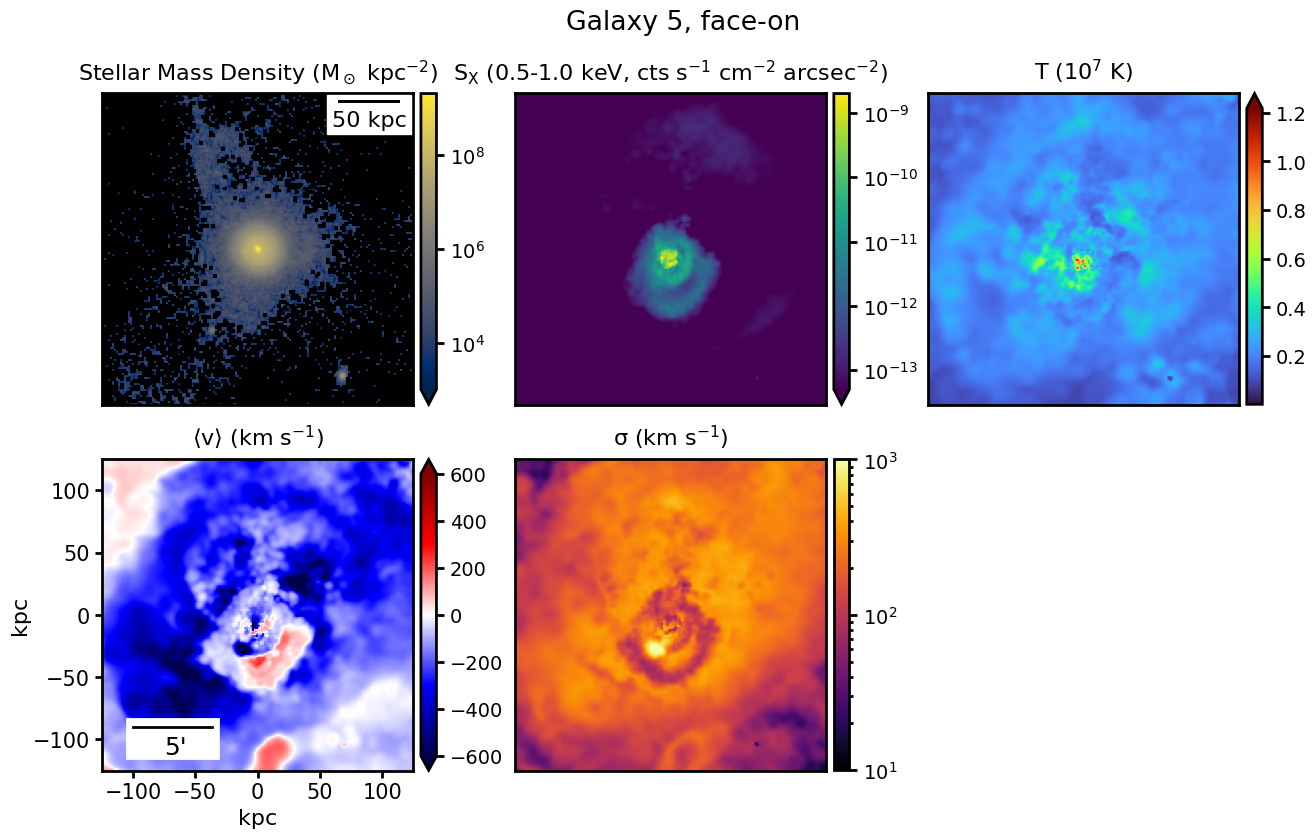}
\caption{Projections of various quantities from galaxy 5, viewed edge-on (upper panels) and face-on (lower panels) with respect to the plane of the galactic disk. Panel descriptions are the same as in Figure \ref{fig:gal1_proj}. Each panel is 250~kpc on a side, or $\sim$20' for the given redshift and cosmology.\label{fig:gal5_proj}}
\end{figure*}
        
\begin{figure*}
\centering
\includegraphics[width=0.47\textwidth]{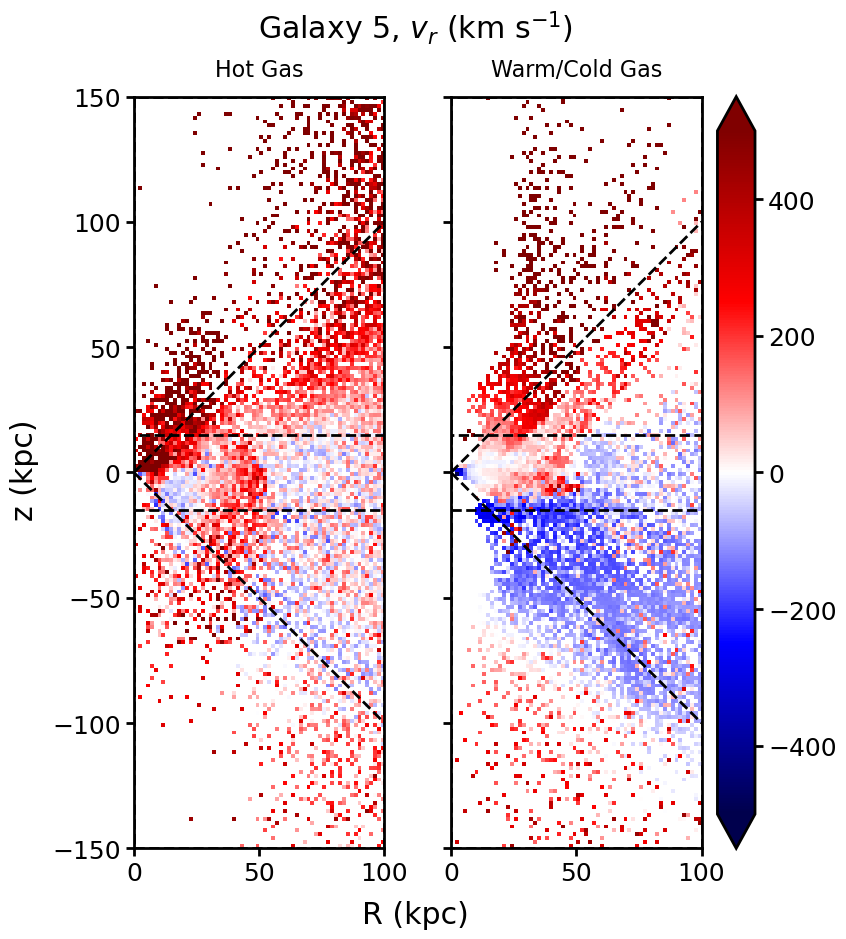}
\includegraphics[width=0.47\textwidth]{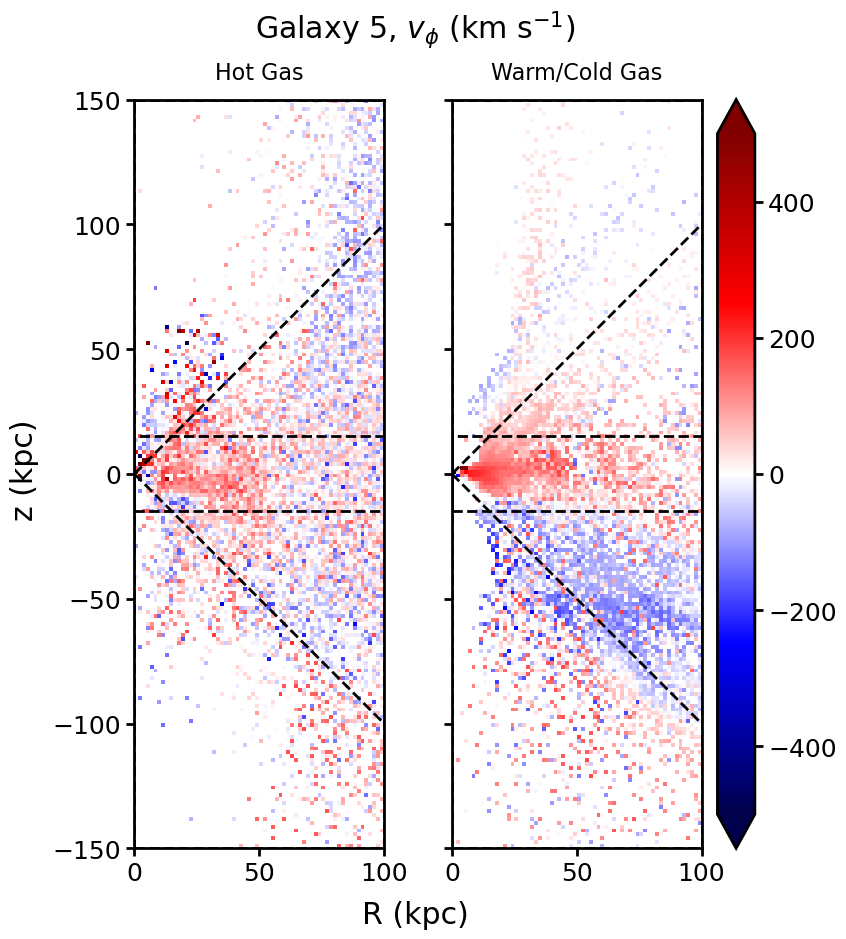}
\caption{Azimuthally averaged mass-weighted profiles in the radial ($R$) and vertical $z$ cylindrical coordinates of the spherical-$r$ and cylindrical-$\phi$ components of the gas velocity in hot (left sub-panels) and warm/cold (right sub-panels) gas phases for galaxy 5. The dashed horizontal lines indicate the region that is used to extract 1D cylindrical radial profiles in Section \ref{sec:1d_profiles}. 45$^\circ$ lines show the approximate boundary between the hot outflow and the slow inflow for the hot gas.\label{fig:gal5_phase}}
\end{figure*}    

\begin{figure*}
\centering
\includegraphics[width=0.82\textwidth]{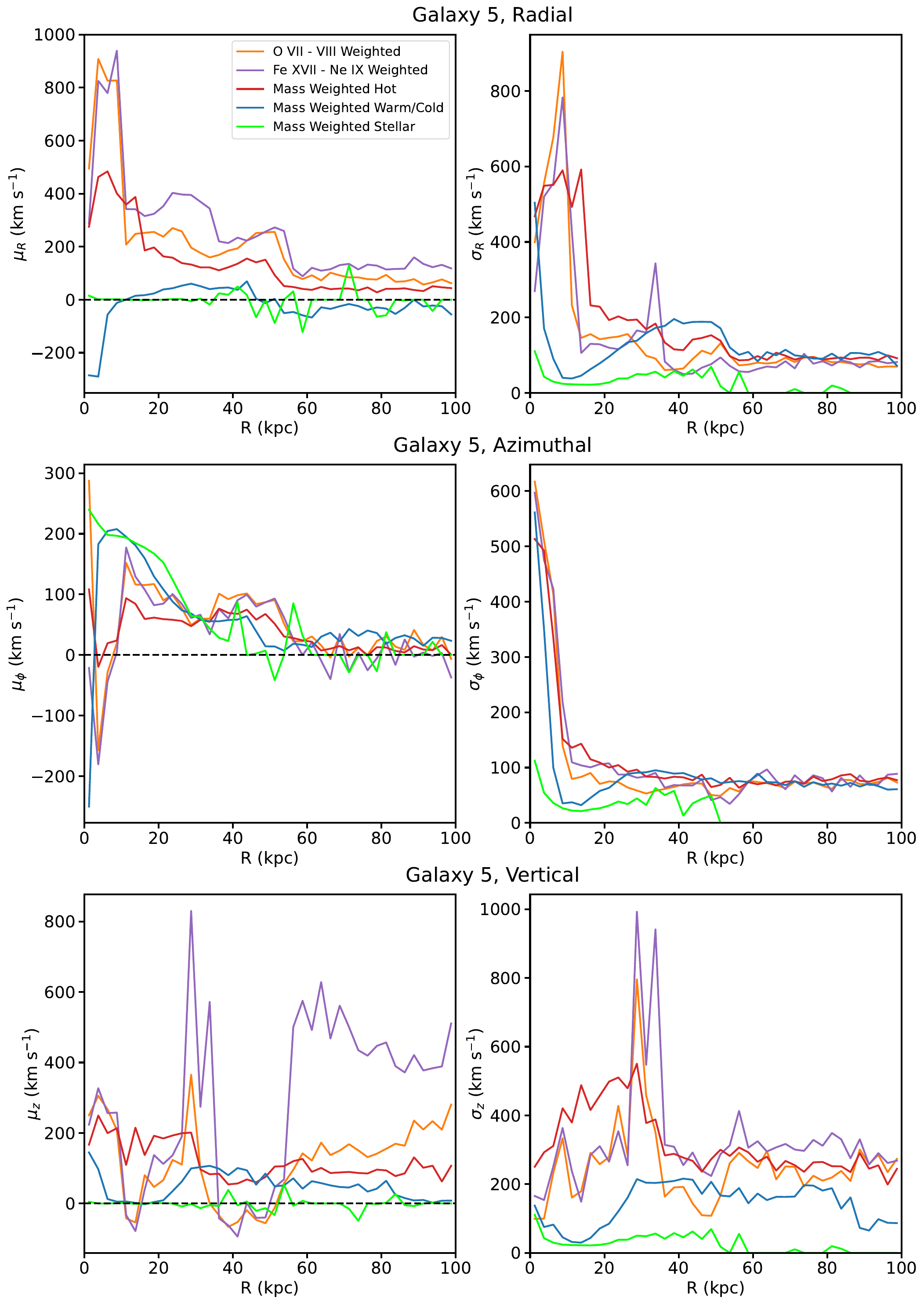}
\caption{Azimuthally and height-averaged mass-weighted and emission-weighted radial profiles of the gas and stellar velocity for galaxy 5. The description of the panels are the same as in Figure \ref{fig:gal1_profiles}.\label{fig:gal5_profiles}}
\end{figure*}
            
\section{The Circumstances of Galaxy 6}\label{sec:AppendixC}

We noted throughout the paper that galaxy 6 is significantly different from the others--its outflow indicators are only above the disk (Figure \ref{fig:gal6_proj}), its mean radial velocity field is almost entirely outflowing (left sub-panels of Figure \ref{fig:gal6_phase}), and its mean azimuthal velocity field is in the opposite direction of the rotation of the stellar disk (right sub-panels of Figure \ref{fig:gal6_phase}, and Figure \ref{fig:gal6_profiles}). To understand why it is so different from the others at a redshift of 0, we investigated its current properties and its prior history in more detail.

\begin{figure*}
\centering
\includegraphics[width=0.95\textwidth]{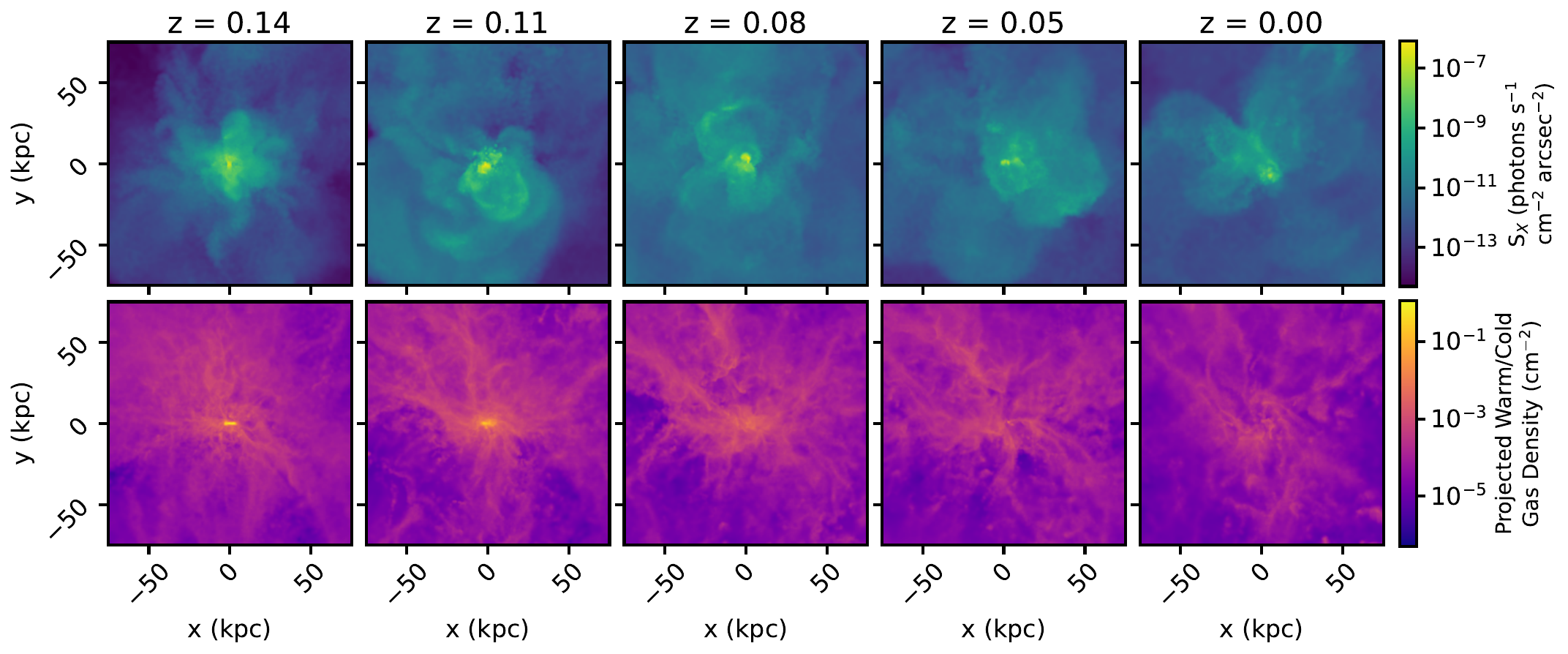}
\caption{Maps of projected X-ray surface brightness (top panels) and projected warm/cold gas density (bottom panels) of galaxy 6, oriented edge-on, for several epochs shortly prior to redshift 0. Signatures of strong feedback are shown at all of these epochs, and between redshifts of 0.11 and 0.08 the warm/cold gas disk is completely disrupted.\label{fig:gal6_evol}}
\end{figure*}    
    
In Figure \ref{fig:gal6_evol} we show the projected X-ray surface brightness in the 0.5-1.0~keV band (top panels) and the projected density of the warm/cold gas (bottom panels) of galaxy 6 for a series of epochs shortly before a redshift of zero. The plots of X-ray surface brightness show strong indications of AGN feedback at all epochs. The bottom panels show that at some point between a redshift of 0.11 and 0.08, the AGN feedback was strong enough to blow the warm/cold disk (comprised mostly of cold ISM) apart. At all later epochs up to redshift 0, the distribution of this warm/cold gas is filamentary and chaotic. This explains the lack of coherent rotation in the warm/cold gas and the fact that on average it is mostly outflowing in the radial direction. After the disruption of the dense and cold gas disk, the isotropically directed kinetic feedback mode in the TNG model (see Section \ref{sec:tng50}) is no longer confined mainly to regions above and below the stellar disk, and thus is able to inject energy more isotropically into the CGM.

\begin{figure*}
\centering
\includegraphics[width=0.95\textwidth]{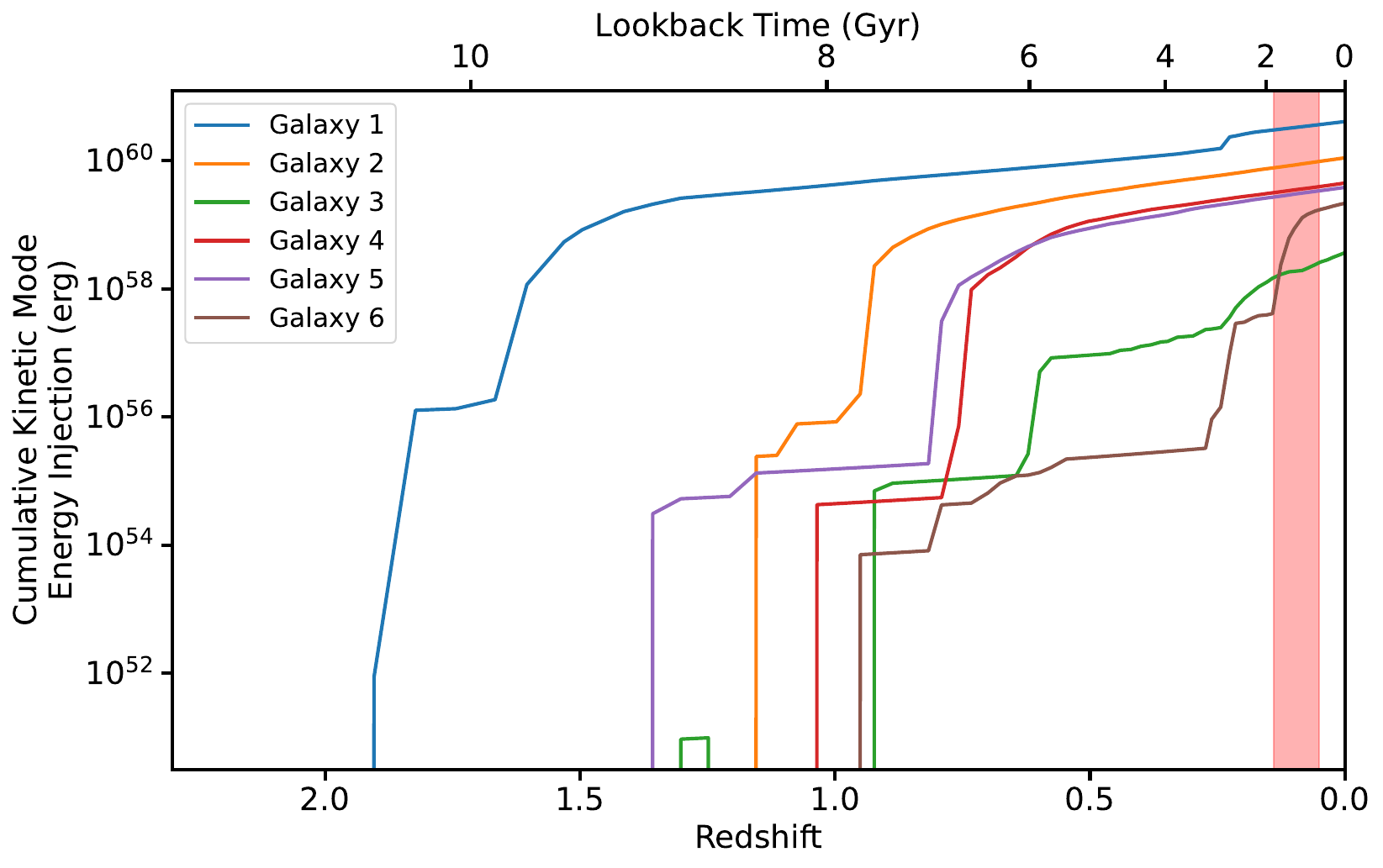}
\caption{History of the cumulative energy imparted to the gas in the form of kinetic AGN feedback (as implemented in the TNG model) as a function of redshift for the six galaxies in our sample. The pink band at later epochs shows the range of redshifts shown in Figure \ref{fig:gal6_evol}.\label{fig:RM_feedback}}
\end{figure*}

Figure \ref{fig:RM_feedback} shows the cumulative kinetic feedback energy imparted to the gas from the central SMBH for all six galaxies in our sample. All of the galaxies in the sample exhibit a particular epoch when the kinetic feedback ``turns on'' and increases rapidly, but in most of them this occurs at a redshift of 0.5 or earlier. In the case of galaxy 6, the cumulative kinetic feedback energy increases rapidly at very late times. The plot highlights in pink the approximate range of epochs spanned by the snapshots shown in Figure \ref{fig:gal6_evol}; this corresponds directly to a strong increase in the kinetic feedback energy imparted to the gas.

\end{document}